\documentclass{aa}

\usepackage{natbib}
\bibpunct{(}{)}{;}{a}{}{,}
\usepackage{array}
\usepackage{newtxtext,newtxmath}
\usepackage[T1]{fontenc}
\usepackage{ae,aecompl}
\usepackage{graphicx}
\usepackage{amsmath}
\usepackage{amssymb}
\usepackage{lscape}
\usepackage{comment}
\usepackage{adjustbox}
\usepackage{morefloats}
\usepackage{tabularx}
\usepackage{endnotes}

\begin{document}

\title{The XXL Survey}
\subtitle{XXXVI. Evolution and black hole feedback of high-excitation and low-excitation radio galaxies in XXL-S}

\author{Andrew Butler\inst{\ref{inst1}}\thanks{E-mail: andrew.butler@icrar.org} \and Minh Huynh\inst{\ref{inst1},\ref{inst2}}
\and Anna Kapi\'{n}ska\inst{\ref{inst1},\ref{inst3}}
\and Ivan Delvecchio\inst{\ref{inst4}}
\and Vernesa Smol\v{c}i\'{c}\inst{\ref{inst4}}
\and Lucio Chiappetti\inst{\ref{inst5}}
\and Elias Koulouridis\inst{\ref{inst6},\ref{inst7}}
\and Marguerite Pierre\inst{\ref{inst7}}}

\institute{International Centre for Radio Astronomy Research (ICRAR), University of Western Australia, 35 Stirling Hwy, Crawley WA 6009, Australia\label{inst1}
\and
CSIRO Astronomy and Space Science, 26 Dick Perry Ave, Kensington WA 6151, Australia\label{inst2}
\and
National Radio Astronomy Observatory, 1003 Lopezville Rd, Socorro NM 87801, USA\label{inst3}
\and
Physics Department, University of Zagreb, Bijeni\v{c}ka cesta 32, 10002 Zagreb, Croatia\label{inst4}
\and
INAF, IASF Milano, via Corti 12, 20133 Milano, Italy\label{inst5}
\and
Institute for Astronomy \& Astrophysics, Space Applications \& Remote Sensing, National Observatory of Athens, GR-15236 Palaia Penteli, Greece\label{inst6}
\and
AIM, CEA, CNRS, Universit\'{e} Paris-Saclay, Universit\'{e} Paris Diderot, Sorbonne Paris Cit\'{e}, F-91191 Gif-sur-Yvette, France\label{inst7}}

\date{Received date / Accepted date }

\abstract{The evolution of the comoving kinetic luminosity densities ($\Omega_{\rm{kin}}$) of the radio loud high-excitation radio galaxies (RL HERGs) and the low-excitation radio galaxies (LERGs) in the ultimate XMM extragalactic survey south (XXL-S) field is presented.  The wide area and deep radio and optical data of XXL-S have allowed the construction of the radio luminosity functions (RLFs) of the RL HERGs and LERGs across a wide range in radio luminosity out to high redshift ($z = 1.3$). The LERG RLFs display weak evolution: $\Phi (z)$ $\propto$ (1+$z$)$^{0.67 \pm 0.17}$ in the pure density evolution (PDE) case and $\Phi (z)$ $\propto$ (1+$z$)$^{0.84 \pm 0.31}$ in the pure luminosity evolution (PLE) case. The RL HERG RLFs demonstrate stronger evolution than the LERGs: $\Phi (z)$ $\propto$ (1+$z$)$^{1.81 \pm 0.15}$ for PDE and $\Phi (z)$ $\propto$ (1+$z$)$^{3.19 \pm 0.29}$ for PLE. Using a scaling relation to convert the 1.4 GHz radio luminosities into kinetic luminosities, the evolution of $\Omega_{\rm{kin}}$ was calculated for the RL HERGs and LERGs and compared to the predictions from various simulations. The prediction for the evolution of radio mode feedback in the Semi-Analytic Galaxy Evolution (SAGE) model is consistent with the $\Omega_{\rm{kin}}$ evolution for all XXL-S RL AGN (all RL HERGs and LERGs), indicating that the kinetic luminosities of RL AGN may be able to balance the radiative cooling of the hot phase of the IGM. Simulations that predict the $\Omega_{\rm{kin}}$ evolution of LERG equivalent populations show similar slopes to the XXL-S LERG evolution, suggesting that observations of LERGs are well described by models of SMBHs that slowly accrete hot gas. On the other hand, models of RL HERG equivalent populations differ in their predictions. While LERGs dominate the kinetic luminosity output of RL AGN at all redshifts, the evolution of the RL HERGs in XXL-S is weaker compared to what other studies have found.  This implies that radio mode feedback from RL HERGs is more prominent at lower redshifts than was previously thought.}


\titlerunning{The XXL Survey XXXVI: Evolution and black hole feedback of HERGs and LERGs in XXL-S}
\authorrunning{Butler et al.}
\maketitle


\section{Introduction}
\label{sec:intro}

Understanding how massive galaxies evolve is an important topic in modern astrophysics.  Massive galaxies make up a large fraction of the total baryonic matter in the universe, and therefore their evolution reflects how the universe as a whole has evolved.  It is now commonly understood that nearly all massive galaxies have supermassive black holes (SMBHs) at their centres (e.g. \citealp{kormendy2013}).  Furthermore, the properties of SMBHs are related to the properties of their host galaxies.  For example, the masses of SMBHs are correlated with the stellar velocity dispersions (e.g. \citealp{magorrian1998,gebhardt2000,graham2008}) and the stellar masses (e.g. Marconi \& Hunt 2003; Haring \& Rix 2004) of the bulges of their host galaxies. In addition, \cite{shankar2009} discovered that the growth curve of black holes and that of stellar mass in galaxies have the same shape. These findings indicate that the evolution of galaxies and SMBHs are closely linked, and therefore SMBHs play an important role in massive galaxy evolution.  In particular, active SMBHs, commonly referred to as active galactic nuclei (AGN), have been recognised as having a major influence on massive galaxy evolution via a process called feedback (e.g. \citealp{bohringer1993,forman2005,fabian2012}).  This feedback is the most likely cause of the link between SMBHs and their host galaxy properties because outflows from the AGN can heat the interstellar medium, which would otherwise collapse to form stars \citep{bohringer1993,binney1995,forman2005,best2006,mcnamara2007,cattaneo2009a,fabian2012}.  Consequently, AGN can affect the stellar and gas content of their host galaxies, fundamentally altering their properties.

AGN feedback is often thought of as existing in two forms: `quasar' mode and `radio' mode \citep{croton2006}.  The quasar mode involves radiatively efficient accretion and feedback in the form of radiative winds, whereas the radio mode involves radiatively inefficient accretion and feedback in the form of radio jets that carry kinetic energy (\citealp{best2012} and references therein).  Radio mode feedback has been identified as the most likely mechanism behind the heating of the interstellar medium because galaxy formation models that include this extra AGN component are able to more accurately reproduce many observed galaxy properties for $z \leq 0.2$ (particularly at the high-mass end), including the optical luminosity function, colours, stellar ages, and morphologies (e.g. \citealp{bower2006,croton2006,croton2016}). Therefore, AGN feedback, and in particular radio mode feedback, is a crucial component to galaxy evolution models and fundamental to overall galaxy evolution. This is likely due to the fact that most of the energy from the radio jets is deposited locally in the systems that generate them, increasing the feedback efficiency compared to quasar mode feedback (\citealp{bohringer1993}; Carilli, Perley \& Harris 1994; McNamara et al. 2000; Fabian et al. 2006).  

Radio mode feedback has been found to manifest in two different AGN populations -- high-excitation radio galaxies (HERGs) and low-excitation radio galaxies (LERGs).  HERGs and LERGs are characterised by different host galaxy properties. HERGs exhibit either strong [\ion{O}{III}] emission (e.g. \citealp{best2012}; \citealp{hardcastle2013a}), high X-ray luminosity ($L_{\rm{X}} > 10^{42}$ erg s$^{-1}$, e.g. \citealp{xue2011,juneau2011}), or redder mid-infrared colours (e.g. \citealp{jarrett2011,mateos2012}) than normal galaxies.  They also tend to have higher radio luminosities (e.g. \citealp{best2012,heckman2014}), are hosted by less massive bluer galaxies (e.g. \citealp{tasse2008,janssen2012,best2012,hardcastle2013a,miraghaei2017,ching2017a}), and fit into the unified AGN model as summarised by \cite{urry1995}.  The dominant form of feedback in HERGs is the quasar mode, but a small fraction of HERGs are radio loud, and therefore they exhibit some radio mode feedback.  This is manifested in some HERGs having red colours that are consistent with passively-evolving galaxies (e.g. \citealp{ching2017a,butler2018_xxl31}, hereafter XXL Paper XXXI).   On the other hand, LERGs show weak or no [\ion{O}{III}] emission (e.g. \citealp{hine1979,laing1994,jackson1997}) and little to no evidence of accretion-related X-ray or MIR emission typical of a conventional AGN (e.g. \citealp{hardcastle2006,hardcastle2009,mingo2014,gurkan2014,ching2017a}), and therefore do not fit into the unified AGN model.  They also have lower radio luminosities and are hosted by more massive redder galaxies (e.g. \citealp{janssen2012,best2012,heckman2014,miraghaei2017,ching2017a}).  LERGs are identified as AGN only at radio wavelengths \citep{hickox2009}, and thus only exhibit radio mode feedback. It has been hypothesised that HERG and LERG differences are driven by a split in their Eddington-scaled accretion rates (e.g. \citealp{best2005a,hardcastle2018b}). LERGs tend to accrete the hot X-ray emitting phase of the intergalactic medium at a rate less than $\sim$1-3\% of Eddington, while HERGs tend to accrete the cold phase at higher accretion rates (e.g. \citealp{narayan1994,narayan1995a,narayan1995b,hardcastle2007,trump2009b,best2012,heckman2014}; \citealp{mingo2014}).  This hypothesis can be used to generally explain their different host galaxy properties, environments, rates of evolution, and the agreement between the energy required to heat cooling flows and the power output of low-luminosity radio galaxies \citep{allen2006,best2006,hardcastle2006,hardcastle2007}.

However, the precise origin of HERG and LERG differences remains unclear. In order to understand the physical driver for their differences and the role of the radio mode feedback in galaxy evolution as a function of time, a full understanding of the HERG and LERG luminosity functions, host galaxies, and cosmic evolution is needed  (\citealp{best2012} and references therein). It is crucial that the evolution of radio mode feedback, and in particular the relative contribution of the LERG and HERG populations to the total radio power emitted at a given epoch, is accurately measured.  The essential tool for measuring this quantity is the radio luminosity function (RLF), which is the most direct and accurate way to measure the cosmic evolution of radio sources (e.g. \citealp{mauch2007}).  The RLF is, for a complete sample of radio sources, the volume density as a function of radio luminosity at a given cosmological epoch \citep{longair1966}.  If RLFs are constructed for different redshift ranges (cosmological epochs), the evolution of these RLFs can be modelled.  In turn, the evolution of the RLF directly measures the changes in volume density in a given population as a function of radio luminosity and redshift.  The contribution of LERGs and HERGs to radio mode feedback, at a given epoch, can be measured by converting the RLFs into kinetic luminosity functions.  This can be done via a scaling relation between the monochromatic radio luminosity and kinetic luminosity (e.g. \citealp{willott1999,cavagnolo2010}) or via dynamical models of radio source evolution (e.g. \citealp{raouf2017,turner2018a,hardcastle2018a,hardcastle2018c}), from which the radio jet powers can be inferred using the radio luminosities and projected linear sizes of the sources.  In this way, the evolution in the HERG and LERG RLFs has a direct impact on their contribution to radio mode feedback throughout cosmic time.

Only a few studies have constructed separate RLFs for HERGs and LERGs and calculated the corresponding radio mode feedback evolution.  The RLF evolution at 1.4 GHz measured by \cite{best2014} and \cite{pracy2016} using FIRST, NVSS, and SDSS data indicate that, for $z \lesssim 1$, HERGs evolve strongly and LERGs exhibit volume densities that are consistent with weak or no evolution.  On the other hand, \cite{williams2018} constructed HERG and LERG RLFs using 150 MHz LOFAR observations of the $\sim$9.2 deg$^2$ Bo\"{o}tes field, and found that the HERG RLFs were consistent with no evolution and the LERG RLFs exhibit negative evolution from $z = 0.5$ to $z = 2$. This demonstrates that separating between LERGs and HERGs is important not only because the two populations have different host galaxies, but because they make different contributions to radio mode feedback at different times and at different observing frequencies.  These different contributions can be linked to the environments of HERGs and LERGs and the different origins of their fuelling gas (e.g. \citealp{ching2017b}), which in turn can be used to constrain the AGN jet launching mechanism and its dependence on accretion mode, which are poorly understood (e.g. \citealp{romero2017} and references therein).

The radio data in the 1.4 GHz studies probed no deeper than $S_{\rm{1.4GHz}} \sim 3$ mJy (over very large areas of $\gtrsim$800 deg$^2$), and so the RLFs are not well-constrained at the low-luminosity end ($L_{\rm{1.4GHz}} \lesssim 10^{24}$ W Hz$^{-1}$) for intermediate to high redshifts ($z > 0.3$).  In addition, the local HERG and star-forming galaxy (SFG) RLFs from these papers disagree with each other for $L_{\rm{1.4GHz}} \lesssim 10^{24}$ W Hz$^{-1}$, indicating that these two populations can be difficult to discriminate at low radio luminosities.  More clarity on this discrepancy requires a deep radio survey over a relatively wide area combined with excellent multi-wavelength data in order to capture the largest possible range of radio luminosities out to $z\sim1$.  In light of this, the 25 deg$^2$ ultimate XMM extragalactic survey (\citealp{pierre2016_xxl1}; XXL Paper I) south field (hereafter XXL-S) was observed with the Australia Telescope Compact Array (ATCA) at 2.1 GHz, achieving a median rms sensitivity of $\sigma$$\approx$41 $\mu$Jy beam $^{-1}$ and a resolution of $\sim$5" (\citealp{butler2018_xxl18}, hereafter XXL Paper XVIII; \citealp{smolcic2016_xxl11}, hereafter XXL Paper XI).  Due to the size and depth of XXL-S, rare luminous objects not found in other fields have been captured and a large population of low-luminosity AGN have been detected simultaneously.  The large area and depth of the radio observations of XXL-S, combined with the excellent multi-wavelength coverage, enables the construction of the RL HERG and LERG RLFs in multiple redshift bins.  It also enables the bright and faint end of the RLF to be probed over a large redshift range, which has been difficult thus far due to small sky coverage (e.g. \citealp{smolcic2009a,smolcic2017b}) or shallow radio observations of previous surveys \citep{best2012,best2014,pracy2016}.  This new capability allows for a new measurement of the cosmic evolution of the radio mode feedback of RL HERGs and LERGs out to high redshift ($z$$\sim$1) that includes a more complete sampling of the radio luminosity distribution of the two populations.

The purpose of this paper is to measure the evolution of the kinetic luminosity densities of the RL HERGs and LERGs in XXL-S and compare the results to the literature, particularly simulations of radio mode feedback.  Section \ref{sec:data} summarises the data used, while Section \ref{sec:RLFs} describes the construction of the RLFs and the comparison to other RLFs in the literature. The measurement of the evolution of the RL HERG and LERG RLFs is discussed in Section \ref{sec:RL_HERG_LERG_evolution}.  Section \ref{sec:IKLDs} details the calculations involved in measuring the RL HERG and LERG kinetic luminosity densities and compares the results to the literature.  Section \ref{sec:conclusion} draws the conclusions. Throughout this paper, the following cosmology is adopted: $H_0 = 69.32$ km s$^{-1}$ Mpc$^{-1}$, $\Omega_{\rm{m}} = 0.287$ and $\Omega_{\Lambda} = 0.713$ \citep{hinshaw2013}.  The following notation for radio spectral index ($\alpha_R$) is used: $S_{\nu} \propto \nu^{\alpha_R}$.

\section{Data}
\label{sec:data}

\subsection{Radio data}

The ATCA 2.1 GHz radio observations of XXL-S reached a median rms of $\sigma$ $\approx$ 41 $\mu$Jy beam$^{-1}$ and a resolution of $\sim$4.8$''$ over 25 deg$^2$.  The number of radio sources extracted above 5$\sigma$ is 6287.  More details of the observations, data reduction and source statistics can be found in XXL Paper XVIII and XXL Paper XI.

\subsection{Cross-matched sample and radio source classifications}
\label{sec:cm_radio_sources}

Out of the 6287 radio sources in the XXL-S catalogue, 4758 were cross-matched to reliable optical counterparts in the XXL-S multi-wavelength catalogue (\citealp{fotopoulou2016_xxl6}; XXL Paper VI) via the likelihood ratio method (\citealp{ciliegi2018_xxl26}; XXL Paper XXVI).  For a discussion of how extended radio sources (for which maximum likelihood methods tend to fail) were treated, see Sections 3.6 and 3.7 in XXL Paper XVIII.  XXL Paper XXXI describes the classification of the 4758 optically-matched radio sources as LERGs, radio loud (RL) HERGs, radio quiet (RQ) HERGs, and star-forming galaxies (SFGs), but some sources (including radio AGN) are unclassified because of a lack of data available for those sources.  In this context, RL HERGs are defined as high-excitation sources with radio emission originating from an AGN, and RQ HERGs are defined as high-excitation sources with radio emission that likely originates from star formation, although there could be some contribution from radio AGN in these sources.
The definition of `radio galaxy' one adopts (whether a galaxy with radio emission from an AGN or a galaxy with detectable radio emission arising from either AGN or star formation) has no bearing on the results of this paper, as RQ HERGs were removed from the RL AGN sample (comprised of RL HERGs and LERGs).  Once the HERGs were identified, the LERGs were separated from the SFGs on the basis of optical spectra, colours, and radio AGN indicators, particularly their radio excesses (the ratio of 1.4 GHz radio luminosity to SFR derived by \textsc{magphys}). See Section 3.7 of XXL Paper XXXI for an overview of the decision tree used to classify the XXL-S radio sources.
Table \ref{tab:class_results} summarises the number of sources classified into each source type.  Tables \ref{tab:example_class_entries1} and \ref{tab:example_class_entries2} display the list of columns in the catalogue containing the optically-matched XXL-S radio sources and the full suite of their radio and associated multi-wavelength data (see XXL Paper XXXI).  The catalogue is available as a queryable database table XXL\_ATCA\_16\_class via the XXL Master Catalogue browser\footnote{http://cosmosdb.iasf-milano.inaf.it/XXL}.  A copy will also be deposited at the Centre de Donn\'{e}s astronomiques de Strasbourg (CDS)\footnote{http://cdsweb.u-strasbg.fr}.

\begin{table}
\centering
\caption[Results of classification of XXL-S radio sources.]{Results of classification of XXL-S radio sources from
XXL Paper XXXI.  Unclassified AGN potentially include LERGs and RL HERGs, while unclassified sources potentially include LERGs, RQ HERGs, and SFGs.}
\begin{tabular}{c c c}
Source type & Number & Fraction of final sample \\
\hline
\hline
LERGs & 1729 & 36.3\% \\
RL HERGs & 1159 & 24.4\% \\
RQ HERGs & 296 & 6.2\% \\
SFGs & 558 & 11.7\% \\
Unclassified AGN & 910 & 19.1\% \\
Unclassified sources & 106 & 2.2\% \\
\end{tabular}
\label{tab:class_results}
\end{table}

\begin{table*}
\centering
\caption[Columns 1-38 in the catalogue containing the optically-matched XXL-S radio sources and the full suite of their radio and associated multi-wavelength data.]{Columns 1-38 in the catalogue containing the optically-matched XXL-S radio sources and the full suite of their radio and associated multi-wavelength data (see XXL Paper XXXI). The table is available online (see text for details). The catalogue help file explains each quantity and their possible values.}
\begin{tabular}{l l c}
Quantity & Description & Units\\
\hline
\hline
IAU name & IAU-registered radio source numeric identifier & --\\
ID & XXL-S radio source catalogue identification number & --\\
RAdeg & Right ascension (J2000) & deg\\
DEdeg & Declination (J2000) & deg\\
redshift & Final redshift of radio source & --\\
zspec\_flag & Spectroscopic redshift flag & --\\
classification & Final source classification & --\\
agn\_radio\_L\_flag & Radio AGN (luminosity) flag & --\\
agn\_radio\_morph\_flag & Radio AGN (morphology) flag & --\\
agn\_radio\_alpha\_flag & Radio AGN (spectral index) flag & --\\
agn\_radio\_excess\_flag & Radio AGN (radio excess) flag & --\\
agn\_xray\_L\_flag & X-ray AGN (luminosity) flag & --\\
agn\_xray\_HR\_flag & X-ray AGN (hardness ratio) flag & --\\
agn\_sed\_flag & SED AGN flag & --\\
agn\_IRAC1\_IRAC2\_flag & MIR AGN (IRAC1+IRAC2) flag & --\\
agn\_W1\_W2\_flag & MIR AGN (W1+W2) flag & --\\
agn\_W1\_W2\_W3\_flag & MIR AGN (W1+W2+W3) flag & --\\
agn\_W1\_W2\_W3\_W4\_flag & MIR AGN (W1+W2+W3+W4) flag & --\\
agn\_bpt\_flag & BPT AGN flag & --\\
agn\_OIII\_flag & [\ion{O}{III}] AGN flag & --\\
agn\_spec\_temp\_flag & AGN spectral template flag & --\\
Sp\_1400MHz & 1.4 GHz peak flux density & mJy beam$^{-1}$\\
S\_1400MHz & 1.4 GHz integrated flux density & mJy\\
SNR\_1400MHz & 1.4 GHz $S/N$ & --\\
alpha\_R & Radio spectral index & --\\
alpha\_R\_err & Radio spectral index error & --\\
L\_R\_1800MHz & 1.8 GHz radio luminosity & W Hz$^{-1}$\\
L\_R\_1400MHz & 1.4 GHz radio luminosity & W Hz$^{-1}$\\
alpha\_X & X-ray spectral index & --\\
gamma\_X & X-ray photon index & --\\
Xray\_HR & X-ray hardness ratio & --\\
L\_X\_2\_10\_keV & 2-10 keV X-ray luminosity & erg s$^{-1}$\\
W1WISEmag & WISE $W1$ apparent magnitude & mag\\
W2WISEmag & WISE $W2$ apparent magnitude & mag\\
W3WISEmag & WISE $W3$ apparent magnitude & mag\\
W4WISEmag & WISE $W4$ apparent magnitude & mag\\
IRAC1mag & $IRAC1$ (3.6 $\mu$m) apparent magnitude & mag\\
IRAC2mag & $IRAC2$ (4.5 $\mu$m) apparent magnitude & mag\\
\end{tabular}
\label{tab:example_class_entries1}
\end{table*}

\begin{table*}
\centering
\caption[Columns 39-83 in the catalogue containing the optically-matched XXL-S radio sources and the full suite of their radio and associated multi-wavelength data.]{Columns 39-83 in the catalogue containing the optically-matched XXL-S radio sources and the full suite of their radio and associated multi-wavelength data (see XXL Paper XXXI). The table is available online (see text for details). The catalogue help file explains each quantity and their possible values.}
\begin{tabular}{l l c}
Quantity & Description & Units\\
\hline
\hline
NUVGALEXMag & NUV (GALEX) absolute magnitude & mag\\
uSDSSMag & $u$ (SDSS) absolute magnitude & mag\\
gBCSMag & $g$ (BCS) absolute magnitude & mag\\
gDECamMag & $g$ (DECam) absolute magnitude & mag\\
rDECamMag & $r$ (DECam) absolute magnitude & mag\\
iBCSMag & $i$ (BCS) absolute magnitude & mag\\
iDECamMag & $i$ (DECam) absolute magnitude & mag\\
zDECamMag & $z$ (DECam) absolute magnitude & mag\\
JVISTAMag & $J$ (VISTA) absolute magnitude & mag\\
KVISTAMag & $K$ (VISTA) absolute magnitude & mag\\
iDECammag & $i$ (DECam) apparent magnitude & mag\\
zDECammag & $z$ (DECam) apparent magnitude & mag\\
sfr\_magphys\_med & Median \textsc{magphys} star formation rate & M$_{\odot}$ yr$^{-1}$\\
stellar\_mass & \textsc{magphys} stellar mass & M$_{\odot}$\\
zphot & Photometric redshift & --\\
opt\_spectrum\_ID & Optical spectrum ID & --\\
zspec & Spectroscopic redshift & --\\
zspec\_qual\_flag & Spectroscopic redshift quality flag & --\\
SNR\_cont & Continuum $S/N$ of optical spectrum & --\\
spectral\_template & Best fit \textsc{marz} spectral template & --\\
OII\_lambda & [\ion{O}{II}] line wavelength & \AA\\
OII\_rel\_flux & [\ion{O}{II}] line relative flux & --\\
OII\_SNR & [\ion{O}{II}] line $S/N$ & --\\
OII\_EW & [\ion{O}{II}] line equivalent width & m\AA\\
OII\_EW\_err & [\ion{O}{II}] line equivalent width error & m\AA\\
Hb\_lambda & H$\beta$ line wavelength & \AA\\
Hb\_rel\_flux & H$\beta$ line relative flux & --\\
Hb\_SNR & H$\beta$ line $S/N$ & --\\
Hb\_EW & H$\beta$ line equivalent width & m\AA\\
Hb\_EW\_err & H$\beta$ line equivalent width error & m\AA\\
OIII\_lambda & [\ion{O}{III}] line wavelength & \AA\\
OIII\_rel\_flux & [\ion{O}{III}] line relative flux & --\\
OIII\_SNR & [\ion{O}{III}] line $S/N$ & --\\
OIII\_EW & [\ion{O}{III}] line equivalent width & m\AA\\
OIII\_EW\_err & [\ion{O}{III}] line equivalent width error & m\AA\\
Ha\_lambda & H$\alpha$ line wavelength & \AA\\
Ha\_rel\_flux & H$\alpha$ line relative flux & --\\
Ha\_SNR & H$\alpha$ line $S/N$ & --\\
Ha\_EW & H$\alpha$ line equivalent width & m\AA\\
Ha\_EW\_err & H$\alpha$ line equivalent width error & m\AA\\
NII\_lambda & [\ion{N}{II}] line wavelength & \AA\\
NII\_rel\_flux & [\ion{N}{II}] line relative flux & --\\
NII\_SNR & [\ion{N}{II}] line $S/N$ & --\\
NII\_EW & [\ion{N}{II}] line equivalent width & m\AA\\
NII\_EW\_err & [\ion{N}{II}] line equivalent width error & m\AA\\
\end{tabular}
\label{tab:example_class_entries2}
\end{table*}

\section{Radio luminosity functions}
\label{sec:RLFs}

\subsection{Construction}
\label{sec:RLF_construction}

The RLFs were constructed using the $1/V_{\rm{max}}$ method \citep{schmidt1968}, which is summarised here.  The maximum distance out to which each source can be detected before it falls below the detection limit of the ATCA XXL-S radio survey was calculated according to 
\begin{equation}
d_{\rm{max}} = d_{\rm{src}} \sqrt{\frac{(S/N)_{\rm{src}}}{(S/N)_{\rm{det}}}},
\end{equation}
where $d_{\rm{src}}$ is the comoving distance of the source at its redshift, $(S/N)_{\rm{src}}$ is the source's $S/N$ at 1.8 GHz (the effective detection frequency) and $(S/N)_{\rm{det}} = 5$ is the 1.8 GHz detection limit.  The corresponding maximum volume $V_{\rm{max}}$ that the source can occupy was calculated via
\begin{equation}
V_{\rm{max}} = \Omega_{\rm{frac}} \frac{4}{3} \pi (d_{\rm{max}}^3 - d_{\rm{min}}^3)
\end{equation}
where $\Omega_{\rm{frac}} \approx 5.579 \times 10^{-3}$ is the fraction of the whole sky that XXL-S covers and $d_{\rm{min}}$ is the comoving distance corresponding to the lower redshift limit of the redshift bin the source is contained in.  It is common practice to also account for the limiting optical magnitude in determining $V_{\rm{max}}$, but including the optical $V_{\rm{max}}$ in the calculation resulted in almost no difference to the RLFs, especially after the $M_i < -22$ optical cut was made to measure the evolution of the RL HERGs and LERGs (see Section \ref{sec:optical_selection}).  Therefore, only the radio $V_{\rm{max}}$ was considered.

For comparison with the literature, the rest-frame 1.4 GHz monochromatic luminosity densities (hereafter luminosities) of each source were calculated.  This was done by converting each 1.8 GHz flux density into a 1.4 GHz flux density ($S_{\rm{1.4GHz}}$) using the radio spectral index $\alpha_R$ for each source (see Section 2.4.3 and Appendix A of XXL Paper XXXI for details).  Then the 1.4 GHz luminosity of each source was computed with the following equation: 
\begin{equation}
\label{eq:L_R}
L_{\rm{1.4GHz}} = 4 \pi d_L^2 S_{\rm{1.4GHz}} (1+z)^{-(1+\alpha_R)},
\end{equation}
where $d_L$ is the luminosity distance in metres and $z$ is the source's best redshift (spectroscopic if available, otherwise photometric).

Each radio source was placed in its corresponding redshift bin, all four of which are listed in Table \ref{tab:z_bins}.  
An upper limit of $z = 1.3$ was chosen for three reasons: 1) the majority of the positive evolution in RL AGN takes place between $0 < z < 1.3$ (e.g. see \citealp{smolcic2017b}, \citealp{ceraj2018}); 2) it allows a more direct comparison between the RLFs of \cite{smolcic2009a} and \cite{smolcic2017b}; 3) almost all ($\sim$93.2\%) of the spectroscopic redshifts available for XXL-S are associated with sources at $z < 1.3$.

In each redshift bin, every source was placed in its corresponding $L_{\rm{1.4GHz}}$ bin, which are $10^{0.4}$ W Hz$^{-1}$ (1 magnitude) wide.  The volume density per $L_{\rm{1.4GHz}}$ bin, $\Phi$, is then:
\begin{equation}
\label{eq:Phi}
\Phi = \sum_{i=1}^N \frac{f_i}{V_{\rm{max}, i}},
\end{equation}
where the sum is over all $N$ galaxies in the $L_{\rm{1.4GHz}}$ bin and $f_i$ is the radio completeness correction factor.  If source had a peak flux density $S_{\rm{p}} < 0.92$ mJy (the flux density regime exhibiting less than $\sim$100\% completeness), $f_i$ was calculated as the inverse of the completeness fraction at the source's flux density, as shown in Figure 11 of XXL Paper XVIII. Otherwise, $f_i = 1$.  The uncertainty in $\Phi$ was calculated according to
\begin{equation}
d\Phi = \sqrt{\sum_{i=1}^N \left(\frac{f_i}{V_{\rm{max},i}}\right)^2}.
\end{equation}

\begin{table}
\centering
\caption[Redshift bins chosen for the XXL-S RLFs.]{Redshift bins chosen for the XXL-S RLFs.  See Section \ref{sec:RLF_construction} for justification of the $z = 1.3$ upper limit.}
\begin{tabular}{c c c c}
Redshift & Redshift  & Median & Number\\
bin number & range & redshift & of sources\\
\hline
\hline
1 & $0.0 < z < 0.3$ & 0.15 & 833\\
2 & $0.3 < z < 0.6$ & 0.45 & 1021\\
3 & $0.6 < z < 0.9$ & 0.75 & 795\\
4 & $0.9 < z < 1.3$ & 1.1 & 840\\
\hline
\end{tabular}
\label{tab:z_bins}
\end{table}

\subsection{XXL-S local (0 < $z$ < 0.3) 1.4 GHz RLFs}

Using the classifications of the radio sources from XXL Paper XXXI, the RLFs were initially constructed for the following source types: all radio sources, all RL AGN (LERGs plus RL HERGs), and SFGs (which includes RQ HERGs because the dominant source of radio emission is likely to be star formation).

\begin{figure}
        \includegraphics[width=\columnwidth]{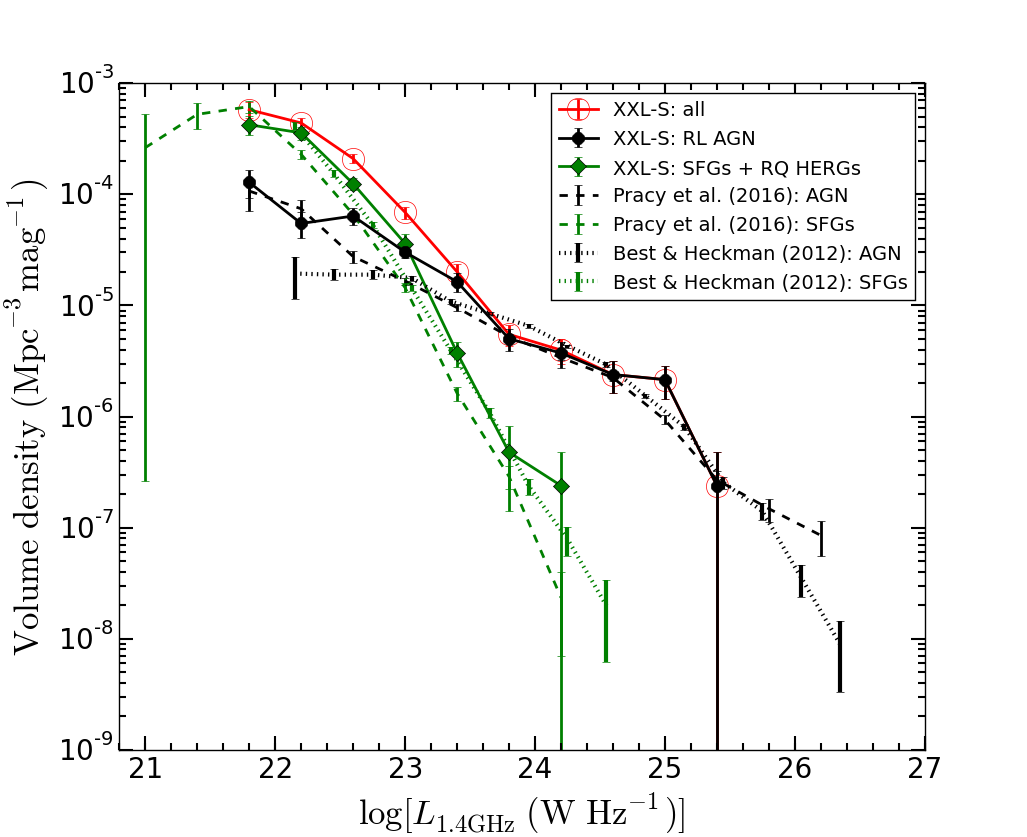}
    \caption[1.4 GHz XXL-S RLFs for all sources, radio loud (RL) AGN, and SFGs in the local universe (0 < $z$ < 0.3).]{1.4 GHz XXL-S RLFs for all sources (red open circles), RL AGN (filled black circles), and SFGs (green diamonds) in the local universe (0 < $z$ < 0.3).  For comparison, the AGN (black lines) and SFG (green lines) RLFs from Pracy16 (long-dashed lines) and \cite{best2012} (short-dashed lines) are shown.}
    \label{fig:rlf_z_bin_1_all_plot}
\end{figure}

Figure \ref{fig:rlf_z_bin_1_all_plot} shows the 1.4 GHz RLF for all radio sources, all RL AGN, and SFGs (including RQ HERGs) in the local universe (0 < $z$ < 0.3) for XXL-S (the data are displayed in Table \ref{tab:rlf_z_bin_1_data}).  Unclassified sources (potential LERGs, RQ HERGs, and SFGs) were ignored, as they form an insignificant population of the optically-matched radio sources in XXL-S (2.2\%) and they are not identifiable as RL AGN, which is the focus of this paper.  Therefore, the XXL-S RL AGN and SFG RLFs plotted in Figure \ref{fig:rlf_z_bin_1_all_plot} represent lower limits.  

\begin{figure}
        \includegraphics[width=\columnwidth]{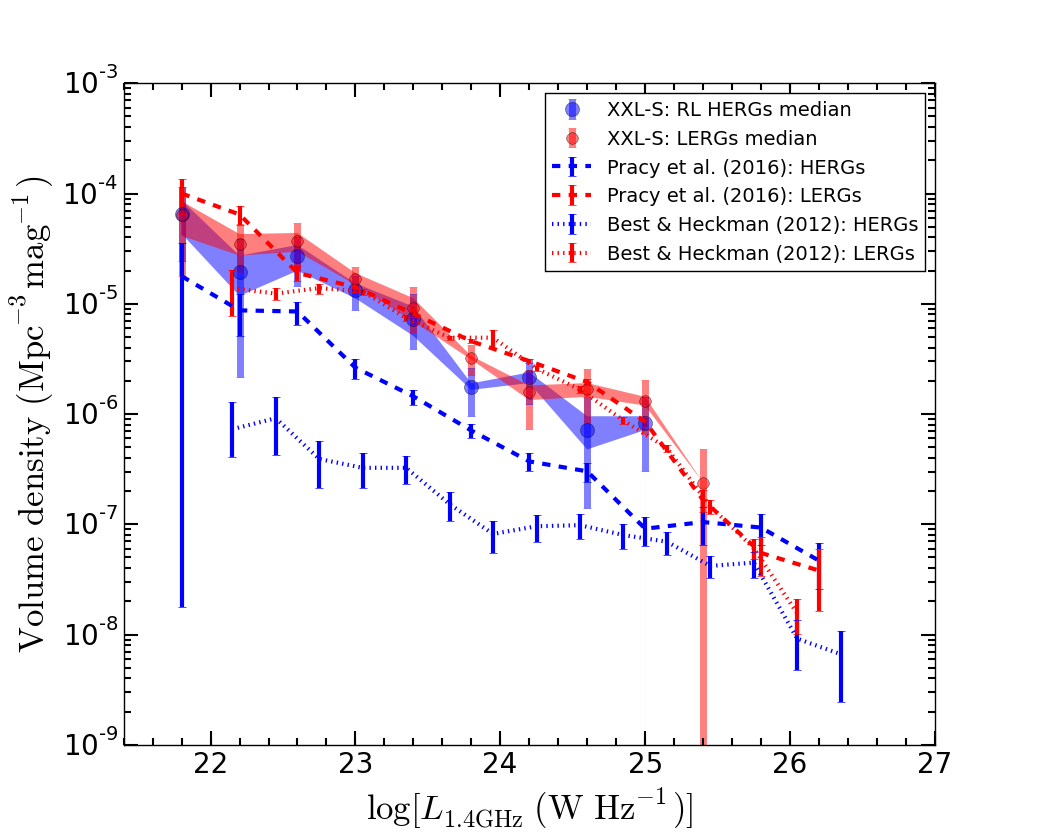}
    \caption[1.4 GHz XXL-S RLFs for RL HERGs and LERGs in the local universe (0 < $z$ < 0.3).]{1.4 GHz XXL-S RLFs for RL HERGs (blue shaded region) and LERGs (red shaded region) in the local universe (0 < $z$ < 0.3).  For comparison, the HERG and LERG RLFs from Pracy16 (long-dashed lines) and \cite{best2012} (short-dashed lines) are shown.}
    \label{fig:rlf_z_bin_1_sep_plot}
\end{figure}

The RL AGN in XXL-S were then separated into RL HERGs, LERGs, and unclassified RL AGN (potential LERGs and RL HERGs). The unclassified RL AGN are a result of a lack of sufficient data that would allow a definite classification.  Since these unclassified RL AGN comprise a large fraction ($\sim$24\%) of the total XXL-S RL AGN population, they significantly contribute to the RL AGN RLF.  Therefore, these unclassified RL AGN were added to both the RL HERG and LERG populations as a way of probing the full possible range of the RL HERG and LERG RLFs.  Figure \ref{fig:rlf_z_bin_1_sep_plot} shows the 1.4 GHz RLF for XXL-S RL HERGs and LERGs in the local universe (0 < $z$ < 0.3).  The blue and red shaded regions represent the range of values the RLFs for RL HERGs and LERGs could possibly have assuming 100\% of the unclassified RL AGN are added to each population.  The blue circles represent, at a given radio luminosity, the median values of $\Phi$ between the definite RL HERG RLF and the RLF that results from combining the definite RL HERGs with all unclassified RL AGN.  The red circles represent the equivalent for LERGs.  The upper error bars represent the upper extrema of the error bars from the RL HERG and LERG plus all unclassified RL AGN RLFs, and the lower error bars represent the lower extrema of the error bars from the definite RL HERG and LERG RLFs.  The blue and red circles, with their corresponding error bars, have been chosen as the data points that represents the final RL HERG and LERG RLFs, respectively.  These data are shown in Table \ref{tab:rlf_z_bin_1_data}.

\begin{table*}
\centering
\caption[RLF data for all XXL-S sources, SFGs (including RQ HERGs), all RL AGN, RL HERGs, and LERGs in the local universe (0 < $z$ < 0.3).] {RLF data for all XXL-S sources, SFGs (including RQ HERGs), all RL AGN, RL HERGs, and LERGs in the local universe (0 < $z$ < 0.3). log($L_{\rm{1.4GHz}}$) represents the median value for each 1.4 GHz radio luminosity bin (which are 0.4 dex, or 1 mag, wide), $N$ is the number of sources in each log($L_{\rm{1.4GHz}}$) bin for the given population, and log($\Phi$) is the median volume density per log($L_{\rm{1.4GHz}}$) bin for the given population. $N$ corresponds to the median number of sources between the definite RL HERG and LERG RLFs and the RLFs that include the definite RL HERGs and LERGs plus all unclassified RL AGN added to each population. For RL HERGs and LERGs, if $N$ is not a whole number in a given log($L_{\rm{1.4GHz}}$) bin, it indicates that there is an odd number of unclassified RL AGN in that log($L_{\rm{1.4GHz}}$) bin.}
\begin{adjustbox}{width=\textwidth}
\begin{tabular}{c c c c c c c c c c c}
\hline
& & All sources & & SFGs & & RL AGN & & RL HERGs (median) & & LERGs (median)\\
log($L_{\rm{1.4GHz}}$) & $N$ & log($\Phi$) & $N$ &  log($\Phi$) & $N$ & log($\Phi$) & $N$ &  log($\Phi$) & $N$ &  log($\Phi$)\\
(W Hz$^{-1}$) & & (mag$^{-1}$ Mpc$^{-3}$) & & (mag$^{-1}$ Mpc$^{-3}$) & & (mag$^{-1}$ Mpc$^{-3}$) & & (mag$^{-1}$ Mpc$^{-3}$) & & (mag$^{-1}$ Mpc$^{-3}$)\\
\hline
\hline
21.8 & 66.0 & -3.24$^{+0.07}_{-0.08}$   & 43.0 & -3.37$^{+0.08}_{-0.10}$    & 21.0 & -3.89$^{+0.11}_{-0.14}$   & 12.0 & -4.19$^{+0.24}_{-0.43}$   & 9.0 & -4.20$^{+0.26}_{-0.55}$\\
22.2 & 147.0 & -3.36$^{+0.05}_{-0.05}$  & 106.0 & -3.45$^{+0.06}_{-0.06}$  & 27.0 & -4.26$^{+0.10}_{-0.13}$  & 7.5 & -4.71$^{+0.30}_{-0.96}$     & 19.5 & -4.45$^{+0.18}_{-0.26}$\\
22.6 & 261.0 & -3.68$^{+0.04}_{-0.04}$  & 146.0 & -3.90$^{+0.05}_{-0.06}$  & 91.0 & -4.20$^{+0.07}_{-0.09}$  & 41.0 & -4.57$^{+0.18}_{-0.27}$   & 50.0 & -4.43$^{+0.17}_{-0.26}$\\
23.0 & 183.0 & -4.16$^{+0.05}_{-0.06}$  & 79.0 & -4.44$^{+0.08}_{-0.11}$    & 96.0 & -4.52$^{+0.05}_{-0.05}$  & 36.5 & -4.88$^{+0.13}_{-0.18}$   & 59.5 & -4.77$^{+0.10}_{-0.12}$\\
23.4 & 69.0 & -4.70$^{+0.07}_{-0.08}$    & 15.0 & -5.43$^{+0.10}_{-0.13}$   & 54.0 & -4.79$^{+0.08}_{-0.09}$   & 23.5 & -5.14$^{+0.22}_{-0.27}$   & 30.5 & -5.04$^{+0.19}_{-0.22}$\\
23.8 & 22.0 & -5.26$^{+0.09}_{-0.11}$    & 2.0 & -6.32$^{+0.23}_{-0.53}$     & 20.0 & -5.30$^{+0.09}_{-0.11}$   & 6.5 & -5.75$^{+0.17}_{-0.27}$     & 13.5 & -5.49$^{+0.12}_{-0.16}$\\
24.2 & 16.0 & -5.40$^{+0.10}_{-0.13}$    & 1.0 & -6.62$^{+0.30}_{-\infty}$    & 15.0 & -5.43$^{+0.10}_{-0.13}$   & 9.0 & -5.67$^{+0.17}_{-0.24}$     & 6.0 & -5.80$^{+0.20}_{-0.34}$\\
24.6 & 10.0 & -5.62$^{+0.12}_{-0.17}$     &      &                                            & 10.0 & -5.62$^{+0.12}_{-0.17}$   & 3.0 & -6.15$^{+0.30}_{-0.71}$    & 7.0 & -5.78$^{+0.19}_{-0.29}$\\
25.0 & 9.0 & -5.67$^{+0.12}_{-0.18}$      &       &                                            & 9.0 & -5.67$^{+0.12}_{-0.18}$     & 3.5 & -6.08$^{+0.23}_{-0.44}$    & 5.5 & -5.88$^{+0.19}_{-0.30}$\\
25.4 & 1.0 & -6.62$^{+0.30}_{-\infty}$     &       &                                             & 1.0 & -6.62$^{+0.30}_{-\infty}$    &       &                                            & 1.0 & -6.62$^{+0.30}_{-\infty}$\\
\hline
\end{tabular}
\end{adjustbox}
\label{tab:rlf_z_bin_1_data}
\end{table*}

\subsection{XXL-S 1.4 GHz RLFs at higher redshifts (0.3 < $z$ < 1.3)}
\label{sec:RLFS_high_z}

\begin{figure}
        \includegraphics[width=\columnwidth]{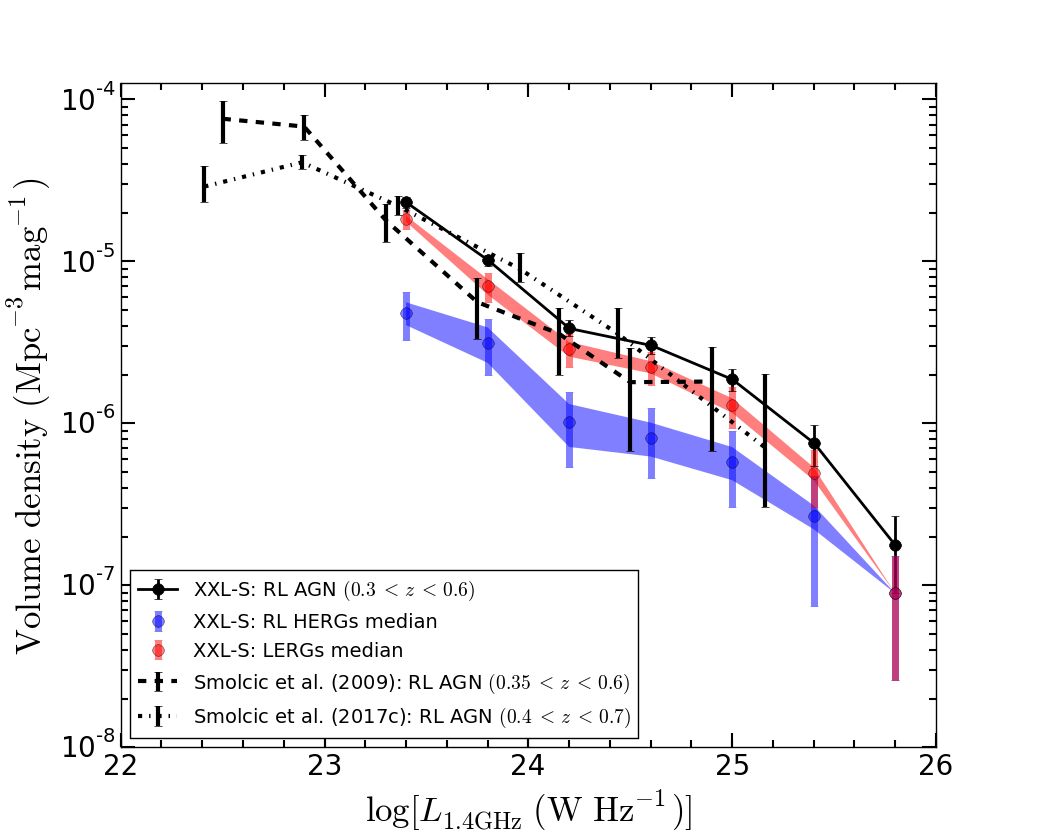}
    \caption[1.4 GHz XXL-S RLFs for RL AGN, RL HERGs and LERGs in 0.3 < $z$ < 0.6.]{1.4 GHz XXL-S RLFs for RL AGN, RL HERGs and LERGs in 0.3 < $z$ < 0.6. For comparison, the corresponding RLFs for COSMOS radio AGN from \cite{smolcic2009a} and \cite{smolcic2017b} are shown.}
    \label{fig:rlf_z_bin_2_sep_plot}
\end{figure}

\begin{figure}
        \includegraphics[width=\columnwidth]{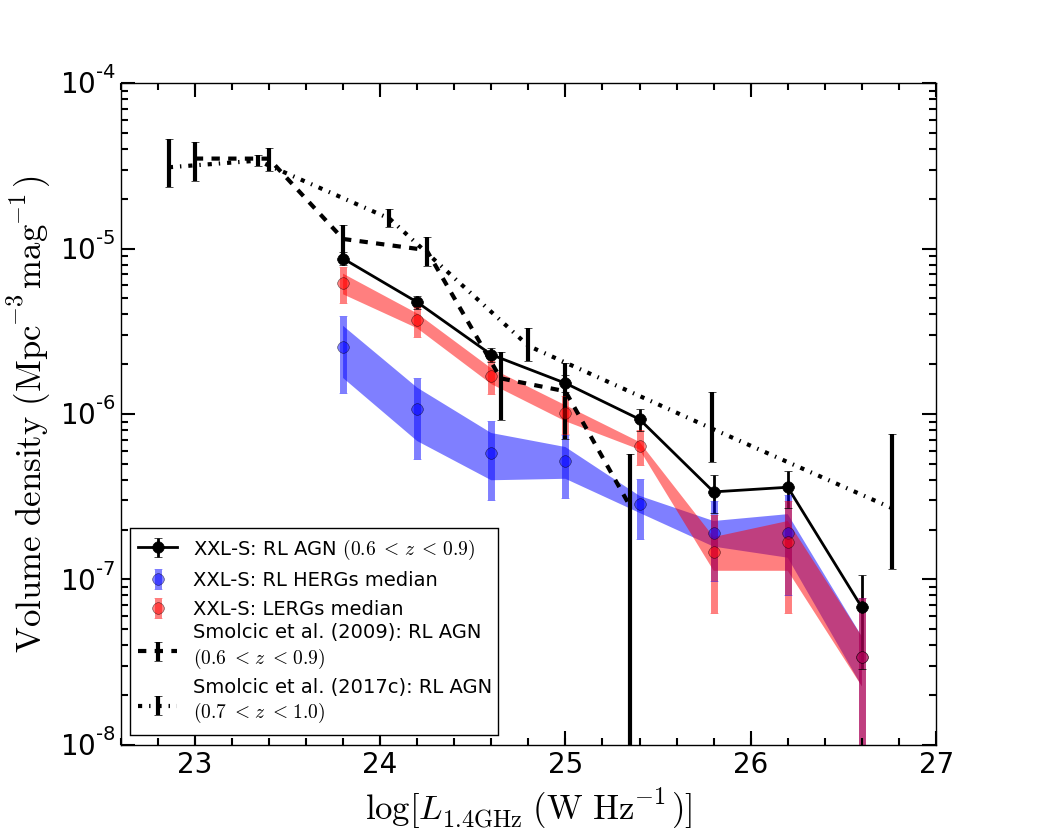}
    \caption[1.4 GHz XXL-S RLFs for RL AGN, RL HERGs and LERGs in 0.6 < $z$ < 0.9.]{1.4 GHz XXL-S RLFs for RL AGN, RL HERGs and LERGs in 0.6 < $z$ < 0.9. For comparison, the corresponding RLFs for COSMOS radio AGN from \cite{smolcic2009a} and \cite{smolcic2017b} are shown.}
    \label{fig:rlf_z_bin_3_sep_plot}
\end{figure}

\begin{figure}
        \includegraphics[width=\columnwidth]{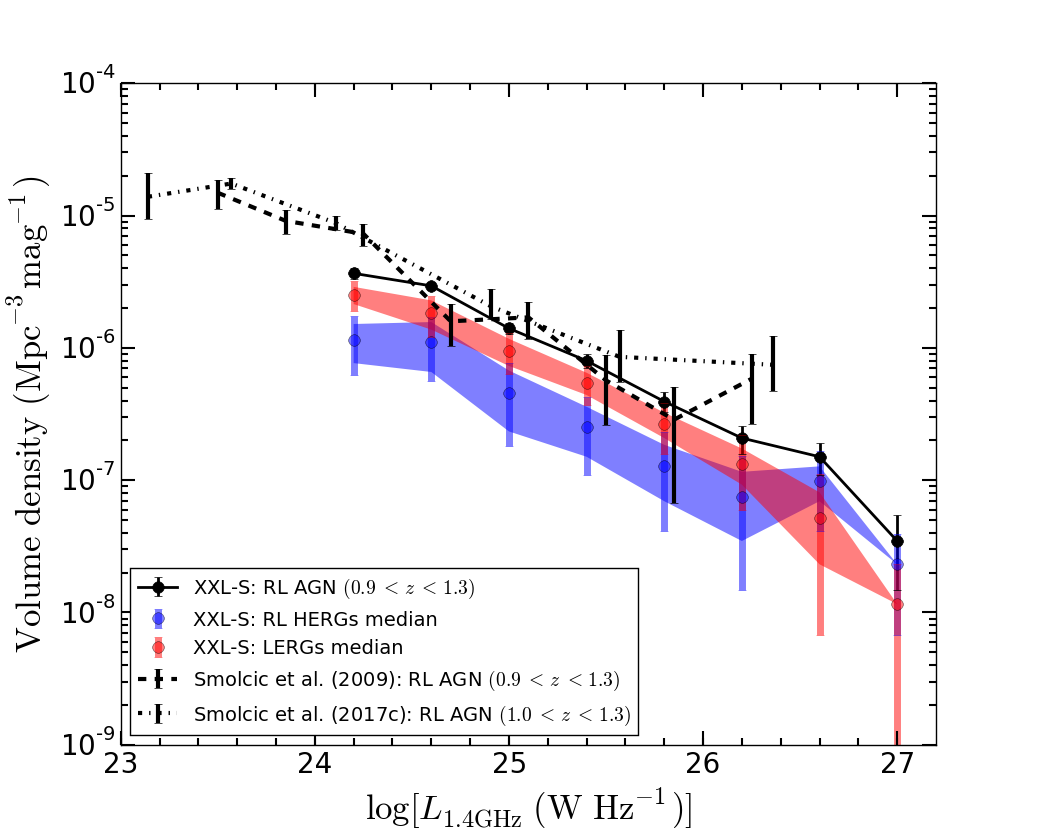}
    \caption[1.4 GHz XXL-S RLFs for RL AGN, RL HERGs and LERGs in 0.9 < $z$ < 1.3.]{1.4 GHz XXL-S RLFs for RL AGN, RL HERGs and LERGs in 0.9 < $z$ < 1.3. For comparison, the corresponding RLFs for COSMOS radio AGN from \cite{smolcic2009a} and \cite{smolcic2017b} are shown.}
    \label{fig:rlf_z_bin_4_sep_plot}
\end{figure}

Figures \ref{fig:rlf_z_bin_2_sep_plot}, \ref{fig:rlf_z_bin_3_sep_plot}, and \ref{fig:rlf_z_bin_4_sep_plot} show the XXL-S 1.4 GHz RLFs in redshift bins 0.3 < $z$ < 0.6, 0.6 < $z$ < 0.9 and 0.9 < $z$ < 1.3 (bins 2, 3, and 4), respectively, for all RL AGN, RL HERGs and LERGs (the data are also shown in Tables \ref{tab:rlf_z_bin_2_data} - \ref{tab:rlf_z_bin_4_data}).  The data points, error bars, and shaded regions represent the same quantities as those shown in Figure \ref{fig:rlf_z_bin_1_sep_plot}.  

\begin{table*}
\centering
\caption[RLF data for all RL AGN, RL HERGs, and LERGs in XXL-S for 0.3 < $z$ < 0.6.]{RLF data for all RL AGN, RL HERGs, and LERGs in XXL-S for 0.3 < $z$ < 0.6.  See the caption for Table \ref{tab:rlf_z_bin_1_data} for an explanation of the columns.}
\begin{adjustbox}{width=13cm}
\begin{tabular}{c c c c c c c}
\hline
& & RL AGN & & RL HERGs (median) & & LERGs (median)\\
log($L_{\rm{1.4GHz}}$) & $N$ & log($\Phi$) & $N$ &  log($\Phi$) & $N$ & log($\Phi$)\\
(W Hz$^{-1}$) & & (mag$^{-1}$ Mpc$^{-3}$) & & (mag$^{-1}$ Mpc$^{-3}$) & & (mag$^{-1}$ Mpc$^{-3}$)\\
\hline
\hline
23.4 & 252.0 & -4.64$^{+0.03}_{-0.04}$ & 54.0 & -5.32$^{+0.13}_{-0.17}$ & 198.0 & -4.74$^{+0.06}_{-0.06}$\\
23.8 & 183.0 & -4.99$^{+0.03}_{-0.04}$ & 54.0 & -5.50$^{+0.15}_{-0.20}$ & 129.0 & -5.15$^{+0.08}_{-0.10}$\\
24.2 & 83.0 & -5.41$^{+0.05}_{-0.05}$ & 22.0 & -5.99$^{+0.19}_{-0.28}$ & 61.0 & -5.54$^{+0.10}_{-0.11}$\\
24.6 & 66.0 & -5.52$^{+0.05}_{-0.06}$ & 17.5 & -6.09$^{+0.18}_{-0.25}$ & 48.5 & -5.65$^{+0.09}_{-0.11}$\\
25.0 & 42.0 & -5.73$^{+0.06}_{-0.07}$ & 13.0 & -6.24$^{+0.19}_{-0.28}$ & 29.0 & -5.89$^{+0.11}_{-0.14}$\\
25.4 & 15.0 & -6.12$^{+0.11}_{-0.14}$ & 4.0 & -6.57$^{+0.25}_{-0.55}$ & 11.0 & -6.31$^{+0.15}_{-0.21}$\\
25.8 & 4.0 & -6.75$^{+0.18}_{-0.30}$ & 2.0 & -7.05$^{+0.23}_{-0.53}$ & 2.0 & -7.05$^{+0.23}_{-0.53}$\\
\hline
\end{tabular}
\end{adjustbox}
\label{tab:rlf_z_bin_2_data}
\end{table*}

\begin{table*}
\centering
\caption[RLF data for all RL AGN, RL HERGs, and LERGs in XXL-S for 0.6 < $z$ < 0.9.]{RLF data for all RL AGN, RL HERGs, and LERGs in XXL-S for 0.6 < $z$ < 0.9. See the caption for Table \ref{tab:rlf_z_bin_1_data} for an explanation of the columns.}
\begin{adjustbox}{width=13cm}
\begin{tabular}{c c c c c c c}
\hline
& & RL AGN & & RL HERGs (median) & & LERGs (median)\\
log($L_{\rm{1.4GHz}}$) & $N$ & log($\Phi$) & $N$ &  log($\Phi$) & $N$ & log($\Phi$)\\
(W Hz$^{-1}$) & & (mag$^{-1}$ Mpc$^{-3}$) & & (mag$^{-1}$ Mpc$^{-3}$) & & (mag$^{-1}$ Mpc$^{-3}$)\\
\hline
\hline
23.8 & 197.0 & -5.06$^{+0.04}_{-0.04}$ & 55.0 & -5.60$^{+0.19}_{-0.28}$ & 142.0 & -5.20$^{+0.10}_{-0.12}$\\
24.2 & 168.0 & -5.32$^{+0.04}_{-0.04}$ & 38.0 & -5.97$^{+0.19}_{-0.30}$ & 130.0 & -5.43$^{+0.08}_{-0.10}$\\
24.6 & 98.0 & -5.64$^{+0.04}_{-0.05}$ & 25.0 & -6.23$^{+0.19}_{-0.29}$ & 73.0 & -5.77$^{+0.09}_{-0.11}$\\
25.0 & 68.0 & -5.81$^{+0.05}_{-0.06}$ & 23.0 & -6.29$^{+0.16}_{-0.22}$ & 45.0 & -5.99$^{+0.10}_{-0.13}$\\
25.4 & 41.0 & -6.03$^{+0.06}_{-0.07}$ & 12.5 & -6.54$^{+0.15}_{-0.21}$ & 28.5 & -6.19$^{+0.10}_{-0.12}$\\
25.8 & 15.0 & -6.47$^{+0.10}_{-0.13}$ & 8.5 & -6.72$^{+0.19}_{-0.29}$ & 6.5 & -6.83$^{+0.22}_{-0.37}$\\
26.2 & 16.0 & -6.44$^{+0.10}_{-0.12}$ & 8.5 & -6.72$^{+0.23}_{-0.38}$ & 7.5 & -6.77$^{+0.24}_{-0.43}$\\
26.6 & 3.0 & -7.17$^{+0.20}_{-0.37}$ & 1.5 & -7.47$^{+0.36}_{-\infty}$ & 1.5 & -7.47$^{+0.36}_{-\infty}$\\
\hline
\end{tabular}
\end{adjustbox}
\label{tab:rlf_z_bin_3_data}
\end{table*}

\begin{table*}
\centering
\caption[RLF data for all RL AGN, RL HERGs, and LERGs in XXL-S for 0.9 < $z$ < 1.3.]{RLF data for all RL AGN, RL HERGs, and LERGs in XXL-S for 0.9 < $z$ < 1.3. See the caption for Table \ref{tab:rlf_z_bin_1_data} for an explanation of the columns.}
\begin{adjustbox}{width=13cm}
\begin{tabular}{c c c c c c c}
\hline
& & RL AGN & & RL HERGs (median) & & LERGs (median)\\
log($L_{\rm{1.4GHz}}$) & $N$ & log($\Phi$) & $N$ &  log($\Phi$) & $N$ & log($\Phi$)\\
(W Hz$^{-1}$) & & (mag$^{-1}$ Mpc$^{-3}$) & & (mag$^{-1}$ Mpc$^{-3}$) & & (mag$^{-1}$ Mpc$^{-3}$)\\
\hline
\hline
24.2 & 178.0 & -5.44$^{+0.04}_{-0.04}$ & 53.0 & -5.94$^{+0.18}_{-0.26}$ & 125.0 & -5.60$^{+0.10}_{-0.12}$\\
24.6 & 210.0 & -5.53$^{+0.03}_{-0.03}$ & 83.0 & -5.95$^{+0.19}_{-0.30}$ & 127.0 & -5.74$^{+0.13}_{-0.18}$\\
25.0 & 118.0 & -5.85$^{+0.04}_{-0.04}$ & 38.0 & -6.34$^{+0.23}_{-0.40}$ & 80.0 & -6.02$^{+0.13}_{-0.17}$\\
25.4 & 69.0 & -6.10$^{+0.05}_{-0.06}$ & 22.0 & -6.59$^{+0.22}_{-0.37}$ & 47.0 & -6.26$^{+0.13}_{-0.17}$\\
25.8 & 34.0 & -6.41$^{+0.07}_{-0.08}$ & 11.0 & -6.90$^{+0.26}_{-0.49}$ & 23.0 & -6.58$^{+0.16}_{-0.22}$\\
26.2 & 18.0 & -6.68$^{+0.09}_{-0.12}$ & 6.5 & -7.12$^{+0.31}_{-0.71}$ & 11.5 & -6.88$^{+0.22}_{-0.35}$\\
26.6 & 13.0 & -6.82$^{+0.11}_{-0.14}$ & 8.5 & -7.01$^{+0.23}_{-0.38}$ & 4.5 & -7.28$^{+0.33}_{-0.89}$\\
27.0 & 3.0 & -7.46$^{+0.20}_{-0.37}$ & 2.0 & -7.64$^{+0.23}_{-0.53}$ & 1.0 & -7.94$^{+0.30}_{-10.06}$\\
\hline
\end{tabular}
\end{adjustbox}
\label{tab:rlf_z_bin_4_data}
\end{table*}

\subsection{Comparison of XXL-S RLFs to literature}
\label{sec:comp_RLFs_to_other_samples}

\subsubsection{Local RL AGN and SFG RLFs}

The XXL-S RLFs for RL AGN and SFGs are similar to those of \cite{pracy2016} (hereafter Pracy16) and \cite{best2012}, although the XXL-S volume densities are higher (by a factor of $\sim$3-4) at some luminosities (particularly 22.5 < log[$L_{\rm{1.4GHz}}$~(W~Hz$^{-1})$] < 23.5).
This is due to the differences between the way the HERGs and SFGs were classified and to the deeper XXL-S radio and optical data (see Section \ref{sec:comp_local_RL_HERG_LERG_RLFs_to_other_samples} for an explanation).  The RLFs from Pracy16 are consistent with the RLFs from \cite{mauch2007}, which are known to be in good agreement with previously constructed RLFs (e.g. \citealp{sadler2002}).  Therefore, the RLFs for RL AGN and SFGs in XXL-S broadly agree with previously constructed RLFs for the local universe, but are different in a way that reflects the unique XXL-S data and radio source classification scheme.

\subsubsection{Local RL HERG and LERG RLFs}
\label{sec:comp_local_RL_HERG_LERG_RLFs_to_other_samples}

The XXL-S LERG RLFs are consistent with (within 1$\sigma$ of) the LERG RLFs from Pracy16 and \cite{best2012}, but the HERG RLF from Pracy16 shows lower volume densities than the XXL-S RL HERG RLF, and the one from \cite{best2012} is lower still.  This is due to two things: the classification method used to distinguish RL HERGs from RQ HERGs and SFGs and the optical and radio depths probed by each sample.

The RLFs constructed by Pracy16 use the sample of radio galaxies in the LARGESS survey classified by \cite{ching2017a}, who primarily employed optical spectroscopic diagnostics to determine the origin of the radio emission in each source.  For example, all radio sources at $z < 0.3$ that had  $L_{\rm{1.4GHz}} \leq 10^{24}$ W Hz$^{-1}$ and that were located in the star-forming galaxy region of the BPT diagram (below the \citealp{kauffmann2003} line) were classified as SFGs.  In addition, all sources in the AGN region of the BPT diagram (above the \citealp{kewley2001b} line) were considered radio-loud AGN, unless their radio luminosity placed them within 3$\sigma$ of the one-to-one relation between the SFR inferred by $L_{\rm{1.4GHz}}$ and the SFR inferred by the H$\alpha$ line luminosity, as found in \cite{hopkins2003}. In the latter case, they were considered radio-quiet AGN (i.e. AGN existing in galaxies in which the origin of the radio emission is predominantly star formation).  All other sources were considered radio-loud AGN and separated into LERGs and HERGs on the basis of their EW([OIII]).  The XXL-S radio sources were classified differently: all XXL-S HERGs were identified before any other sources (no matter where they lied in the BPT diagram), whereas the SFGs in the LARGESS sample were identified first and assumed to all lie in the star-forming galaxy region of the BPT diagram.  However, Figure 15 of XXL Paper XXXI shows that some XXL-S galaxies in this region have EW([OIII]) > 5 $ $\AA$ $ (some of which are radio-loud on the basis of the radio AGN indicators used for XXL-S), which means that the corresponding sources in the LARGESS survey would be classified as RL HERGs according to the XXL-S classification scheme, not SFGs.  Another difference is that the XXL-S RL AGN were identified by three radio-only indicators (luminosity, spectral index, and morphology) and one radio-optical SFR ratio (radio excess), while the LARGESS RL AGN were identified by whether or not they were located in the AGN region of the BPT diagram or found at $z > 0.3$.  These are two very different ways of classifying radio sources and evidently lead to different RLFs, especially for RL HERGs.

The effect that different classification techniques have on the final classification results is also reflected in the differences between the HERG and SFG RLFs from Pracy16 and \cite{best2012}.  \cite{best2012} were more strict in identifying HERGs than Pracy16 because the authors had access to more spectroscopic diagnostics. The main difference between the two classification methods is that \cite{best2012} employed a method comparing the strength of the 4000$\ \AA$ break $(D_{4000})$ to the ratio of radio luminosity to stellar mass $(L_{\rm{rad}} / M_*)$.  Pracy16 did not employ this method because, as they point out, \cite{herbert2010} showed that a sample of high luminosity HERGs clearly exhibiting radio emission from an AGN have a range of $D_{4000}$ values that are spread among both the AGN and SFG regions of the $D_{4000}$ vs $L_{\rm{rad}} / M_*$ plot.  In addition, Figure 9 in \cite{best2005b} demonstrates that some sources identified as AGN in the BPT diagram fall in the SFG region of this technique. Therefore, some of the sources that \cite{best2012} classified as SFGs Pracy16 would have classified as HERGs, which caused the volume densities of the Pracy16 HERGs to increase relative to the \cite{best2012} HERGs.  This is evident from Figure \ref{fig:rlf_z_bin_1_sep_plot}. At the same time, the volume densities of the \cite{best2012} SFGs are increased relative to the Pracy16 SFG RLF. This is reflected in Figure \ref{fig:rlf_z_bin_1_all_plot}, which shows the \cite{best2012} SFG RLF slightly offset above the Pracy16 SFG RLF. 

In addition, XXL-S probed deeper in the optical than either of these two other samples, and so more optical sources were available to be cross-matched to the radio sources. The sample of Pracy16 probed down to $i$-band\footnote{The central wavelength of the $i$-band (DECam) is 784 nm \citep{flaugher2015}.} magnitude $m_i < 20.5$, whereas XXL-S probed down to $m_i=25.6$ (XXL Paper VI; \citealp{desai2012,desai2015}; XXL Paper XXXI).  The difference this made can be seen in the fainter absolute magnitudes present in the XXL-S RL HERG population.  Figure 6 in Pracy16 shows that, in their local redshift bin, virtually no HERGs with $L_{\rm{1.4GHz}}$~>~10$^{22}$ W Hz$^{-1}$ are fainter than $M_i$~$\approx$~$-20$, but $\sim$44\% (45/102) of the XXL-S RL HERGs with $L_{\rm{1.4GHz}}$~>~10$^{22}$ W Hz$^{-1}$ in the local redshift bin have $M_i$~>~$-20$.  Clearly, the deeper optical data available for XXL-S detected faint RL HERGs at $z < 0.3$ missed by other surveys, which contributed to an increase in the volume densities of the RL HERGs compared to the Pracy16 and \cite{best2012} samples.

\begin{figure}
        \includegraphics[width=\columnwidth]{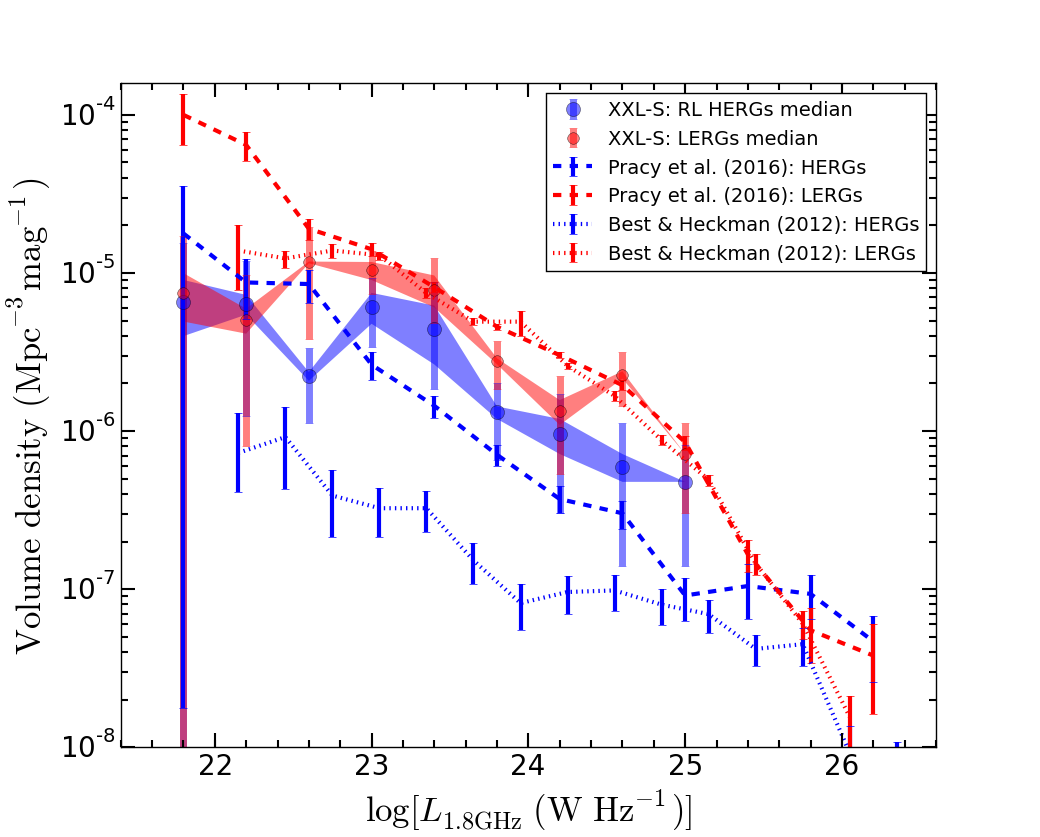}
    \caption[RL HERG and LERG RLFs for XXL-S in $0 < z < 0.3$ for the sub-sample that corresponds to the cut that Pracy16 applied to their local HERG and LERG RLFs ($m_i < 20.5$ and $S_{\rm{1.4GHz}}$ > 2.8 mJy).]{RL HERG and LERG RLFs for XXL-S in $0 < z < 0.3$ for the sub-sample that corresponds to the cut that Pracy16 applied to their local HERG and LERG RLFs ($m_i < 20.5$ and $S_{\rm{1.4GHz}}$ > 2.8 mJy).  See Section \ref{sec:comp_local_RL_HERG_LERG_RLFs_to_other_samples} for details.  The XXL-S RL HERG RLF is now consistent (within 1-2$\sigma$) with the HERG RLF from Pracy16 at all radio luminosities sampled in XXL-S.}
    \label{fig:rlf_z_bin_1_sep_sample_match_plot}
\end{figure}

Furthermore, Pracy16 applied a radio flux density cut of $S_{\rm{1.4GHz}}$ > 2.8 mJy, which is the flux density down to which the NVSS survey is complete, and an optical cut of $m_i < 20.5$ to their local RLFs.  In order to properly compare the XXL-S sample to the Pracy16 sample, these cuts should be applied to the XXL-S data.  However, applying the $S_{\rm{1.4GHz}}$ > 2.8 mJy cut would leave too few sources available for the construction of the XXL-S RLFs.  A flux density cut that is high enough to select a similar radio population but low enough to include enough sources to generate a local RLF is needed. Since XXL-S is complete down to $S_{\rm{1.8GHz}}$$\sim$0.5 mJy, selecting radio sources brighter than that is sufficient for the sake of this comparison.  Another issue is that the XXL-S $L_{\rm{1.4GHz}}$ values were calculated using the flux densities of the effective frequency ($\sim$1.8 GHz; see XXL Paper XVIII), not 1.4 GHz, and a wide range of radio spectral indices.  By contrast, the sample of Pracy16 started with 1.4 GHz flux densities and applied the same $\alpha=-0.7$ spectral index to all sources to calculate the 1.4 GHz luminosities.  Accordingly, applying a 1.4 GHz flux density cut to a 1.8 GHz sample effectively redistributes the $L_{\rm{1.4GHz}}$ values of the sources, changing the shape of the 1.4 GHz RLF.  Therefore, for the purposes of comparing the Pracy16 sample and the XXL-S sample as fairly as possible, a cut of $S_{\rm{1.8GHz}}$ > 0.5 mJy was applied to the local XXL-S RL HERG and LERG sources, in addition to the $m_i < 20.5$ cut.  The RLFs were then reconstructed using the 1.8 GHz luminosities.  The resulting 1.8 GHz RL HERG and LERG RLFs are shown in Figure \ref{fig:rlf_z_bin_1_sep_sample_match_plot}.  This time, the RL HERG RLF for XXL-S is consistent with that of Pracy16 within 1-2$\sigma$ at all luminosities, albeit lower for $L_{\rm{1.8GHz}} \lesssim 10^{23}$ W Hz$^{-1}$.  The remaining differences are probably due to the classification methods and differences in the $L_{\rm{1.4GHz}}$ values.  It is likely that more low luminosity RL HERGs would have been able to be identified if more XXL-S sources had a spectrum available.  The $m_i < 20.5$ cut simultaneously lowered the XXL-S LERG RLF for $L_{\rm{1.8GHz}} \lesssim 10^{23}$ W Hz$^{-1}$, but this flattening at low luminosities is also evident in the LERG RLF from \cite{best2012}, which was constructed using a relatively bright sample of optical counterparts (with $r$-band\footnote{The central wavelength of the $r$-band (DECam) is 642 nm \citep{flaugher2015}.} magnitudes between $14.5 \leq m_r \leq 17.77$).  The differences between XXL-S RL HERG and LERG RLFs and those from Pracy16 and \cite{best2012} for the local universe ($0 < z < 0.3$) can be confidently attributed to differences in the optical and radio depth probed by each sample and to the classification criteria used to identify RL HERGs and LERGs.

\subsubsection{High redshift (0.3 < $z$ < 1.3) RL AGN RLFs}

The RLFs for radio AGN in the COSMOS field from \cite{smolcic2009a} and \cite{smolcic2017b} are shown in Figures \ref{fig:rlf_z_bin_2_sep_plot}-\ref{fig:rlf_z_bin_4_sep_plot} as the black dashed lines and black dash dot lines, respectively.  The XXL-S and COSMOS RL AGN RLFs are consistent (within $3\sigma$, where $\sigma$ is the uncertainty in the COSMOS RLFs) at all radio luminosities plotted in all redshift bins, except for redshift bin 3 ($0.6 < z < 0.9$).  The XXL-S RL AGN RLF in this bin has a lower normalisation (at a >3$\sigma$ level) than the \cite{smolcic2017b} COSMOS RLF for $L_{\rm{1.4GHz}}$ < 10$^{24}$ W Hz$^{-1}$.  However, this is probably due to the fact that the COSMOS RLF was binned between $0.7 < z < 1.0$, so the median redshift in that bin is higher than for XXL-S.  The offset is likely due to evolution of the sources, and therefore does not constitute a major discrepancy. 

Overall, the XXL-S RLFs for all RL AGN are in good agreement with (within 3$\sigma$ of) the COSMOS RLFs \citep{smolcic2009a,smolcic2017b} for all radio AGN in redshift bins 2-4 ($0.3 < z < 1.3$).  The similarity between the three RLFs is probably due to the fact that both fields probed similar limiting magnitudes in the $i$-band (COSMOS probed $m_i \leq 26$ and XXL-S probed $m_i \leq 25.6$).  Despite the similarities, the high redshift ($0.9 < z < 1.3$) XXL-S results are more significant than the high redshift results for COSMOS because XXL-S probed a larger volume.  The remaining differences between the COSMOS and XXL-S RLFs are likely due to cosmic variance in the survey areas, different median redshifts, and the different radio depths (the VLA-COSMOS 1.4 GHz Large/Deep Projects reached an rms noise of $\sim$10-15/$\sim$7-12 $\mu$Jy beam$^{-1}$, respectively, and the VLA-COSMOS 3 GHz Large Project reached $\sim$2.3 $\mu$Jy beam$^{-1}$).

\subsubsection{High redshift (0.3 < $z$ < 1.3) RL HERG and LERG RLFs}

In order to ensure near 100\% optical and radio completeness for the analysis of HERG and LERG evolution, Pracy16 constructed the RLFs using a sub-sample of sources with $M_i < -23$ and $S_{\rm{1.4GHz}}$ > 2.8 mJy at all redshifts.  Therefore, only the brightest of optical galaxies with high radio flux densities were included.  A large fraction of the XXL-S sample is fainter than this in both the optical and radio.  Therefore, in order to properly compare the XXL-S RLFs to those of Pracy16, a sub-sample of XXL-S sources with $M_i < -23$ and $S_{\rm{1.4GHz}}$ > 2.8 mJy was used to match the two samples as closely as possible.  This cut was not able to be made for the first redshift bin ($0 < z < 0.3$) for XXL-S because it left virtually no sources available for the construction of the local RLFs.  However, this cut was able to be made for the higher redshift bins.

\begin{figure}
        \includegraphics[width=\columnwidth]{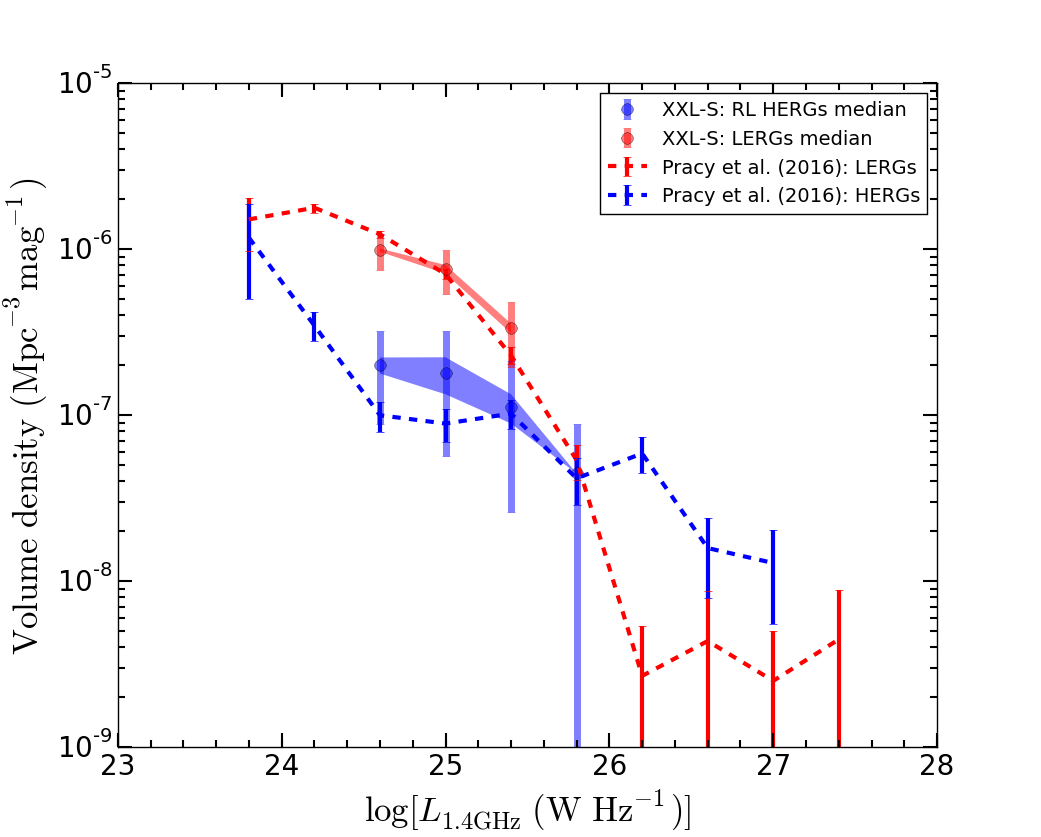}
    \caption[1.4 GHz RLFs for XXL-S RL HERGs and LERGs with $M_i < -23$ and $S_{\rm{1.4GHz}}$ > 2.8 mJy in redshift bin 2 ($0.3 < z < 0.6$).]{1.4 GHz RLFs for XXL-S RL HERGs (blue shaded region) and LERGs (red shaded region) with $M_i < -23$ and $S_{\rm{1.4GHz}}$ > 2.8 mJy in redshift bin 2 ($0.3 < z < 0.6$).  There are only four log($L_{\rm{1.4GHz}}$) bins for the XXL-S data because of the $M_i$ and $S_{\rm{1.4GHz}}$ cuts.  For comparison, the RLFs for HERGs and LERGs from Pracy16 for $0.3 < z < 0.5$ are shown as the blue and red dashed lines, respectively.}
    \label{fig:rlf_z_bin_2_sep_Pracy_sample_match_plot}
\end{figure}

Figure \ref{fig:rlf_z_bin_2_sep_Pracy_sample_match_plot} shows the RLFs in XXL-S for redshift bin 2 ($0.3 < z < 0.6$) for the sub-sample of RL HERGs and LERGs with $M_i < -23$ and $S_{\rm{1.4GHz}}$ > 2.8 mJy, as well as the HERG and LERG RLFs from Pracy16 for their second redshift bin ($0.3 < z < 0.5$).  Despite small differences in the redshift ranges of the samples, the XXL-S RL HERG and LERG RLFs for sources with $M_i < -23$ and $S_{\rm{1.4GHz}}$ > 2.8 mJy in $0.3 < z < 0.6$ are consistent with (within 1$\sigma$ of) the HERG and LERG RLFs from Pracy16 for $0.3 < z < 0.5$.

\begin{figure}
        \includegraphics[width=\columnwidth]{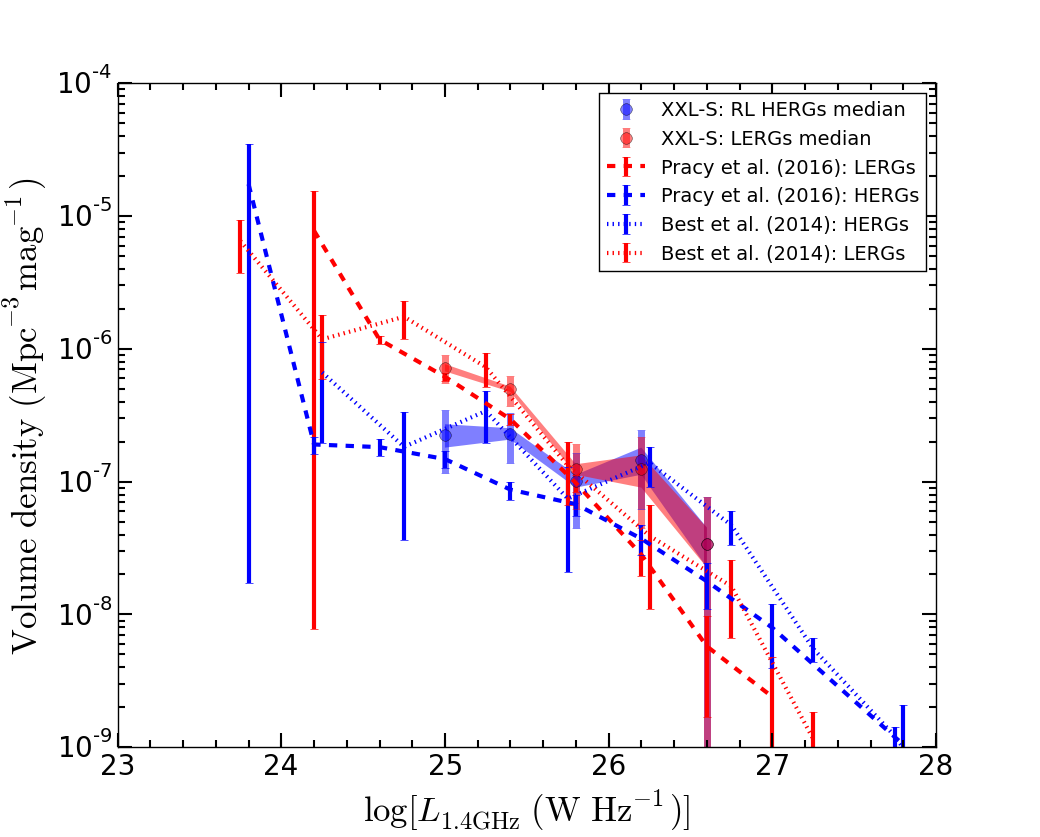}
    \caption[1.4 GHz RLFs for XXL-S RL HERGs and LERGs with $M_i < -23$ and $S_{\rm{1.4GHz}}$ > 2.8 mJy in redshift bin 3 ($0.6 < z < 0.9$).]{1.4 GHz RLFs for XXL-S RL HERGs (blue shaded region) and LERGs (red shaded region) with $M_i < -23$ and $S_{\rm{1.4GHz}}$ > 2.8 mJy in redshift bin 3 ($0.6 < z < 0.9$). The RLFs for HERGs and LERGs from Pracy16 for $0.5 < z < 0.75$ are shown as the blue and red long-dashed lines, respectively.  The HERG and LERG RLFs from \cite{best2014} for $0.5 < z < 1.0$ are shown as the blue and red short-dashed lines, respectively.}
    \label{fig:rlf_z_bin_3_sep_Pracy_sample_match_plot}
\end{figure}

Figure \ref{fig:rlf_z_bin_3_sep_Pracy_sample_match_plot} shows the RLFs in XXL-S for redshift bin 3 ($0.6 < z < 0.9$) for the sub-sample of RL HERGs and LERGs with $M_i < -23$ and $S_{\rm{1.4GHz}}$ > 2.8 mJy.  It also shows the HERG and LERG RLFs from Pracy16 for their third redshift bin ($0.5 < z < 0.75$) and from \cite{best2014} for their second redshift bin ($0.5 < z < 1$).  \cite{best2014} used eight different samples to construct their final sample, but given that 90\% of their radio sources have $S_{\rm{1.4GHz}}$ > 2 mJy, the radio flux density and optical magnitude distribution of their sample is not expected to be significantly different to that from Pracy16. Again, even with small differences in the redshift ranges, the XXL-S RL HERG and LERG RLFs for sources with $M_i < -23$ and $S_{\rm{1.4GHz}}$ > 2.8 mJy in $0.6 < z < 0.9$ are consistent with (within $\sim$1-2$\sigma$ of) the HERG and LERG RLFs from Pracy16 for $0.5 < z < 0.75$ and from \cite{best2014} for $0.5 < z < 1$.

The fact that the XXL-S LERG and (particularly) RL HERG RLFs are consistent with that of Pracy16 and \cite{best2014} when the samples are matched in optical magnitude depth and radio flux density as closely as possible validates the construction of the RLFs made using the full XXL-S sample.  It is also strong evidence that RL HERGs at all radio luminosities exist in galaxies with a wide range of optical luminosities, some of which are missed in shallow surveys.  Raising the optical magnitude limit of the XXL-S sources to $M_i < -23$ and increasing the radio flux density limit to $S_{\rm{1.4GHz}}$~>~2.8 mJy lowered the measured volume densities of the XXL-S RL HERG population across the full range of radio luminosities measured for $z>0.3$ ($L_{\rm{1.4GHz}} > 10^{23}$ W Hz$^{-1}$).  

\section{Evolution of RL HERGs and LERGs}
\label{sec:RL_HERG_LERG_evolution}

\subsection{Optical selection}
\label{sec:optical_selection}

It is possible that a number of the radio sources without optical counterparts exist at $z < 1.3$ (the maximum redshift out to which the RLFs in this paper are constructed).  If this is the case, the RLFs would be missing galaxies that should be included, which would affect the measurement of the evolution of the RL HERGs and LERGs.

\begin{figure}
        \includegraphics[width=\columnwidth]{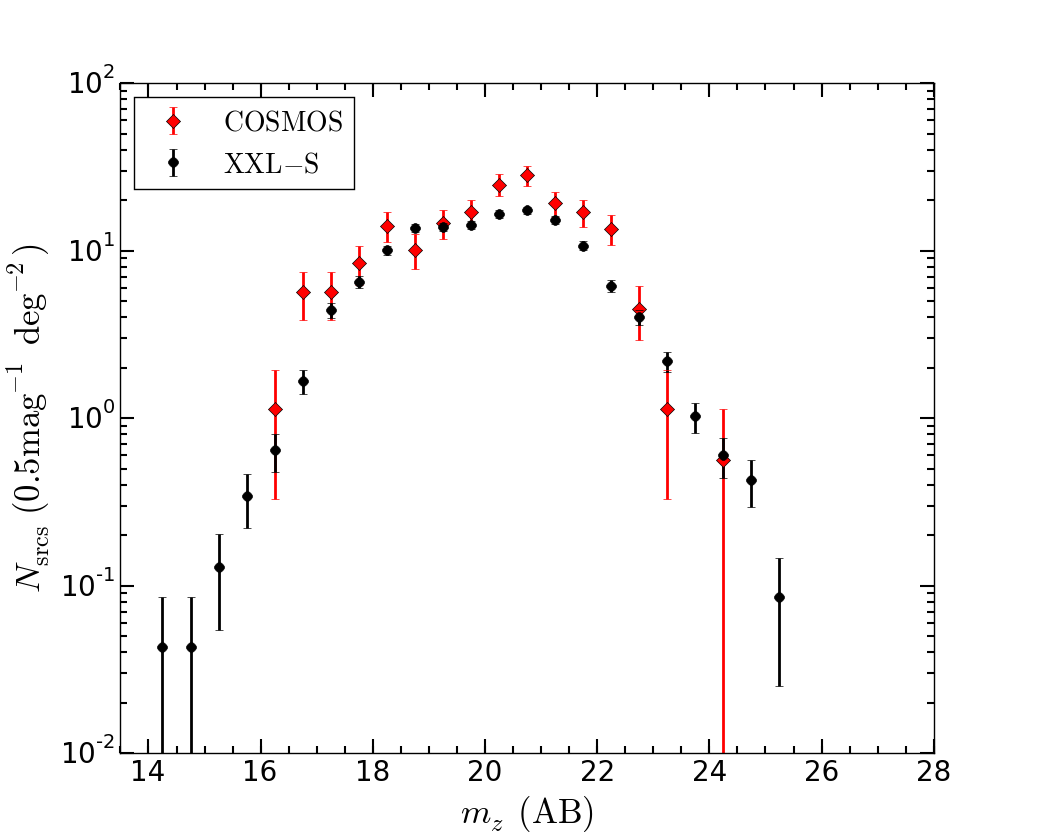}
    \caption[$z$-band (AB) source counts from COSMOS and XXL-S for sources with $S_{\rm{1.8GHz}}$ > 200 $\mu$Jy and at $z < 1.3$ in both surveys.]{$z$-band (AB) source counts from COSMOS and XXL-S for sources with $S_{\rm{1.8GHz}}$ > 200 $\mu$Jy and at $z < 1.3$ in both surveys.  The y-axis is the number of sources per 0.5 magnitude per square degree and the x-axis is $z$-band apparent magnitude (AB) in bins of 0.5 magnitude.  The error bars for each bin are calculated as $\sigma = \sqrt{N_{\rm{srcs}}}$ 0.5mag$^{-1}$ deg$^{-2}$.  The COSMOS source counts are within 3$\sigma$ of the XXL-S source counts for each magnitude bin, but for $20 < m_z < 22.5$ the COSMOS counts are systematically higher the XXL-S counts, indicating that XXL-S potentially has less than $\sim$100\% optical completeness for $m_z > 20$.}
    \label{fig:z-band_src_counts_plot}
\end{figure}

In order to assess the optical counterpart completeness of the optically-matched radio sources in XXL-S, the $z$-band\footnote{The central wavelength of the $z$-band (DECam) is $\lambda$ = 926 nm \citep{flaugher2015}.} source counts for these sources was constructed for 0~<~$z$~<~1.3 and compared to the $z$-band source counts for the $\sim$1.77 deg$^2$ COSMOS field \citep{schinnerer2007}, which has $\sim$100\% optical counterpart completeness for $i^+_{\rm{AB}}<25$ (see \citealp{laigle2016}).  The COSMOS $z$-band source counts were constructed by selecting a sample of optically-matched COSMOS radio sources with $S_{\rm{1.8GHz}} \geq $ 200 $\mu$Jy ($S/N \geq$ 5 for XXL-S) over the same redshift range.  Figure \ref{fig:z-band_src_counts_plot} shows the $z$-band source counts, defined as the number of $z$-band sources per 0.5 magnitude bin per square degree, for XXL-S and COSMOS.  The XXL-S source counts are within 3$\sigma$ of the COSMOS source counts at all $m_z$ values, but for $m_z > 20$ the COSMOS counts are systematically higher than the XXL-S counts (excluding the $m_z>22.5$ bins with large uncertainties).  Since the COSMOS field is missing virtually none of the optical counterparts for the radio sources corresponding to the XXL-S field ($S_{\rm{1.8GHz}}$ > 200 $\mu$Jy), this may indicate that the optical counterpart completeness for XXL-S radio sources with $m_z > 20$ is less than $\sim$100\%.  

\begin{figure}
        \includegraphics[width=\columnwidth]{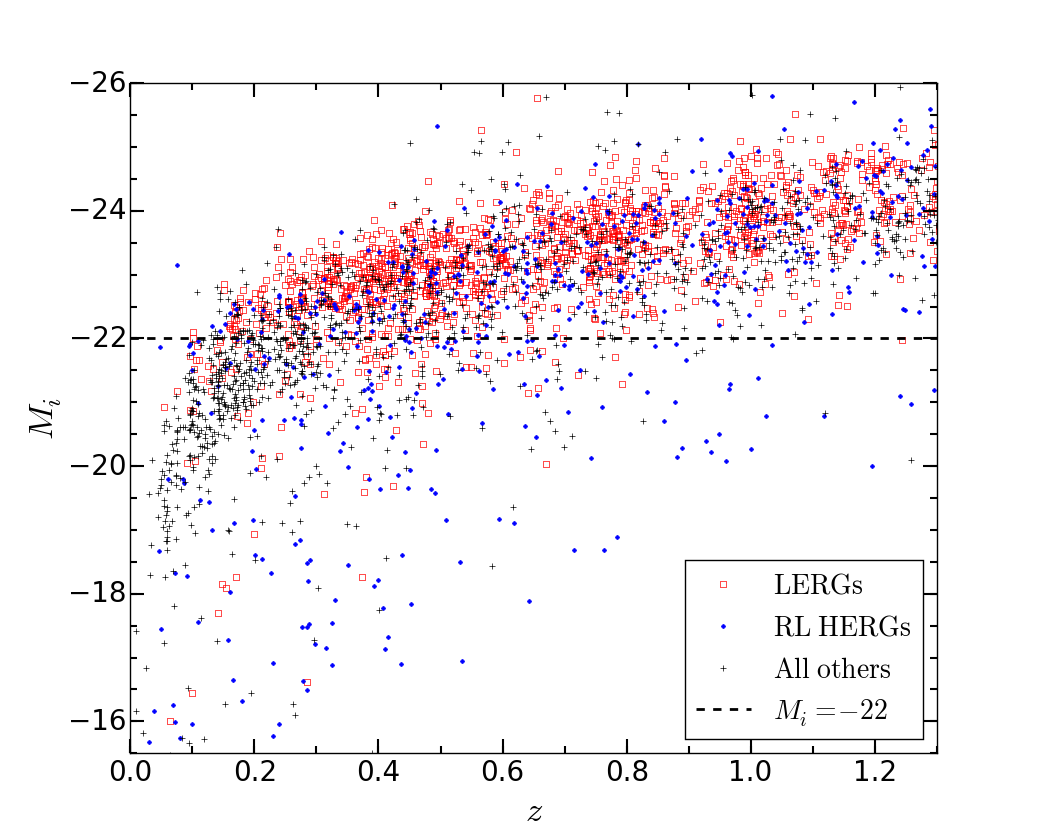}
    \caption[$M_i$ as a function of redshift for all XXL-S radio sources.]{$M_i$ as a function of redshift for all XXL-S radio sources.  At the high redshift end, the faintest LERG reaches down to $M_i \approx -22$, as shown by the dashed black line.  Therefore, only sources with $M_i < -22$ are included in analysis of the RL HERG and LERG evolution.}
    \label{fig:M_i_decam_vs_z}
\end{figure}

In order to mitigate this potential incompleteness, an absolute magnitude cut is applied to the 
RL HERG and LERG samples.  This ensures that 100\% of the galaxies in the sample are detectable out to $z=1.3$.  Figure \ref{fig:M_i_decam_vs_z} shows $M_i$ as a function of redshift for XXL-S RL HERGs and LERGs.  In the highest redshift bin ($0.9 < z < 1.3$), the faintest LERG has $M_i \approx -22$.  Therefore, in order to probe the same optical luminosity distribution for both the LERGs and RL HERGs and to minimise Malmquist bias, a cut of $M_i < -22$ was chosen.  A brighter optical cut would leave too few sources in the local redshift bin to construct an RLF of sufficient precision. 

\subsection{RLFs used for measuring RL HERG and LERG evolution}
\label{sec:final_RLFs}

\begin{figure}
        \includegraphics[width=\columnwidth]{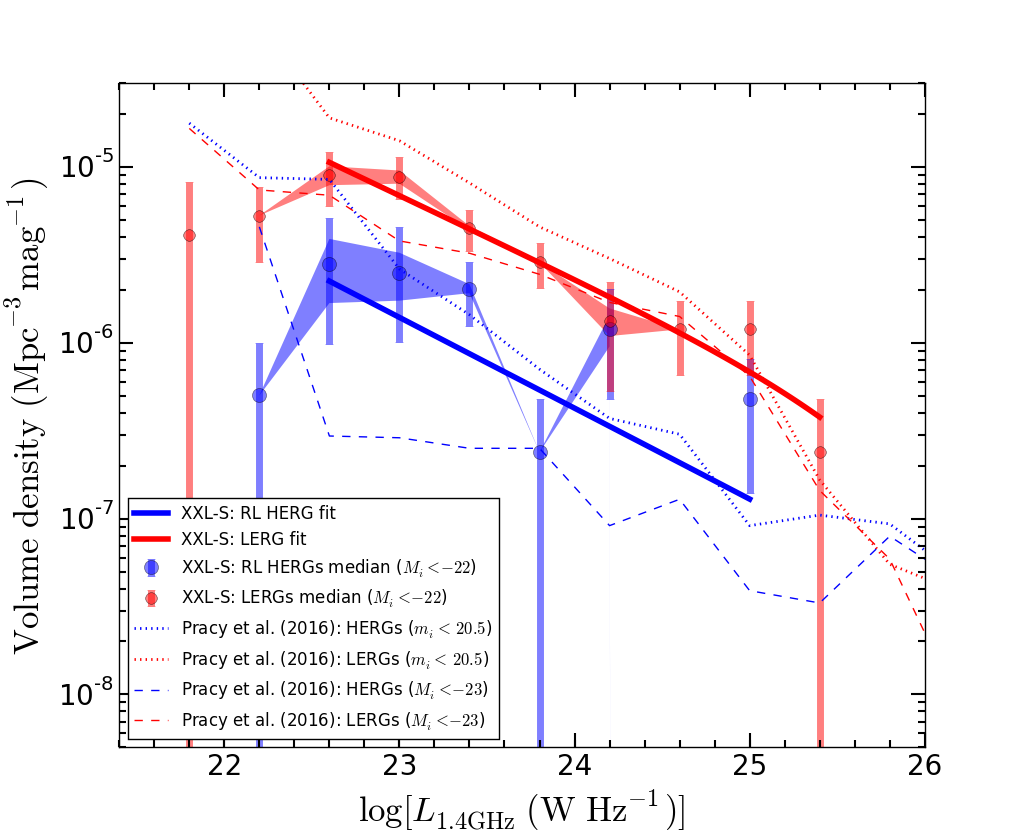}
    \caption[Local (0 < $z$ < 0.3) 1.4 GHz XXL-S RLFs and best fit functions for RL HERGs and LERGs with $M_i < -22$.]{Local (0 < $z$ < 0.3) 1.4 GHz XXL-S RLFs for RL HERGs (blue shaded region) and LERGs (red shaded region) with $M_i < -22$. The best fit function for the RL HERGs is the solid blue line and the best fit function for the LERGs is the solid red line.  The fits exclude the data points with log[$L_{\rm{1.4GHz}}$~(W~Hz$^{-1}$)]~<~22.6.  See Table \ref{tab:best_fit_params_local_RLF_HERG_LERG} for the best fit parameters.  For comparison, the Pracy16 HERG and LERG RLFs with $m_i < 20.5$ (including all sources) and $M_i < -23$ are shown.}
    \label{fig:RLF_z_bin_1_fits_M_i_cut}
\end{figure}

\begin{figure}
        \includegraphics[width=\columnwidth]{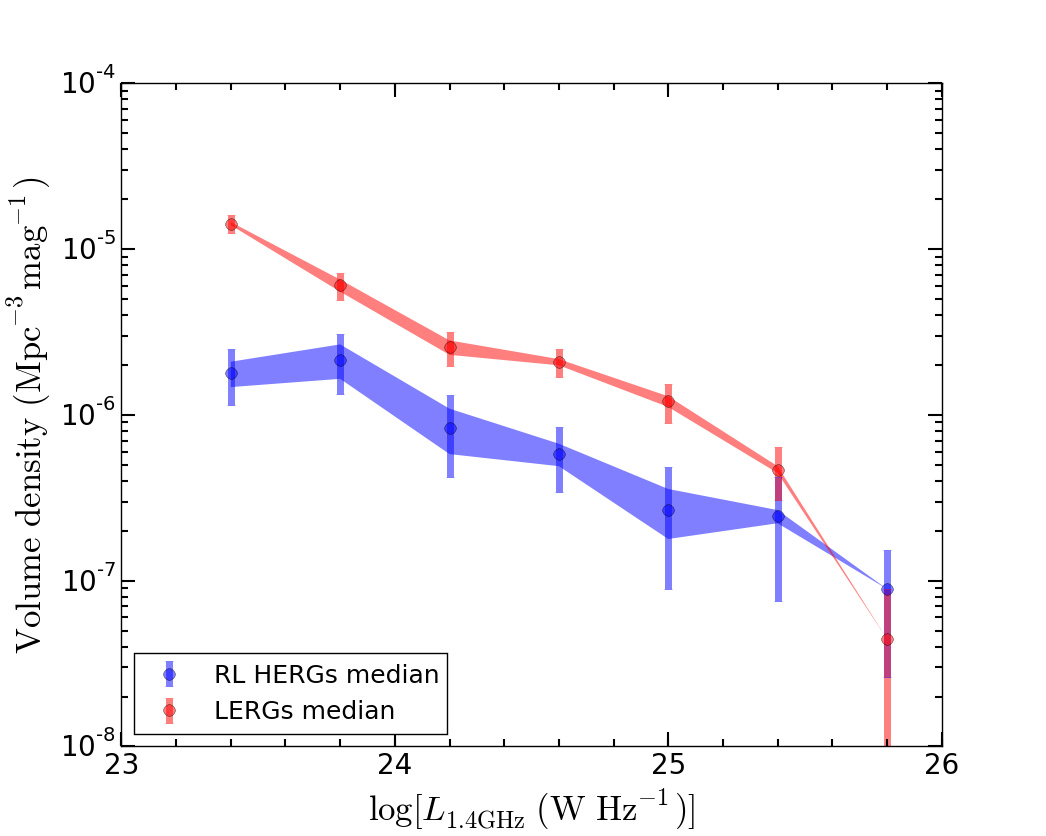}
    \caption[1.4 GHz XXL-S RLFs for RL HERGs and LERGs with $M_i < -22$ in 0.3 < $z$ < 0.6.]{1.4 GHz XXL-S RLFs for RL HERGs (blue shaded region) and LERGs (red shaded region) with $M_i < -22$ in 0.3 < $z$ < 0.6.}
    \label{fig:RLF_z_bin_2_M_i_cut}
\end{figure}

\begin{figure}
        \includegraphics[width=\columnwidth]{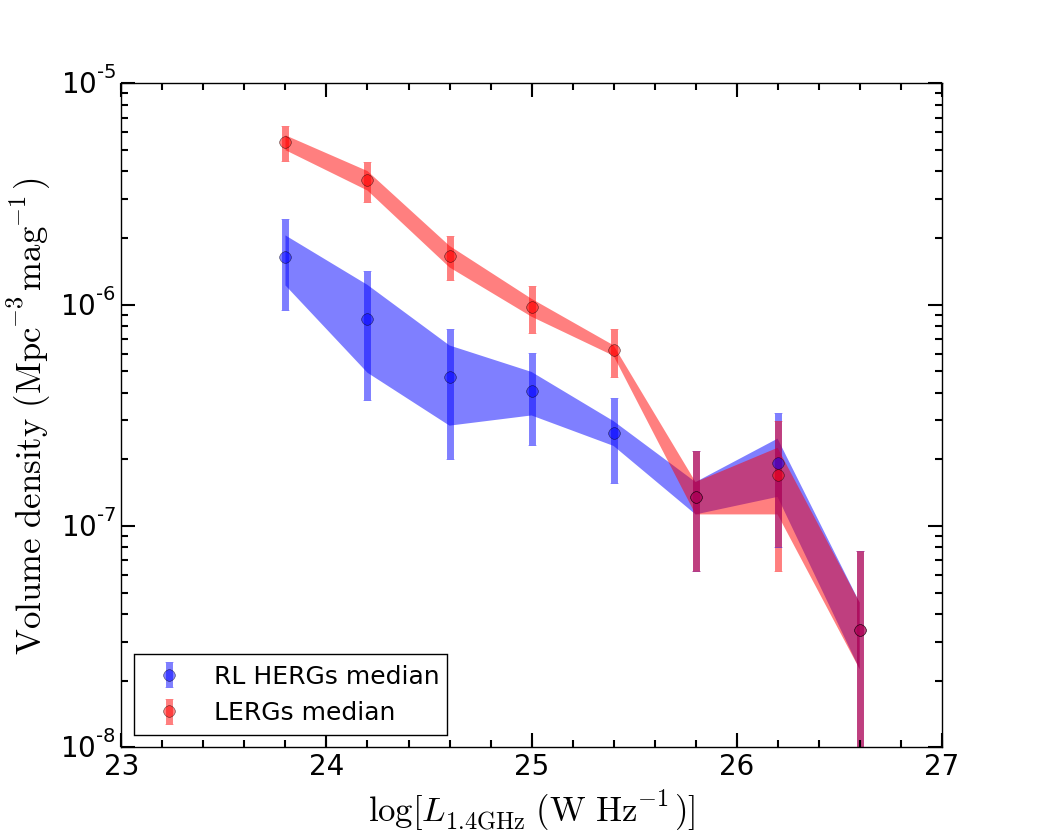}
    \caption[1.4 GHz XXL-S RLFs for RL HERGs and LERGs with $M_i < -22$ in 0.6 < $z$ < 0.9.]{1.4 GHz XXL-S RLFs for RL HERGs (blue shaded region) and LERGs (red shaded region) with $M_i < -22$ in 0.6 < $z$ < 0.9.}
    \label{fig:RLF_z_bin_3_M_i_cut}
\end{figure}

\begin{figure}
        \includegraphics[width=\columnwidth]{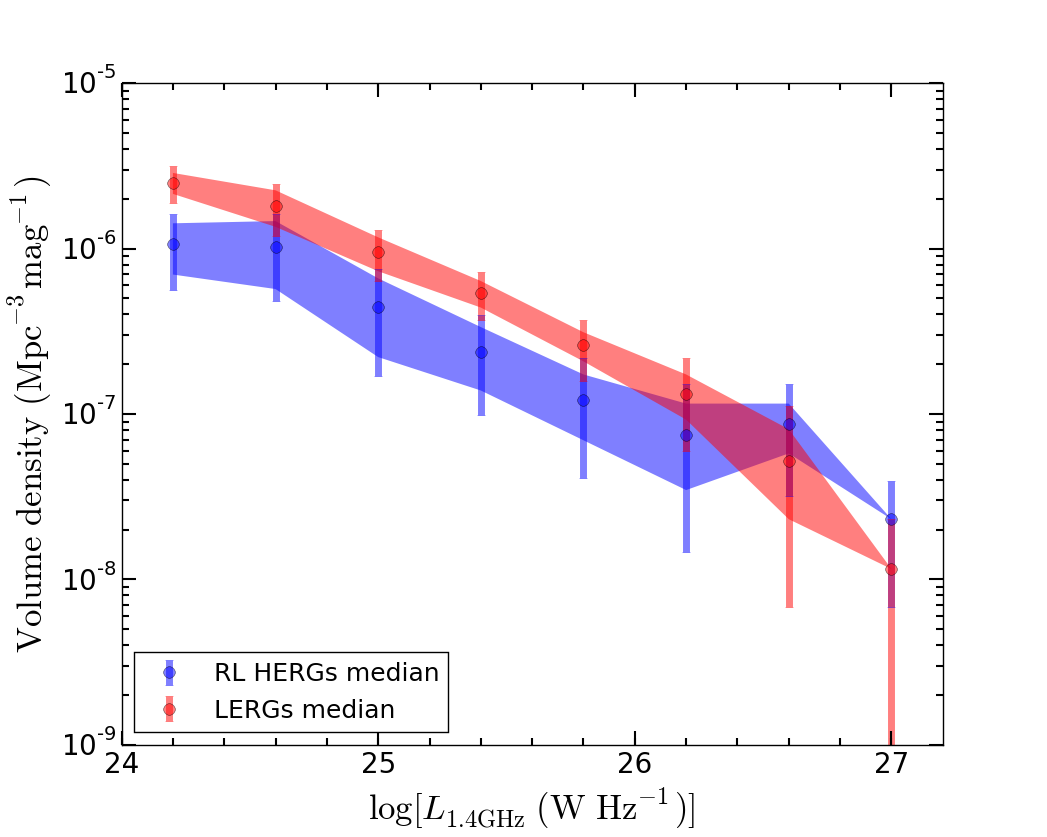}
    \caption[1.4 GHz XXL-S RLFs for RL HERGs and LERGs with $M_i < -22$ in 0.9 < $z$ < 1.3.]{1.4 GHz XXL-S RLFs for RL HERGs (blue shaded region) and LERGs (red shaded region) with $M_i < -22$ in 0.9 < $z$ < 1.3.}
    \label{fig:RLF_z_bin_4_M_i_cut}
\end{figure}

For the remainder of this paper, the sample that is used for analysis is the subset of XXL-S RL HERGs and LERGs that have $M_i < -22$ (unless otherwise specified).  Figure \ref{fig:RLF_z_bin_1_fits_M_i_cut} shows the local RLF for RL HERGs and LERGs when the $M_i < -22$ cut has been applied, and Figures \ref{fig:RLF_z_bin_2_M_i_cut}, \ref{fig:RLF_z_bin_3_M_i_cut}, and \ref{fig:RLF_z_bin_4_M_i_cut} show the RLFs for RL HERGs and LERGs with $M_i < -22$ in redshift bins 2, 3, and 4, respectively.  Tables \ref{tab:rlf_z_bin_1_herg_lerg_M_i_cut_data}-\ref{tab:rlf_z_bin_4_herg_lerg_M_i_cut_data} show the RLF data for all RL AGN, RL HERGs and LERGs with $M_i < -22$ in redshift bins 1-4, respectively.  The RLF data in these tables are used to measure the evolution of the XXL-S RL HERGs and LERGs and their kinetic luminosity densities.

\begin{table*}
\centering
\caption[RLF data for all RL AGN, RL HERGs and LERGs in XXL-S with $M_i < -22$ for $0~<~z~<~0.3$.]{RLF data for all RL AGN, RL HERGs and LERGs in XXL-S with $M_i < -22$ for $0~<~z~<~0.3$. See the caption for Table \ref{tab:rlf_z_bin_1_data} for an explanation of the columns.}
\begin{adjustbox}{width=13cm}
\begin{tabular}{c c c c c c c}
\hline
& & RL AGN & & RL HERGs (median) & & LERGs (median)\\
log($L_{\rm{1.4GHz}}$) & $N$ & log($\Phi$) & $N$ &  log($\Phi$) & $N$ & log($\Phi$)\\
(W Hz$^{-1}$) & & (mag$^{-1}$ Mpc$^{-3}$) & & (mag$^{-1}$ Mpc$^{-3}$) & & (mag$^{-1}$ Mpc$^{-3}$)\\
\hline
\hline
21.8 & 1.0 & -5.39$^{+0.30}_{-\infty}$ & 0.0 &  & 1.0 & -5.39$^{+0.30}_{-\infty}$\\
22.2 & 6.0 & -5.24$^{+0.15}_{-0.24}$ & 1.0 & -6.30$^{+0.30}_{-\infty}$ & 5.0 & -5.28$^{+0.16}_{-0.26}$\\
22.6 & 32.0 & -4.93$^{+0.08}_{-0.09}$ & 8.5 & -5.55$^{+0.27}_{-0.45}$ & 23.5 & -5.05$^{+0.13}_{-0.18}$\\
23.0 & 40.0 & -4.95$^{+0.07}_{-0.08}$ & 7.5 & -5.60$^{+0.26}_{-0.39}$ & 32.5 & -5.06$^{+0.11}_{-0.13}$\\
23.4 & 27.0 & -5.18$^{+0.08}_{-0.09}$ & 8.5 & -5.69$^{+0.15}_{-0.22}$ & 18.5 & -5.35$^{+0.10}_{-0.13}$\\
23.8 & 13.0 & -5.51$^{+0.11}_{-0.14}$ & 1.0 & -6.62$^{+0.30}_{-\infty}$ & 12.0 & -5.54$^{+0.11}_{-0.15}$\\
24.2 & 10.0 & -5.60$^{+0.12}_{-0.17}$ & 5.0 & -5.92$^{+0.23}_{-0.40}$ & 5.0 & -5.88$^{+0.22}_{-0.40}$\\
24.6 & 5.0 & -5.92$^{+0.16}_{-0.26}$ & 0.0 &  & 5.0 & -5.92$^{+0.16}_{-0.26}$\\
25.0 & 7.0 & -5.78$^{+0.14}_{-0.21}$ & 2.0 & -6.32$^{+0.23}_{-0.53}$ & 5.0 & -5.92$^{+0.16}_{-0.26}$\\
25.4 & 1.0 & -6.62$^{+0.30}_{-\infty}$ & 0.0 &  & 1.0 & -6.62$^{+0.30}_{-\infty}$\\
\hline
\end{tabular}
\end{adjustbox}
\label{tab:rlf_z_bin_1_herg_lerg_M_i_cut_data}

\vspace{0.5cm}

\centering
\caption[RLF data for all RL AGN, RL HERGs and LERGs in XXL-S with $M_i < -22$ for $0.3~<~z~<~0.6$.]{RLF data for all RL AGN, RL HERGs and LERGs in XXL-S with $M_i < -22$ for $0.3~<~z~<~0.6$. See the caption for Table \ref{tab:rlf_z_bin_1_data} for an explanation of the columns.}
\begin{adjustbox}{width=13cm}
\begin{tabular}{c c c c c c c}
\hline
& & RL AGN & & RL HERGs (median) & & LERGs (median)\\
log($L_{\rm{1.4GHz}}$) & $N$ & log($\Phi$) & $N$ &  log($\Phi$) & $N$ & log($\Phi$)\\
(W Hz$^{-1}$) & & (mag$^{-1}$ Mpc$^{-3}$) & & (mag$^{-1}$ Mpc$^{-3}$) & & (mag$^{-1}$ Mpc$^{-3}$)\\
\hline
\hline
23.4 & 192.0 & -4.79$^{+0.04}_{-0.04}$ & 27.0 & -5.75$^{+0.15}_{-0.19}$ & 165.0 & -4.85$^{+0.05}_{-0.05}$\\
23.8 & 153.0 & -5.09$^{+0.04}_{-0.04}$ & 38.0 & -5.67$^{+0.16}_{-0.21}$ & 115.0 & -5.22$^{+0.08}_{-0.09}$\\
24.2 & 74.0 & -5.47$^{+0.05}_{-0.05}$ & 18.0 & -6.08$^{+0.20}_{-0.30}$ & 56.0 & -5.59$^{+0.09}_{-0.11}$\\
24.6 & 59.0 & -5.57$^{+0.05}_{-0.06}$ & 13.0 & -6.24$^{+0.16}_{-0.23}$ & 46.0 & -5.68$^{+0.08}_{-0.09}$\\
25.0 & 33.0 & -5.83$^{+0.07}_{-0.08}$ & 6.0 & -6.57$^{+0.26}_{-0.48}$ & 27.0 & -5.92$^{+0.11}_{-0.13}$\\
25.4 & 14.0 & -6.15$^{+0.11}_{-0.15}$ & 3.5 & -6.61$^{+0.24}_{-0.51}$ & 10.5 & -6.33$^{+0.13}_{-0.19}$\\
25.8 & 3.0 & -6.87$^{+0.20}_{-0.37}$ & 2.0 & -7.05$^{+0.23}_{-0.53}$ & 1.0 & -7.35$^{+0.30}_{-10.65}$\\
\hline
\end{tabular}
\end{adjustbox}
\label{tab:rlf_z_bin_2_herg_lerg_M_i_cut_data}

\vspace{0.5cm}

\centering
\caption[RLF data for all RL AGN, RL HERGs and LERGs in XXL-S with $M_i < -22$ for $0.6~<~z~<~0.9$.]{RLF data for all RL AGN, RL HERGs and LERGs in XXL-S with $M_i < -22$ for $0.6~<~z~<~0.9$. See the caption for Table \ref{tab:rlf_z_bin_1_data} for an explanation of the columns.}
\begin{adjustbox}{width=13cm}
\begin{tabular}{c c c c c c c}
\hline
& & RL AGN & & RL HERGs (median) & & LERGs (median)\\
log($L_{\rm{1.4GHz}}$) & $N$ & log($\Phi$) & $N$ &  log($\Phi$) & $N$ & log($\Phi$)\\
(W Hz$^{-1}$) & & (mag$^{-1}$ Mpc$^{-3}$) & & (mag$^{-1}$ Mpc$^{-3}$) & & (mag$^{-1}$ Mpc$^{-3}$)\\
\hline
\hline
23.8 & 168.0 & -5.15$^{+0.04}_{-0.04}$ & 37.5 & -5.79$^{+0.17}_{-0.24}$ & 130.5 & -5.27$^{+0.08}_{-0.09}$\\
24.2 & 160.0 & -5.35$^{+0.04}_{-0.04}$ & 31.5 & -6.07$^{+0.22}_{-0.36}$ & 128.5 & -5.44$^{+0.08}_{-0.10}$\\
24.6 & 91.0 & -5.67$^{+0.04}_{-0.05}$ & 20.0 & -6.33$^{+0.22}_{-0.37}$ & 71.0 & -5.78$^{+0.09}_{-0.11}$\\
25.0 & 61.0 & -5.86$^{+0.05}_{-0.06}$ & 18.0 & -6.39$^{+0.17}_{-0.24}$ & 43.0 & -6.01$^{+0.10}_{-0.12}$\\
25.4 & 39.0 & -6.05$^{+0.06}_{-0.08}$ & 11.5 & -6.58$^{+0.16}_{-0.23}$ & 27.5 & -6.21$^{+0.10}_{-0.12}$\\
25.8 & 12.0 & -6.57$^{+0.11}_{-0.15}$ & 6.0 & -6.87$^{+0.21}_{-0.34}$ & 6.0 & -6.87$^{+0.21}_{-0.34}$\\
26.2 & 16.0 & -6.44$^{+0.10}_{-0.12}$ & 8.5 & -6.72$^{+0.23}_{-0.38}$ & 7.5 & -6.77$^{+0.24}_{-0.43}$\\
26.6 & 3.0 & -7.17$^{+0.20}_{-0.37}$ & 1.5 & -7.47$^{+0.36}_{-\infty}$ & 1.5 & -7.47$^{+0.36}_{-\infty}$\\
\hline
\end{tabular}
\end{adjustbox}
\label{tab:rlf_z_bin_3_herg_lerg_M_i_cut_data}

\vspace{0.5cm}

\centering
\caption[RLF data for all RL AGN, RL HERGs and LERGs in XXL-S with $M_i < -22$ for $0.9~<~z~<~1.3$.]{RLF data for all RL AGN, RL HERGs and LERGs in XXL-S with $M_i < -22$ for $0.9~<~z~<~1.3$. See the caption for Table \ref{tab:rlf_z_bin_1_data} for an explanation of the columns.}
\begin{adjustbox}{width=13cm}
\begin{tabular}{c c c c c c c}
\hline
& & RL AGN & & RL HERGs (median) & & LERGs (median)\\
log($L_{\rm{1.4GHz}}$) & $N$ & log($\Phi$) & $N$ &  log($\Phi$) & $N$ & log($\Phi$)\\
(W Hz$^{-1}$) & & (mag$^{-1}$ Mpc$^{-3}$) & & (mag$^{-1}$ Mpc$^{-3}$) & & (mag$^{-1}$ Mpc$^{-3}$)\\
\hline
\hline
24.2 & 175.0 & -5.45$^{+0.04}_{-0.04}$ & 50.5 & -5.97$^{+0.19}_{-0.28}$ & 124.5 & -5.60$^{+0.10}_{-0.12}$\\
24.6 & 201.0 & -5.55$^{+0.03}_{-0.03}$ & 75.5 & -5.99$^{+0.20}_{-0.33}$ & 125.5 & -5.74$^{+0.13}_{-0.18}$\\
25.0 & 117.0 & -5.86$^{+0.04}_{-0.04}$ & 37.0 & -6.36$^{+0.23}_{-0.41}$ & 80.0 & -6.02$^{+0.13}_{-0.17}$\\
25.4 & 67.0 & -6.11$^{+0.05}_{-0.06}$ & 20.5 & -6.63$^{+0.22}_{-0.38}$ & 46.5 & -6.27$^{+0.13}_{-0.16}$\\
25.8 & 33.0 & -6.42$^{+0.07}_{-0.08}$ & 10.5 & -6.92$^{+0.25}_{-0.47}$ & 22.5 & -6.58$^{+0.16}_{-0.21}$\\
26.2 & 18.0 & -6.68$^{+0.09}_{-0.12}$ & 6.5 & -7.12$^{+0.31}_{-0.71}$ & 11.5 & -6.88$^{+0.22}_{-0.35}$\\
26.6 & 12.0 & -6.86$^{+0.11}_{-0.15}$ & 7.5 & -7.06$^{+0.24}_{-0.43}$ & 4.5 & -7.28$^{+0.33}_{-0.89}$\\
27.0 & 3.0 & -7.46$^{+0.20}_{-0.37}$ & 2.0 & -7.64$^{+0.23}_{-0.53}$ & 1.0 & -7.94$^{+0.30}_{-10.06}$\\
\hline
\end{tabular}
\end{adjustbox}
\label{tab:rlf_z_bin_4_herg_lerg_M_i_cut_data}
\end{table*}

\subsection{RLF functional form}

The RLF of a radio source population can be parametrised using the following double power law as the functional form \citep{dunlop1990,mauch2007}:
\begin{equation}
\label{eq:RLF_func_form}
\Phi (L) = \frac{\Phi^*}{(L^*/L)^{\alpha} + (L^*/L)^{\beta}}, 
\end{equation}
where $\Phi^*$ is the RLF normalisation, $L^*$ is the luminosity at which $\Phi$$(L)$ starts decreasing more rapidly (the `knee' in the RLF), $\alpha$ is the slope at low luminosities (i.e. luminosities lower than $L^*$), and $\beta$ is the slope at high luminosities (i.e. luminosities higher than $L^*$).  Other functional forms have been used in previous work (e.g. \citealp{saunders1990,sadler2002,smolcic2009a}), but many recent authors (e.g. \citealp{best2014,heckman2014,smolcic2017b}; Pracy16) have used Equation \ref{eq:RLF_func_form}.  In order to be able to compare the results of this work to theirs more directly, Equation \ref{eq:RLF_func_form} is used to model the XXL-S RLFs.

\subsection{Local RLF fitting for RL HERGs and LERGs}
\label{sec:local_RLF_fitting}

Similar to the HERG RLF in Pracy16, the XXL-S RL HERGs in the local redshift bin have very few objects at high radio luminosities ($L_{\rm{1.4GHz}} > 10^{25}$ W Hz$^{-1}$).  This results in a poor constraint on the slope of the RL HERG RLF beyond these luminosities, which means that the $\beta$ parameter can approach -$\infty$ with minimal impact on the $\chi^2$ statistic of the fit.  Pracy16 approached this by setting an upper limit on the parameter of $\beta < 0$.  However, since the local XXL-S RL HERG RLF does not probe luminosities as high as that of Pracy16 because of the smaller area of XXL-S, merely setting an upper limit for $\beta$ for the local XXL-S RL HERG RLF resulted in a good fit for the local RL HERG RLF and, simultaneously, an unrealistically sharp decrease in the volume density of RL HERGs in redshift bin 4 ($0.9 < z < 1.3$) at $L_{\rm{1.4GHz}} =10^{27}$ W Hz$^{-1}$.  Therefore, in order to avoid this dramatic cut off at high luminosities while still minimising $\chi^2$ for the local RL HERG RLF, a value of $\beta$ = $-2.0$ was chosen.  Values of $\beta$ significantly below this (even orders of magnitude) do not alter the final results for the comoving kinetic luminosity densities for RL HERGs (see Section \ref{sec:IKLDs}).  In fact, the best fit values that Pracy16 found for $\beta$ while modelling the evolution of their HERG RLFs are $\beta=-1.75$ for pure density evolution and $\beta=-2.17$ for pure luminosity evolution, which further justifies the choice of $\beta=-2.0$.  \cite{ceraj2018} also chose $\beta = -2.0$ to fit their local HLAGN (HERG equivalent) RLFs in the COSMOS field.

The Pracy16 local HERG RLF parameters and their uncertainties (log[$\Phi^*$] = $-7.87^{+0.19}_{-0.70}$, log[$L^*$] = $26.47^{+1.18}_{-0.23}$, $\alpha$ = $-0.66^{+0.05}_{-0.04}$) with $\beta$ = $-2.0^{+0.04}_{-0.04}$ were used as constraints in the fit to the local XXL-S RL HERG RLF (including all optical luminosities) using the \texttt{lmfit} python module \citep{newville2016}.  Once the $M_i < -22$ cut was made, however, the parameters from Pracy16 no longer provided a good fit to the data because the normalisation and slope were now different. In addition, there are fewer sources in the $M_i < -22$ local RLF, making its best fit slope ($\alpha$) more uncertain.  The volume densities start turning over for log[$L_{\rm{1.4GHz}}$ (W Hz$^{-1}$)] = 22.6, so data points below this luminosity were discarded. Furthermore, only one source exists in the log[$L_{\rm{1.4GHz}}$ (W Hz$^{-1}$)] = 23.8 bin, which steepens the best fit value for $\alpha$ to $-0.7$.  This steeper $\alpha$ did not produce a good fit for the higher redshift RL HERG RLFs.  Therefore, in order to use the local RL HERG RLF to describe the evolution of the RL HERGs while still minimising $\chi^2$, the local RL HERG RLF with the $M_i < -22$ cut was refit in the following way.  The values for $L^*$ and $\alpha$ were determined by fitting the RL HERG RLFs in all four redshift bins simultaneously (keeping only $\beta$ fixed at $-2.0$) for two scenarios: pure density and pure luminosity evolution (see Section \ref{sec:RLF_evolution}).  The average values for log($L^*$) and $\alpha$ between the pure density and pure luminosity fits (26.78 and $-0.52$, respectively) were used as the values for the local RL HERG RLF fit. The best fit normalisation for the local RLF was then found by repeating the fitting process, allowing $\Phi^*$ to be a free parameter and keeping $\alpha$ fixed at $-0.52$, $L^*$ fixed at log($L^*$) = 26.78, and $\beta$ fixed at $-2.0$.  The final results for the best fit parameters for the local RL HERG RLF are shown in Table \ref{tab:best_fit_params_local_RLF_HERG_LERG}, and the corresponding best fit function for the $M_i < -22$ local RLF is shown as the solid blue line in Figure \ref{fig:RLF_z_bin_1_fits_M_i_cut}.  The slope of the fit ($\alpha = -0.52$) is steeper than the slope of the fit for the local HERG RLF with $M_i < -23$ from Pracy16 ($\alpha = -0.35$), which is the result expected for a fainter optical cut.  However, the slope is very close to the average between the latter slope ($\alpha = -0.35$) and the slope of the local Pracy16 HERG RLF with the $m_i < 20.5$ cut ($\alpha = -0.66$; i.e. their HERG RLF including all sources), as evidenced in Figure \ref{fig:RLF_z_bin_1_fits_M_i_cut}.  This indicates that the method of fitting the local XXL-S RL HERG RLF generates a sufficiently accurate model for evolution measurement purposes, given the different classification methods, optical selections, and survey areas of XXL-S and the Pracy16 sample.  However, see Appendix \ref{sec:appendix_rebinning_local_RL_HERG_RLF} for a description of the effect that rebinning has on the local XXL-S RL HERG RLF.

A similar procedure was needed for the LERGs because the knee in the local LERG RLF from Pracy16 occurs at a luminosity (log[$L^*$] = 25.21) that is too low to accurately model the XXL-S LERG RLFs at all redshifts (even if all optical luminosities are included).  In other words, the volume density in the local LERG RLF from Pracy16 decreases too rapidly for log[$L^*$] > 25.21, preventing the higher redshift XXL-S LERG RLFs from being accurately modelled.  Therefore, the XXL-S LERG RLFs at all redshifts were considered in order to pinpoint the location of the knee in the local LERG RLF.  The Pracy16 local LERG RLF parameters and their uncertainties (log[$\Phi^*$] = $-6.05^{+0.07}_{-0.07}$, log[$L^*$] = $25.21^{+0.06}_{-0.07}$, $\alpha$ = $-0.53^{+0.03}_{-0.07}$, $\beta$ = $-2.67^{+0.42}_{-0.62}$) were initially used as constraints in the fit to the local XXL-S LERG RLF (for all optical luminosities) using the \texttt{lmfit} python module. The best fit parameters found by \texttt{lmfit} were then used as the initial (free) parameter values to simultaneously fit the LERG RLFs at all redshifts for pure density and pure luminosity evolution (see Section \ref{sec:RLF_evolution}).  The average value for each parameter ($\Phi^*$, $L^*$, $\alpha$, $\beta$) between the pure density and pure luminosity fits were used as the values for the parameters describing the local LERG RLF fit (including all optical luminosities). Like the RL HERG RLFs, the parameters for the LERG RLF from Pracy16 did not provide a good fit to the XXL-S LERG RLF with the $M_i < -22$ cut.  Therefore, the $M_i < -22$ local LERG RLF was refit by allowing $\Phi^*$ and $\alpha$ to be free parameters, keeping $L^*$ fixed at log($L^*$) = 25.91 and $\beta$ fixed at $-1.38$ (the same values used for the local LERG RLF that included all optical luminosities).  For consistency with the RL HERGs, the data points below log[$L_{\rm{1.4GHz}}$ (W Hz$^{-1}$)] = 22.6 were discarded.  The final results for the best fit parameters for the local LERG RLF are shown in Table \ref{tab:best_fit_params_local_RLF_HERG_LERG}, and the corresponding best fit function for the $M_i < -22$ local RLF is shown as the solid red line in Figure \ref{fig:RLF_z_bin_1_fits_M_i_cut}.  The slope of the fit ($\alpha = -0.47$), like the RL HERG RLF slope, is steeper than the slope of the fit for the local Pracy16 LERG RLF with $M_i < -23$ ($\alpha = -0.28$), but is similar to the local Pracy16 LERG RLF with $m_i < 20.5$ ($\alpha = -0.53$).  This result is consistent with the fact that the majority of LERGs are optically bright galaxies: the $M_i < -23$ cut selects only the brightest of LERGs and the $M_i < -22$ selects more, but only a small fraction in the local redshift bin are missed by the latter cut.  Therefore, this suggests that the fit to the local XXL-S LERG RLF can be used to accurately model the evolution of the LERGs.

\begin{table}
\centering
\caption{Best-fitting double power law parameters for the local 1.4 GHz XXL-S LERG and RL HERG RLFs.}
\begin{adjustbox}{width=\columnwidth}
\begin{tabular}{c c c c c}
\label{tab:best_fit_params_local_RLF_HERG_LERG}
 & LERGs & RL HERGs & LERGs & RL HERGs\\
  & (all) & (all) & ($M_i < -22$) & ($M_i < -22$)\\
\hline
\hline
log($\Phi^*_0$) & $-6.655$ & $-7.902$ & $-6.535$ & $-7.805$\\
log($L^*_0$) & 25.910 & 27.212 & 25.910 & 26.776\\
$\alpha$ & $-0.627$ & $-0.683$ & $-0.472$ & $-0.516$\\
$\beta$ & $-1.382$ & $-2.040$ & $-1.382$ & $-2.000$\\
\hline
\end{tabular}
\end{adjustbox}
\label{tab:best_fit_params_herg_lerg_rlfs_z_bin_1}
\end{table}

\subsection{Evolution of RL HERG and LERG RLFs}
\label{sec:RLF_evolution}

\begin{figure}
        \includegraphics[width=\columnwidth]{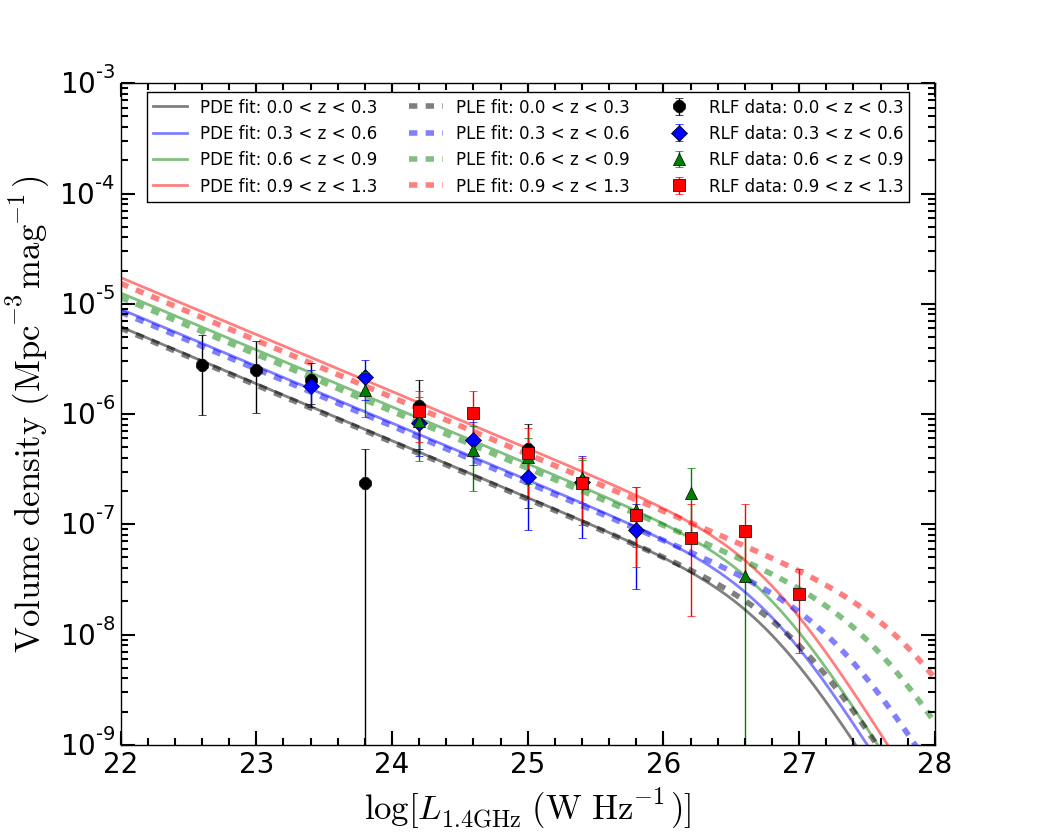}
    \caption[1.4 GHz XXL-S RLF pure density evolution (PDE) and pure luminosity evolution (PLE) fits for RL HERGs with $M_i < -22$ for each redshift bin.]{1.4 GHz XXL-S RLF pure density evolution (PDE) and pure luminosity evolution (PLE) fits for RL HERGs with $M_i < -22$ for each redshift bin. 
    See Table \ref{tab:best_fit_PDE_PLE_params} for the best fit parameters.}
    \label{fig:RLF_PDE_PLE_fit_z_bins_1-4_HERG_RL_plot}
\end{figure}

\begin{figure}
        \includegraphics[width=\columnwidth]{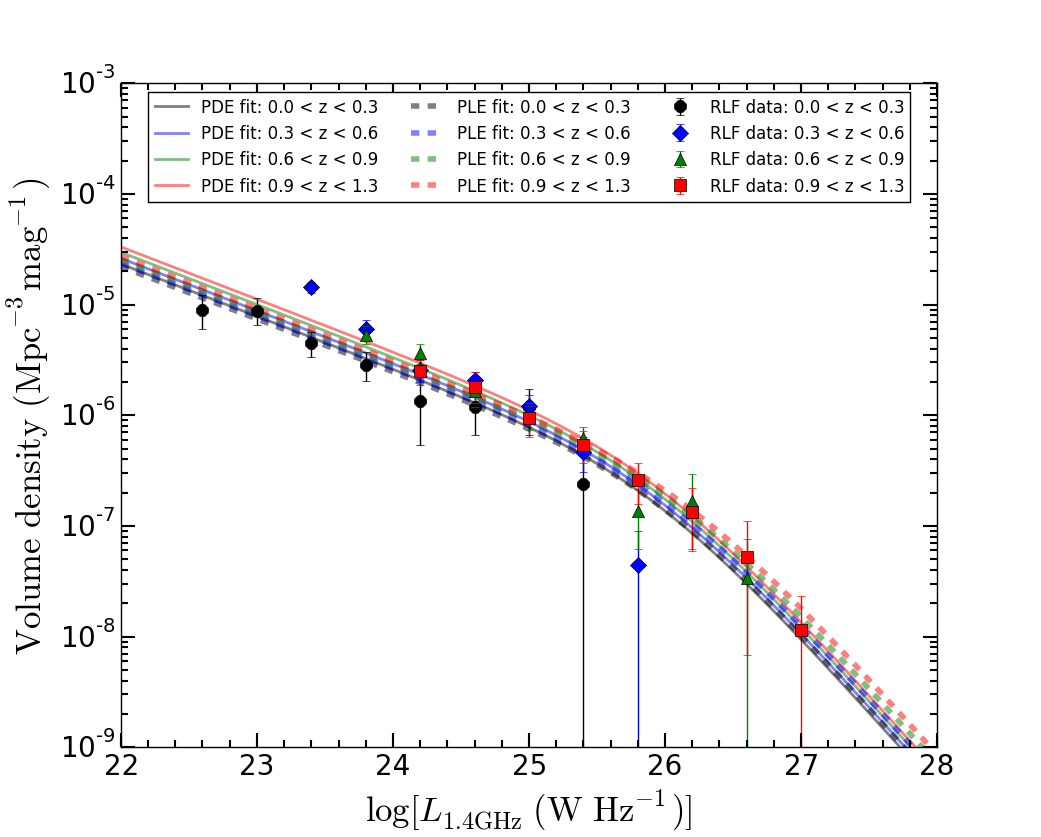}
    \caption[1.4 GHz XXL-S RLF pure density evolution (PDE) and pure luminosity evolution (PLE) fits for LERGs with $M_i < -22$ for each redshift bin.]{1.4 GHz XXL-S RLF pure density evolution (PDE) and pure luminosity evolution (PLE) fits for LERGs with $M_i < -22$ for each redshift bin.  See Table \ref{tab:best_fit_PDE_PLE_params} for the best fit parameters.}
    \label{fig:RLF_PDE_PLE_fit_z_bins_1-4_LERG_plot}
\end{figure}

The evolution of a radio source population is usually expressed via changes only in volume density (pure density evolution, `PDE') or changes only in luminosity (pure luminosity evolution, `PLE').  PDE results in a change in the RLF normalisation ($\Phi^*$) as a function of redshift as
\begin{equation}
\label{eq:Phi_PDE}
\Phi^* (z) =  \Phi^*_0 (1 + z)^{K_{\rm{D}}},
\end{equation}
where $\Phi^*_0$ is the local RLF normalisation and $K_{\rm{D}}$ is a parameter that defines how rapidly the volume density changes.  On the other hand, PLE results in a change in the luminosity knee ($L^*$) as a function of redshift as
\begin{equation}
\label{eq:Phi_PLE}
L^* (z) = L^*_0 (1 + z)^{K_{\rm{L}}},
\end{equation}
where $L^*_0$ is the luminosity knee for the local RLF and $K_{\rm{L}}$ is a parameter that defines how rapidly the sources evolve in luminosity.  Inserting Equation \ref{eq:Phi_PDE} into Equation \ref{eq:RLF_func_form} gives for PDE:
\begin{equation}
\label{eq:Phi_Lz_PDE}
\Phi (L, z) = \frac{\Phi^*_0 (1+z)^{K_{\rm{D}}}}{(L^*_0/L)^{\alpha} + (L^*_0/L)^{\beta}}.
\end{equation}
Inserting Equation \ref{eq:Phi_PLE} into Equation \ref{eq:RLF_func_form} yields for PLE:
\begin{equation}
\label{eq:Phi_Lz_PLE}
\Phi (L, z) = \frac{\Phi^*_0}{(L^*_0 (1+z)^{K_{\rm{L}}} /L)^{\alpha} + (L^*_0 (1+z)^{K_{\rm{L}}} /L)^{\beta}}.
\end{equation}
Using Equations \ref{eq:Phi_Lz_PDE} and \ref{eq:Phi_Lz_PLE}, and fixing the local RLF parameters for each population to be the $M_i < -22$ values listed in Table \ref{tab:best_fit_params_herg_lerg_rlfs_z_bin_1}, the RLFs for RL HERGs and LERGs with $M_i < -22$ across all redshift bins (Tables \ref{tab:rlf_z_bin_1_herg_lerg_M_i_cut_data}-\ref{tab:rlf_z_bin_4_herg_lerg_M_i_cut_data}) were fit using the default $\chi^2$-minimisation method of the \texttt{lmfit} python module.  For the RL HERGs, this procedure gave best fit parameters and 1$\sigma$ uncertainties of $K_{\rm{D}}$ = 1.812 $\pm$ 0.151 and $K_{\rm{L}}$ = 3.186 $\pm$ 0.290.  For the LERGs, it gave $K_{\rm{D}}$ = 0.671 $\pm$ 0.165 and $K_{\rm{L}}$ = 0.839 $\pm$ 0.308.  These parameters are listed in Table \ref{tab:best_fit_PDE_PLE_params}.  Figure \ref{fig:RLF_PDE_PLE_fit_z_bins_1-4_HERG_RL_plot} shows the best fit PDE and PLE fits for RL HERGs in all four redshift bins, and Figure \ref{fig:RLF_PDE_PLE_fit_z_bins_1-4_LERG_plot} shows the corresponding fits for LERGs. 

\begin{table}
\centering
\caption[Best-fitting PDE and PLE parameters ($K_{\rm{D}}$ and $K_{\rm{L}}$, respectively) for the evolution of XXL-S RL HERG and LERG RLFs with $M_i < -22$. For comparison, the $K_{\rm{D}}$ and $K_{\rm{L}}$ parameters for the HERG and LERG RLFs from Pracy16 are shown.]{Best-fitting PDE and PLE parameters ($K_{\rm{D}}$ and $K_{\rm{L}}$, respectively) for the evolution of XXL-S RL HERG and LERG RLFs with $M_i < -22$.  For comparison, the $K_{\rm{D}}$ and $K_{\rm{L}}$ parameters for the HERG and LERG RLFs from Pracy16 are shown.}
\begin{adjustbox}{width=\columnwidth}
\begin{tabular}{c c c c c}
\label{tab:best_fit_PDE_PLE_params}
 & XXL-S & XXL-S & Pracy16 & Pracy16\\
Parameter & LERGs & RL HERGs & LERGs & HERGs\\
\hline
\hline
$K_{\rm{D}}$ & 0.671 $\pm$ 0.165 & 1.812 $\pm$ 0.151 &  0.06$^{+0.17}_{-0.18}$ & 2.93$^{+0.46}_{-0.47}$\\
$K_{\rm{L}}$ & 0.839 $\pm$ 0.308 & 3.186 $\pm$ 0.290 & 0.46$^{+0.22}_{-0.24}$ & 7.41$^{+0.79}_{-1.33}$\\
\hline
\end{tabular}
\end{adjustbox}
\end{table}

\subsection{Luminosity densities of RL HERGs and LERGs}

\begin{figure}
        \includegraphics[width=\columnwidth]{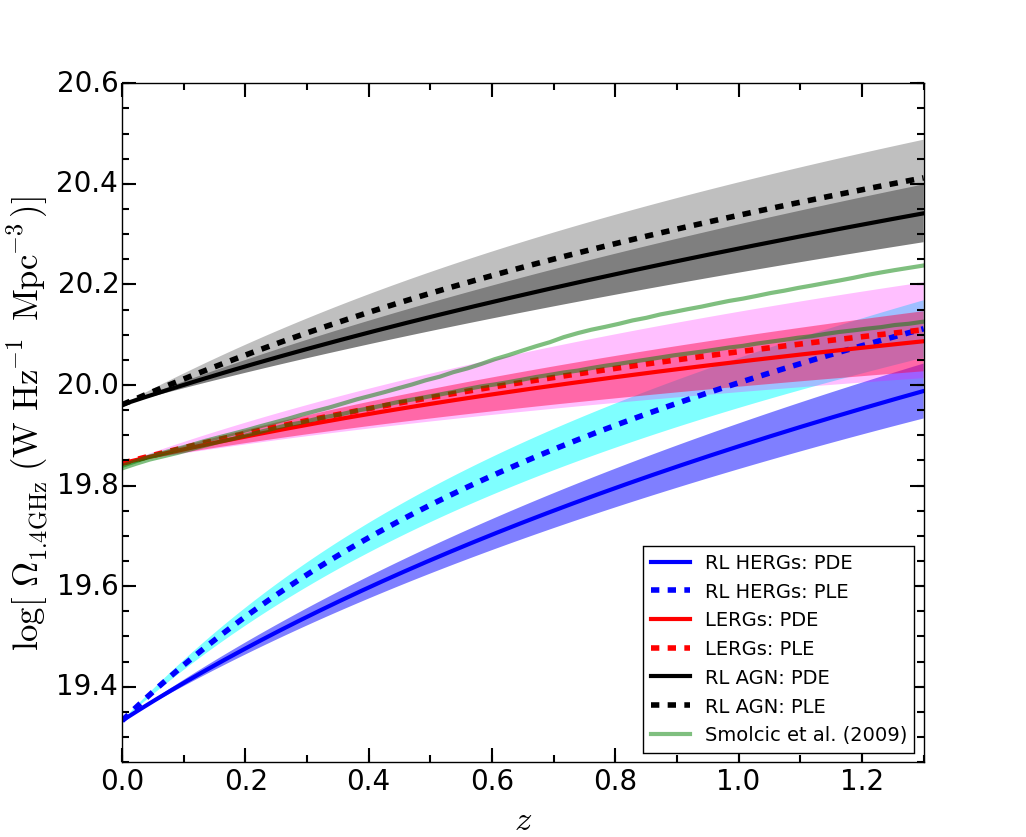}
    \caption[Evolution of the 1.4 GHz comoving luminosity density ($\Omega_{\rm{1.4GHz}}$) for RL HERGs, LERGs, and all RL AGN in XXL-S.]{Evolution of the 1.4 GHz comoving luminosity density ($\Omega_{\rm{1.4GHz}}$) for RL HERGs (blue lines), LERGs (red lines), and all RL AGN (black lines) in XXL-S, integrated from log[$L_{\rm{1.4GHz}}$~(W~Hz$^{-1}$)]~=~22.4 to 27.2 (the full range of luminosities probed in the RLFs) at each redshift for best fit PDE (solid lines) and PLE (dashed lines) models.  The red and magenta shaded areas represent the uncertainties for the LERG PDE and PLE fits, the blue and cyan shaded areas represent the uncertainties for the RL HERG PDE and PLE fits, and the black and grey shaded areas represent the uncertainties for the RL AGN PDE and PLE fits, respectively.  The light green lines represent $\Omega_{\rm{1.4GHz}}$ for the low luminosity ($L_{\rm{1.4GHz}}$ < 5 $\times$ 10$^{25}$ W Hz$^{-1}$) radio AGN in COSMOS from \cite{smolcic2009a}.}
    \label{fig:evolving_comoving_int_lum_density_plot}
\end{figure}

The comoving luminosity density ($\Omega_{\rm{1.4GHz}}$) of a given radio source population represents its total radio luminosity per unit comoving volume as a function of time.  At a given redshift between $0 < z < 1.3$, $\Omega_{\rm{1.4GHz}}$ was calculated for RL HERGs and LERGs for both PDE (Equation \ref{eq:Phi_Lz_PDE}) and PLE (Equation \ref{eq:Phi_Lz_PLE}) by evaluating 
\begin{equation}
\Omega_{\rm{1.4GHz}}(z) = \int L_{\rm{1.4GHz}} \times \Phi(L_{\rm{1.4GHz}}, z) d(\textrm{log}[L_{\rm{1.4GHz}}])
\end{equation}
over the full range of radio luminosities probed at all redshifts (22.4~<~$d$log[$L_{\rm{1.4GHz}}$~(W~Hz$^{-1}$)] < 27.2).  Figure \ref{fig:evolving_comoving_int_lum_density_plot} shows the evolution of $\Omega_{\rm{1.4GHz}}$ for RL HERGs, LERGs, and all RL AGN in XXL-S. The shaded regions represent the uncertainties in the $K_{\rm{D}}$ and $K_{\rm{L}}$ parameters for PDE and PLE, respectively, for each population.

The $\Omega_{\rm{1.4GHz}}$ values for the XXL-S LERGs (red lines in Figure \ref{fig:evolving_comoving_int_lum_density_plot}) are very similar to $\Omega_{\rm{1.4GHz}}$ for the low luminosity radio AGN ($L_{\rm{1.4GHz}} < 5 \times 10^{25}$ W Hz$^{-1}$) in the COSMOS field studied by \cite{smolcic2009a}, shown as the light green lines.  This is a reflection of the fact that LERGs dominate the RL AGN population at low luminosities.

\section{Cosmic evolution of RL HERG and LERG kinetic luminosity densities}
\label{sec:IKLDs}

\subsection{Kinetic luminosities of RL HERGs and LERGs}
\label{sec:kin_lum_scaling_relation}

As SMBHs accrete matter from infalling gas, energy is released that can be transformed into radiation via an accretion disc or converted into kinetic form via jets of relativistic particles, which can reach up to hundreds of kpc beyond the host galaxy and are detectable in the radio \citep{mcnamara2012}.  In the latter scenario (radio mode feedback), the jet structures are able to do mechanical work on the surrounding environment, which can heat the ISM or IGM, and therefore prevent cooling flows from adding stellar mass to the host galaxy (e.g. \citealp{fabian2012}).

Some observations of nearby resolved radio galaxies indicate that they create cavities in the surrounding hot X-ray emitting ICM via the mechanical work done by their radio lobes (e.g. \citealp{bohringer1993}).  These studies have enabled the derivation of various scaling relations between $L_{\rm{1.4GHz}}$ and kinetic luminosity, $L_{\rm{kin}}$ (e.g. \citealp{merloni2007,birzan2008,cavagnolo2010,osullivan2011,daly2012,godfrey2016}).  However, there are large uncertainties associated with each relation, including the one that is arguably the most sophisticated \citep{willott1999}.  The very large ($\sim$2 dex) uncertainty range for this relation originates from the fact that it includes all sources of uncertainty in the conversion between radio luminosity and $L_{\rm{kin}}$ (e.g. deviation from the conditions of minimum energy, uncertainty in the energy of non-radiating particles, and the composition of the jet).  One parameter, $f_W$, represents all these uncertainties and has a range of 1-20, with different values corresponding to different RL AGN populations.  A value of $f_W$ = 15 produces kinetic luminosities close to those calculated via observations of X-ray cavities (surface brightness depressions) in galaxy clusters induced by FRI radio jets and lobes (e.g. \citealp{birzan2004,birzan2008,merloni2007,cavagnolo2010,osullivan2011}), and $f_W$ = 4 produces $L_{\rm{kin}}$ values that closely agree with the results of \cite{daly2012}, who derived a relationship between radio luminosity and $L_{\rm{kin}}$ for some of the most powerful FRII sources using strong shock physics.

Recent simulations, which focus mostly on FRII sources, have produced varying results.  \cite{english2016} used relativistic magnetohydrodynamics to model the dynamical evolution of RL AGN with bipolar supersonic relativistic jets (i.e. FRII sources) in poor cluster environments and found that that \cite{willott1999} relation with $f_W=15$ closely matches the results of their simulations of the evolution of $L_{\rm{178MHz}}$ as a function of radio lobe length.  On the other hand, \cite{hardcastle2018a} modelled the evolution of the shock fronts around the lobes of FRII RL AGN and found that the \cite{willott1999} relation with $f_W=5$ reproduces the $L_{\rm{kin}}$ values for their simulated galaxies existing at $z < 0.5$, but for all galaxies in their sample (which have $z < 4$), high $f_W$ values (10-20) produced a better fit.  The difference is due to higher inverse Compton losses at higher redshift rather than intrinsic evolution of the scaling relation with redshift.  These results illustrate the uncertainty regarding which $f_W$ parameter should be used to compare to observations.

In addition, no observational study has yet developed distinct scaling relations for low-power (FRI) and high-power (FRII) sources, despite theoretical expectations to the contrary.  One of the latest studies \citep{godfrey2016}, which incorporates theoretical considerations such as the composition and age of the radio lobes, is inconclusive about whether FRI and FRII sources actually differ in their scaling relations.  They found a shallower slope for the correlation between radio luminosity and $L_{\rm{kin}}$ for FRI sources than other scaling relations have found, but no correlation for FRII sources.  However, their sample only extends out to $z \leq 0.23$, and therefore it is not clear how applicable this new result is to sources at higher redshift, where most of the XXL-S sources lie. In fact, \cite{smolcic2017b} demonstrated that for $z \gtrsim 0.3$, the \cite{godfrey2016} relation results in $L_{\rm{kin}}$ values that are over an order of magnitude higher than those calculated by other scaling relations, which further demonstrates the uncertainty in how broadly it can be applied.

In light of this uncertainty regarding which scaling relation best applies to a given category of RL AGN, the relation chosen for this paper should be the one that is most appropriate for the majority of RL AGN in XXL-S (LERGs). The \cite{cavagnolo2010} relation is based on FRI galaxies that exist in gas rich cluster environments, where LERGs are expected to exist.  Although this relation has been shown to suffer from Malmquist bias \citep{godfrey2016}, it is, within the uncertainties, consistent with the \cite{willott1999} relation (for $f_W=15$), which does not suffer from distance effects. Furthermore, the studies involving 1.4 GHz radio data that have separated between LERGs and HERGs (\citealp{best2012,best2014}; Pracy16) used the \cite{cavagnolo2010} relation.  Moreover, the simulations to which the XXL-S results are compared in Section \ref{sec:comp_IKLDs_to_sims} all exhibit kinetic luminosity densities that are relatively high (for various reasons, one being the use of the \citealp{merloni2007} scaling relation, which produces higher $L_{\rm{kin}}$ values than \citealp{cavagnolo2010}), implying that a positive scale factor would have to be applied to the XXL-S data for the comparison to the simulations regardless.  Considering all these factors, the \cite{cavagnolo2010} relation is used for the primary results of this paper, although the \cite{willott1999} relation is applied where relevant.  A comparison between the results obtained using these and other scaling relations is found in Appendix \ref{sec:appendix_scaling_relations}.

\begin{figure}
        \includegraphics[width=\columnwidth]{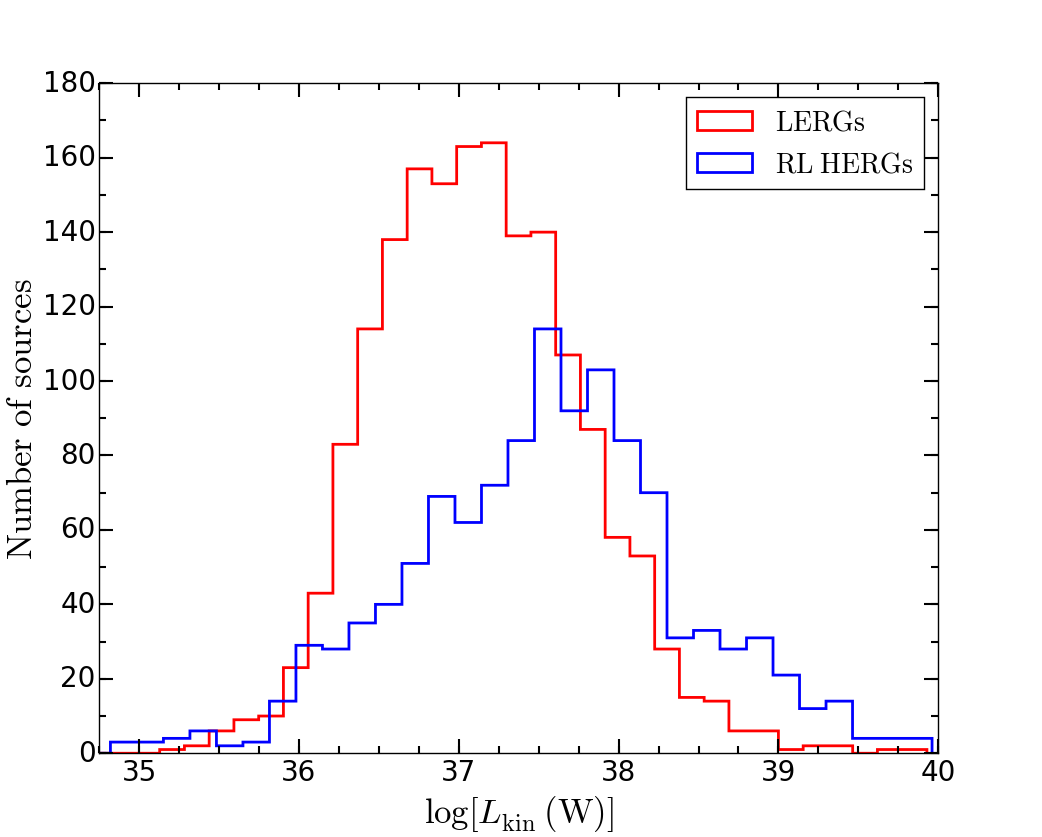}
    \caption[Distribution of jet kinetic luminosities ($L_{\rm{kin}}$) for RL HERGs and LERGs in XXL-S.]{Distribution of $L_{\rm{kin}}$ for RL HERGs and LERGs in XXL-S.  $L_{\rm{kin}}$ for each source was calculated according to the scaling relation from \cite{cavagnolo2010}.}
    \label{fig:kl_accrete_dist_plot}
\end{figure}

The relationship between X-ray cavity power induced by the radio lobes ($P_{\rm{cav}}$) and 1.4 GHz radio power ($P_{\rm{1.4}} = \nu L_{\rm{1.4GHz}}$) found by \cite{cavagnolo2010} is given by their Equation 1.  Converting that relation into units of W and replacing the $P_{\rm{cav}}$ symbol with $L_{\rm{kin}}$ results in 
\begin{equation}
\label{eq:kin_lum_def}
L_{\rm{kin}} (L_{\textrm{1.4GHz}}) = (10^{35} \textrm{ W}) 10^{0.75[\textrm{log}(\nu L_{\textrm{1.4GHz}})] - 22.84},
\end{equation}
where $\nu = 1.4 \times 10^9$ Hz and the corresponding uncertainty range for a given $L_{\rm{kin}}$ is given by: 
\begin{equation}
\label{eq:kin_lum_def_unc}
\Delta L_{\rm{kin}} (L_{\textrm{1.4GHz}}) = (10^{35} \textrm{ W}) 10^{(0.75 \pm 0.14) [\textrm{log}(\nu L_{\textrm{1.4GHz}}) - 33] + (1.91 \pm 0.18)}.
\end{equation}
Figure \ref{fig:kl_accrete_dist_plot} shows the distribution of $L_{\rm{kin}}$ for RL HERGs and LERGs in XXL-S calculated according to Equation \ref{eq:kin_lum_def}. 

\subsection{Measurement of XXL-S comoving kinetic luminosity densities}

\begin{figure}
        \includegraphics[width=\columnwidth]{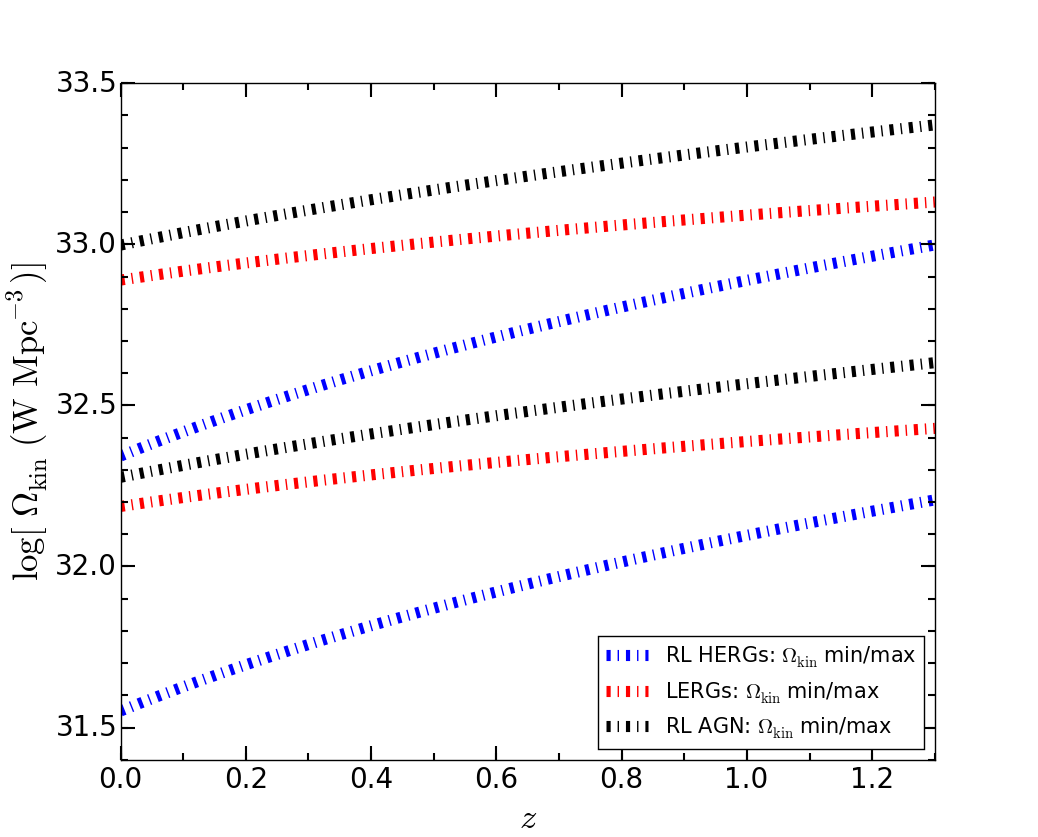}
    \caption[Evolution of the comoving kinetic luminosity density ($\Omega_{\rm{kin}}$) for RL HERGs, LERGs, and all RL AGN in XXL-S using the \cite{cavagnolo2010} scaling relation.]{Evolution of the comoving kinetic luminosity density ($\Omega_{\rm{kin}}$) for RL HERGs (blue lines), LERGs (red lines), and all RL AGN (black lines) in XXL-S using the \cite{cavagnolo2010} scaling relation, integrated from log[$L_{\rm{1.4GHz}}$ (W Hz$^{-1}$)] = 22.4 to 27.2 (the full range of luminosities probed in the RLFs) at each redshift for best fit PDE models.  The PLE models do not differ significantly from the PDE models on this scale.  The upper and lower lines for each population represent the range of uncertainties in the \cite{cavagnolo2010} scaling relation.}
    \label{fig:evolving_kin_lum_per_vol_plot}
\end{figure}

The comoving kinetic luminosity density ($\Omega_{\rm{kin}}$) of a given radio source population represents its total kinetic luminosity per unit comoving volume throughout cosmic time.  Thus, in order to constrain the evolution of radio mode feedback of the XXL-S RL HERGs and LERGs, $\Omega_{\rm{kin}}$ was calculated for each population for both PDE (Equation \ref{eq:Phi_Lz_PDE}) and PLE (Equation \ref{eq:Phi_Lz_PLE}) at a given redshift value between $0 < z < 1.3$ by evaluating
\begin{equation}
\label{eq:omega_kin_RL_HERGs_LERGs}
\Omega_{\rm{kin}}(z) = \int L_{\rm{kin}}(L_{\rm{1.4GHz}}) \times \Phi(L_{\rm{1.4GHz}}, z) d(\textrm{log}[L_{\rm{1.4GHz}}])
\end{equation}
over the full range of radio luminosities probed at all redshifts (22.4~<~$d$log[$L_{\rm{1.4GHz}}$~(W~Hz$^{-1}$)] < 27.2). Figure \ref{fig:evolving_kin_lum_per_vol_plot} shows the cosmic evolution of $\Omega_{\rm{kin}}$ for RL HERGs and LERGs in XXL-S calculated according to Equation \ref{eq:omega_kin_RL_HERGs_LERGs}, where $L_{\rm{kin}}$ and its uncertainty range are calculated using Equations \ref{eq:kin_lum_def} and \ref{eq:kin_lum_def_unc}, respectively.

The average value of the total $\Omega_{\rm{kin}}$ weakly increases from log[$\Omega_{\rm{kin}}$ (W Mpc$^{-3}$)]~$\approx$~32.6 to $\sim$33.0 between $0 < z < 1.3$.  The average LERG $\Omega_{\rm{kin}}$ also shows weak positive evolution, ranging from log[$\Omega_{\rm{kin}}$~(W~Mpc$^{-3}$)]~$\approx$~32.5 to $\sim$32.7.  On the other hand, the RL HERG $\Omega_{\rm{kin}}$ evolves more strongly, starting at log[$\Omega_{\rm{kin}}$ (W Mpc$^{-3}$)]~$\approx$~32.0 at $z=0$ and increasing to 32.6 by $z = 1.3$.  In previous studies, higher luminosity radio sources have been found to evolve even more strongly (e.g. \citealp{dunlop1990,willott2001,best2014}; Pracy16). The difference between those results and the XXL-S results for RL HERGs is a reflection of the increased optical and radio depths probed by XXL-S.

\subsection{Comparison of RL HERG and LERG comoving kinetic luminosity densities to other samples}
\label{sec:comp_IKLDs_to_other_samples}

The evolution of $\Omega_{\rm{kin}}$ for RL HERGs and LERGs in XXL-S can be compared to the results from other samples.  Four of the main studies that have measured the $\Omega_{\rm{kin}}$ evolution for radio AGN are \cite{smolcic2009a}, \cite{best2014}, Pracy16, and \cite{smolcic2017b}.  The XXL-S $\Omega_{\rm{kin}}$ results are compared to each of these.

\begin{figure}
    \includegraphics[width=\columnwidth]{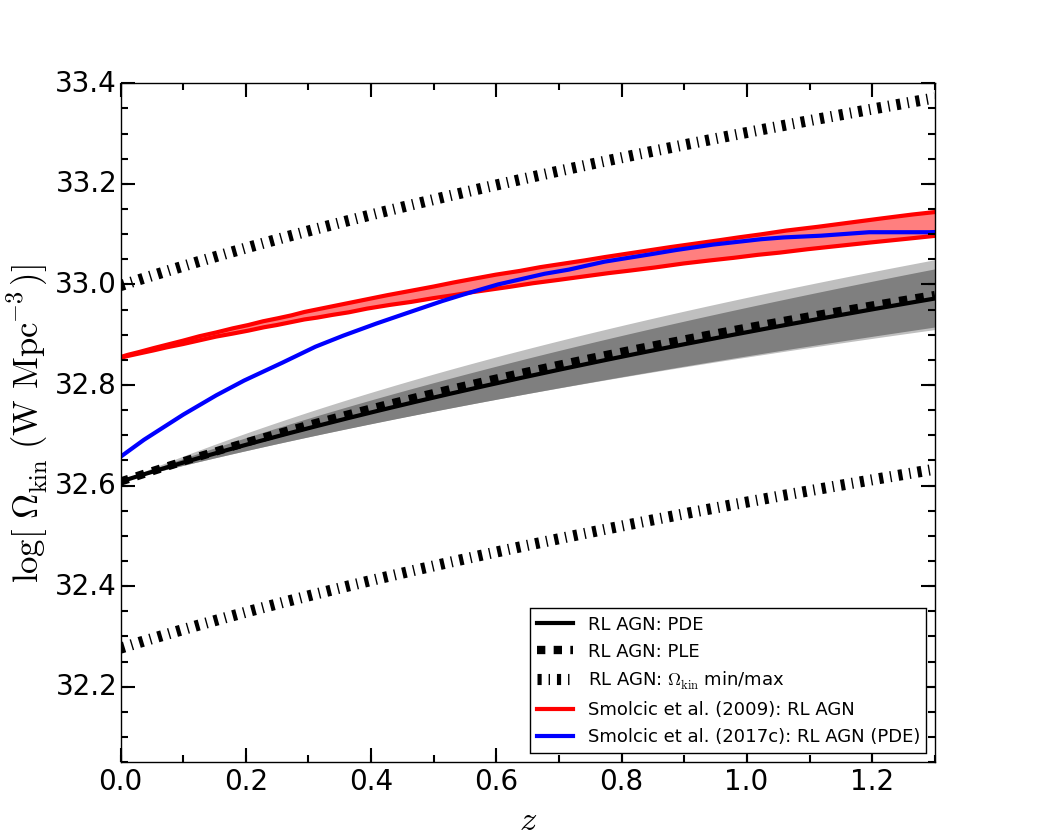}
    \caption[Evolution of the kinetic luminosity density ($\Omega_{\rm{kin}}$) for all RL AGN in XXL-S for PDE and PLE fits.]{Evolution of the kinetic luminosity density ($\Omega_{\rm{kin}}$) for all RL AGN in XXL-S for PDE (solid black line) and PLE (dashed black line) fits, integrated from log[$L_{\rm{1.4GHz}}$ (W Hz$^{-1}$)] = 22.4 to 27.2 (the full range of luminosities probed in the RLFs).  The black and grey shaded areas represent the uncertainties for the RL AGN PDE and PLE fits, respectively. For comparison, $\Omega_{\rm{kin}}$ for the RL AGN from \cite{smolcic2009a} and \cite{smolcic2017b} are displayed as the red shaded region and the blue line, respectively.  The uncertainties in the \cite{cavagnolo2010} scaling relation for the XXL-S data are shown as the black dash dot lines.  The evolution of the RL AGN in XXL-S is broadly consistent with the evolution of the RL AGN in the samples from \cite{smolcic2009a} and \cite{smolcic2017b}.}
    \label{fig:XXL-S_int_kin_lum_dens_smolcic_2009-2017_comp}
\end{figure}

\cite{smolcic2017b} extended the \cite{smolcic2009a} sample out to $z$$\sim$5 by constructing a deeper sample of $\sim$1800 radio AGN using the source catalogues from the VLA-COSMOS 3 GHz Large Project \citep{smolcic2017c} and the VLA-COSMOS 1.4 GHz Large and Deep Projects \citep{schinnerer2004,schinnerer2007,schinnerer2010}.  They did not split between HERGs and LERGs, so the results from \cite{smolcic2017b}, along with the results from \cite{smolcic2009a}, are compared to the total XXL-S RL AGN contribution to radio mode feedback in Figure \ref{fig:XXL-S_int_kin_lum_dens_smolcic_2009-2017_comp}.  \cite{smolcic2017b} primarily used the \cite{willott1999} scaling relation, so in order to properly compare their results to the XXL-S results, the \cite{cavagnolo2010} scaling relation was applied to the \cite{smolcic2017b} sample (\citealp{smolcic2009a} used the scaling relation from \citealp{birzan2008}, which is what \citealp{cavagnolo2010} is based on).  The XXL-S $\Omega_{\rm{kin}}$ at $z=0$ is log[$\Omega_{\rm{kin}}$ (W Mpc$^{-3}$)]~$\approx$~32.6 and rises to $\sim$33.0 at $z=1.3$ for both PDE and PLE. This is below both the \cite{smolcic2009a} and \cite{smolcic2017b} samples, but they are still within the uncertainties of the \cite{cavagnolo2010} scaling relation for XXL-S.  Therefore, for the same $L_{\rm{1.4GHz}}$-$L_{\rm{kin}}$ scaling relation, the $\Omega_{\rm{kin}}$ evolution result for RL AGN in XXL-S is consistent with the $\Omega_{\rm{kin}}$ evolution result for radio AGN in the \cite{smolcic2017b} and \cite{smolcic2009a} samples.

\cite{best2014} measured $\Omega_{\rm{kin}}$ for jet-mode AGN (LERG equivalent) from $0.5 < z < 1.0$ using a sample of 211 RL AGN, which they constructed by combining data from eight different surveys.  Their results (see their Figure 8) are consistent with a model in which $\Omega_{\rm{kin}}$ rises by a factor of $\sim$2 (compared to the $z=0$ value) out to $z$$\sim$0.55 and then falls to $\sim$0.7 times the local $\Omega_{\rm{kin}}$ value by $z$$\sim$0.85.  This is not consistent with the LERG $\Omega_{\rm{kin}}$ evolution seen in XXL-S, which steadily rises monotonically with redshift.  However, the sample in \cite{best2014} is more than 20 times smaller than the XXL-S sample, and most of their sample is much brighter in the radio (90\% of their radio sources have $S_{\rm{1.4GHz}} > 2$ mJy).  Therefore, the \cite{best2014} sample is not able to probe the $\Omega_{\rm{kin}}$ evolution as well as the XXL-S sample, which has allowed a more accurate $\Omega_{\rm{kin}}$ measurement due to the larger sample size and extension out to higher redshifts.

Pracy16 measured $\Omega_{\rm{kin}}$ for LERGs and HERGs for a sample of $\sim$5000 optically-matched radio galaxies with $S_{\rm{1.4GHz}} > 2.8$ mJy and $M_i < -23$ out to $z=0.75$.  Their LERG $\Omega_{\rm{kin}}$ stays constant at log[$\Omega_{\rm{kin}}$ (W Mpc$^{-3}$)]~$\approx$~32.2 for $0 < z < 1$. This is partially influenced by a redshift dependent $e$-correction, which decreases the $i$-band magnitude of each source in order to account for the fading of stellar populations with time.  Without the $e$-correction, the LERGs evolve as $K_{\rm{D}}$ = 0.81$^{+0.15}_{-0.16}$, which is within 1$\sigma$ of the XXL-S LERG value ($K_{\rm{D}}$ = 0.671 $\pm$ 0.165).  Therefore, the XXL-S LERG $\Omega_{\rm{kin}}$ evolution is in good agreement with that found by Pracy16 if no $e$-correction is applied.  However, when the $e$-correction is applied, the uncertainties in the \cite{cavagnolo2010} scaling relation for the LERGs in Pracy16 range from 32.0~$\lesssim$~log[$\Omega_{\rm{kin}}$~(W Mpc$^{-3}$)]~$\lesssim$~32.5.  This is in rough agreement with the lower uncertainty bound of the XXL-S LERG $\Omega_{\rm{kin}}$ evolution (i.e. there is $\sim$0.1-0.3 dex of overlap).  The HERG $\Omega_{\rm{kin}}$ evolution measured by Pracy16, however, is fundamentally different to the XXL-S RL HERG $\Omega_{\rm{kin}}$ evolution.  The HERGs from Pracy16 exhibit strong positive redshift evolution, contributing an average log[$\Omega_{\rm{kin}}$ (W Mpc$^{-3}$)]~$\approx$~31.5 at $z=0$ and increasing up to $\sim$32.5 at $z=1$.  The XXL-S RL HERGs, on the other hand, evolve more weakly, exhibiting average log[$\Omega_{\rm{kin}}$ (W Mpc$^{-3}$)] values of 32.0 at $z$=0 and $\sim$32.5 at $z$=1.  The difference is, again, due to the increased optical and radio depths probed by XXL-S.  In other words, the Pracy16 sample simply measured the evolution allowed by the $M_i < -23$ and $S_{\rm{1.4GHz}} > 2.8$ mJy cuts. Nevertheless, the range of $\Omega_{\rm{kin}}$ evolution of the Pracy16 HERGs (31.5~$\lesssim$~log[$\Omega_{\rm{kin}}$~(W~Mpc$^{-3}$)]~$\lesssim$~32.5) is within the range of uncertainties of the $\Omega_{\rm{kin}}$ evolution of the XXL-S HERGs.

\subsection{Comparison of RL HERG and LERG comoving kinetic luminosity densities to simulations}
\label{sec:comp_IKLDs_to_sims}

The correspondence (or lack thereof) between observations of galaxies and models of their formation and evolution is a powerful indication of how well the underlying physics involved in the models is understood.  A number of authors have made various predictions for the cosmic evolution of radio mode feedback.  A selection of these is compared to the $\Omega_{\rm{kin}}$ calculations of the XXL-S RL HERGs and LERGs.

\cite{croton2006} predicted that the black hole mass accretion rate density ($\dot{m}_{\rm{BH}}$) associated with AGN exhibiting radio mode feedback would be relatively flat at log[$\dot{m}_{\rm{BH}}$ (M$_{\odot}$ yr$^{-1}$ Mpc$^{-3}$)] $\approx$ $-5.8$ out to $z$$\sim$1.5 and decrease by an order of magnitude by $z$$\sim$4 (see their Figure 3).  The $\dot{m}_{\rm{BH}}$ values can be translated into $\Omega_{\rm{kin}}$ values via the mass-to-energy conversion of $L_{\rm{kin}} = \eta \dot{m}_{\rm{BH}} c^2$, where $\eta=0.1$ is the canonical efficiency of gravitational accretion \citep{frank1992} and $c$ is the speed of light. The low redshift ($z<1.5$) $\dot{m}_{\rm{BH}}$ value translates into log[$\Omega_{\rm{kin}}$ (W Mpc$^{-3}$)] $\approx 33.0$.  The $\Omega_{\rm{kin}}$ for all RL AGN in XXL-S weakly increases from log[$\Omega_{\rm{kin}}$ (W Mpc$^{-3}$)]~$\approx$~32.6 to $\sim$33.0 between $0 < z < 1.3$, as seen in Figure \ref{fig:XXL-S_int_kin_lum_dens_smolcic_2009-2017_comp}.  Therefore, given the uncertainties in the \cite{cavagnolo2010} scaling relation, the XXL-S $\Omega_{\rm{kin}}$ evolution for all RL AGN is in good agreement with the \cite{croton2006} prediction for the evolution of radio mode feedback for $0 < z < 1.3$.  However, if another scaling relation is used, the agreement is poorer.

\begin{figure}
        \includegraphics[width=\columnwidth]{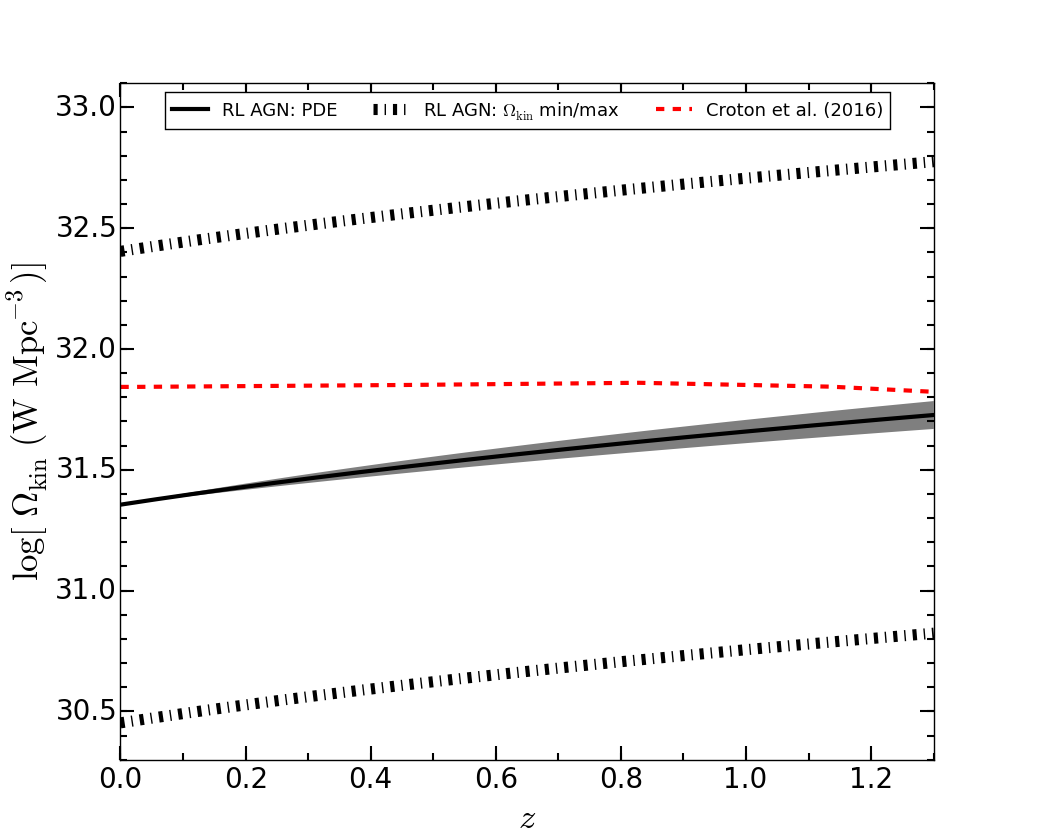}
    \caption[Evolution of the kinetic luminosity density ($\Omega_{\rm{kin}}$) for all RL AGN in XXL-S using the scaling relation from \cite{willott1999} with $f_W=4$.]{Evolution of $\Omega_{\rm{kin}}$ for all RL AGN in XXL-S (black line) using the scaling relation from \cite{willott1999} with $f_W=4$, integrated from log[$L_{\rm{1.4GHz}}$ (W Hz$^{-1}$)] = 22.4 to 27.2 (the full range of luminosities probed in the RLFs).  Only the PDE evolution is displayed for clarity (the PLE results are nearly indistinguishable from the PDE results on this scale).  The prediction from \cite{croton2016} over this redshift range (dashed red line) is within the uncertainties of the XXL-S RL AGN $\Omega_{\rm{kin}}$ calculated using the \cite{willott1999} relation (defined by using $f_W$=1 and $f_W$=20, shown as the black dash-dot lines).}
    \label{fig:IKLD_vs_z_Willott}
\end{figure}

\cite{croton2016} updated the \cite{croton2006} prediction with the addition of a `radio mode efficiency' parameter, $\kappa_{\rm{R}}$, for which the authors adopted a value of $\kappa_{\rm{R}} = 0.08$.  The motivation behind this modification is that the \cite{croton2006} model used an upper limit to the cooling rate of infalling gas by assuming that the cooling and heating are independent.  However, in the real universe, AGN heating would have a lasting effect on the gas.  Therefore, less AGN heating is required to offset the cooling flows.  The \cite{croton2016} model, called the Semi-Analytic Galaxy Evolution (SAGE) model, predicts that the $\Omega_{\rm{kin}}$ from radio mode feedback for $z < 1.5$ is log[$\Omega_{\rm{kin}}$~(W~Mpc$^{-3}$)]~$\approx~31.8$ (approximately ten times less than the previous prediction).  The XXL-S measurement of the $\Omega_{\rm{kin}}$ evolution of RL AGN, using the scaling relation from \cite{cavagnolo2010}, is inconsistent with this value, even considering the uncertainties.  However, if the scaling relation from \cite{willott1999} with an uncertainty parameter of $f_W=4$ is used for the XXL-S $\Omega_{\rm{kin}}$ calculation, then the $\Omega_{\rm{kin}}$ for RL AGN in XXL-S is log[$\Omega_{\rm{kin}}$ (W Mpc$^{-3}$)] $\approx$ 31.4 at $z=0$ and reaches $\sim$31.7 by $z=1.3$ for PDE, as shown in Figure \ref{fig:IKLD_vs_z_Willott}.  This result is within $\sim$0.3 dex of the \cite{croton2016} prediction, which is well within the full extent of the uncertainties in the \cite{willott1999} relation.  The $\Omega_{\rm{kin}}$ evolution for radio AGN in the \cite{smolcic2017b} sample is also most consistent with the prediction from \cite{croton2016} if the \cite{willott1999} scaling relation with $f_W=4$ is used (especially for $z>1$).

\begin{figure}
        \includegraphics[width=\columnwidth]{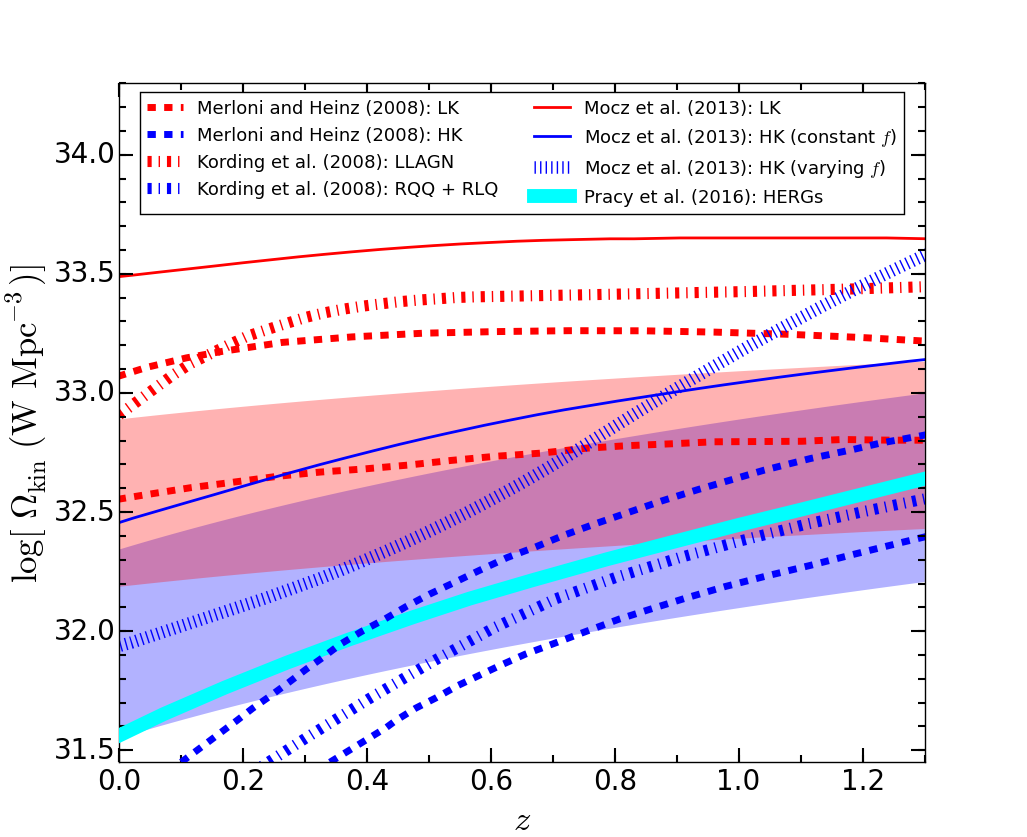}
    \caption[Comparison between the evolution of the kinetic luminosity densities ($\Omega_{\rm{kin}}$) for LERGs and RL HERGs in XXL-S using the $L_{\rm{1.4GHz}}$-$L_{\rm{kin}}$ scaling relation from \cite{cavagnolo2010} and the prediction for LERG and RL HERG equivalent sources from \cite{merloni2008}, \cite{kording2008}, and \cite{mocz2013}.]{Comparison between the evolution of $\Omega_{\rm{kin}}$ for XXL-S LERGs and RL HERGs (red and blue shaded regions, respectively) and the prediction for LERG and RL HERG equivalent sources from \cite{merloni2008}, \cite{kording2008}, and \cite{mocz2013}.  The hashed blue line is the scenario for RL HERG equivalent sources from \cite{mocz2013} when a varying radio AGN duty cycle ($f$) is utilised.  The scaling relation from \cite{cavagnolo2010} was used to calculate the XXL-S $\Omega_{\rm{kin}}$ values, and the RLFs were integrated from log[$L_{\rm{1.4GHz}}$~(W~Hz$^{-1}$)]~=~22.4 to 27.2 (the full range of luminosities probed in the RLFs).  The thick cyan curve shows the evolution of $\Omega_{\rm{kin}}$ for the HERGs from Pracy16 (the values for $z > 1$ have been linearly extrapolated).}
    \label{fig:XXL-S_int_kin_lum_density_RL_HERG_LERG_vs_z_sims_comp}
\end{figure}

More predictions of the evolution of radio mode feedback have been made by \cite{merloni2008}, \cite{kording2008} and \cite{mocz2013}, who all separately modelled the $\Omega_{\rm{kin}}$ for their RL HERG and LERG equivalent populations.   Figure \ref{fig:XXL-S_int_kin_lum_density_RL_HERG_LERG_vs_z_sims_comp} shows these predictions alongside the $\Omega_{\rm{kin}}$ measurements for RL HERGs and LERGs in XXL-S (using the \citealp{cavagnolo2010} scaling relation), the full uncertainty ranges of which are shown as the semi-transparent blue and red shaded regions, respectively.  \cite{merloni2008} called their LERG equivalent population `low kinetic mode', or LK, and their RL HERG equivalent population `radio loud high kinetic mode', or HK.  Their scenario in which both flat- and steep-spectrum radio sources are included in the simulations is considered here.  \cite{kording2008} labelled their LERG equivalent population `low luminosity AGN', or LLAGN.  Their RL HERG equivalent population was obtained by combining radio quiet (low radio-to-optical luminosity ratio) and radio loud quasars (`RQQ' and `RLQ', respectively).  Like \cite{merloni2008}, \cite{mocz2013} designated their LERG and RL HERG equivalent populations as LK and HK.  Their HK prediction involves two scenarios: one in which the radio AGN duty cycle (the fraction of HK sources with radio jets switched on) is fixed at $f=0.1$ and one in which $f$ evolves with redshift.

All three simulations mentioned previously make very similar predictions for the $\Omega_{\rm{kin}}$ evolution of the LERG equivalent populations: their slopes are consistent with the observed weak evolution of the XXL-S LERGs, but their normalisations are systematically higher.  This is due to the simulations using different $L_{\rm{1.4GHz}}$-$L_{\rm{kin}}$ scaling relations, which results in different kinetic luminosity distributions.  For example, \cite{merloni2008} used the scaling relation from \cite{merloni2007}, which generates $\Omega_{\rm{kin}}$ values that are higher than the \cite{cavagnolo2010} scaling relation by $\sim$0.2-0.3 dex for a given redshift and log($L_{\rm{1.4GHz}}$) integration range (see Figure A.2 in \citealp{smolcic2017b}).  This is consistent with the offset seen in Figure \ref{fig:XXL-S_int_kin_lum_density_RL_HERG_LERG_vs_z_sims_comp}.  Nevertheless, the consistency between the slope of the $\Omega_{\rm{kin}}$ evolution of the XXL-S LERGs and the predictions from the simulations for their LERG equivalent populations indicates that the current understanding of the physics of the evolution of slowly accreting SMBHs is well-matched to LERG observations.

On the other hand, the \cite{merloni2008} and \cite{kording2008} predict that their HK and RQQ+RLQ populations evolve strongly.  As seen in Figure \ref{fig:XXL-S_int_kin_lum_density_RL_HERG_LERG_vs_z_sims_comp}, these predictions agree with the evolution of the HERGs from Pracy16 and the XXL-S RL HERGs for $z \gtrsim 0.5$, given the uncertainties in the \cite{cavagnolo2010} scaling relation.  The disagreement at lower redshifts reflects the assumption in the simulations that the HERG population is dominated by high luminosity sources that evolve strongly.  The slope of the HK predictions from \cite{mocz2013} are more consistent with the slope of the XXL-S RL HERG evolution for $z \lesssim 0.5$.  Beyond this redshift, the $\Omega_{\rm{kin}}$ slope for the constant $f$ scenario remains consistent with the XXL-S RL HERGs, but the $\Omega_{\rm{kin}}$ slope for the varying $f$ scenario increases, becoming inconsistent with the XXL-S RL HERGs by $z \approx 0.75$.  This may suggest that an evolving AGN duty cycle does not accurately reflect how RL HERGs accrete through cosmic time.  Regardless, the \cite{mocz2013} prediction for the $\Omega_{\rm{kin}}$ evolution of their HK population with a constant AGN duty cycle is most closely aligned with the evolution of the XXL-S RL HERGs out of all three simulations, with only a slight ($\sim$0.1 dex) normalisation discrepancy across all redshifts.

\subsection{Impact of potential HERG misclassifications}
\label{sec:impact_of_HERG_misclassifications}

It is possible that some RL HERGs were misclassified due to the lack of far-infrared (FIR) photometric data, which aids in constraining SED fits and the corresponding derived SFRs (e.g. \citealp{delvecchio2014}).  In order to test this, a comparison between SED fitting results with and without FIR constraints for optically faint sources was performed.  The COSMOS field has FIR \emph{Herschel} data available, and therefore a sub-sample of the 3 GHz COSMOS radio sources with faint optical counterparts was constructed.  In order to match the XXL-S radio sample as closely as possible, the following cuts were applied to the COSMOS source catalogue: $m_z > 24$ and $S_{\rm{1.8GHz}} > 200$ $\mu$Jy (where $S_{\rm{1.8GHz}}$ was converted from the $S_{\rm{3GHz}}$ value by using a spectral index of $\alpha = -0.7$).  These cuts resulted in a sub-sample of 411 COSMOS sources, some of which have 3$\sigma$ \emph{Herschel} detections and some of which do not (upper limits to the FIR flux densities were used for the latter sources).  The SED fitting was performed twice for each source: once with the FIR data and once without it.  For $\sim$92\% of the sources, the SED classification remained the same (AGN were still classified as AGN and SFGs were still classified as SFGs after the FIR data was removed).  For the remaining $\sim$8\%, the classifications changed (previously identified AGN were classified as SFGs and previously identified SFGs were classified as AGN).  Therefore, it is expected that no more than approximately $\sim$8\% of the RL HERGs would be reclassified as RQ HERGs if FIR data became available for XXL-S.  Even if this percentage of expected potential misclassifications was reached, any change in classification would be spread out among the different redshift and $L_{\rm{1.4GHz}}$ bins.  Therefore, potential misclassifications of the RL HERGs are expected to have minimal impact on the overall results of their evolution. In terms of the SFRs, there was a scatter of 0.3 dex between the SED fits with the FIR data and the fits without the FIR data, but no significant offset between the two runs was found.  Therefore, the radio excess parameter that identified low radio luminosity ($L_{\rm{1.4GHz}} < 10^{24.5}$ W Hz$^{-1}$) RL HERGs would not be strongly affected by the presence of FIR \emph{Herschel} data.

In addition, the percentage of optically bright ($M_i \lesssim -23.8$) and optically faint ($M_i \gtrsim -23.8$) XXL-S RL HERGs that are X-ray AGN, SED AGN, and MIR AGN were compared.  The percentages of optically bright RL HERGs that are X-ray, SED, and MIR AGN (28.9\%, 84.6\%, and 19.6\%, respectively) are similar (within 1$\sigma$ of the $\sqrt{N}/N$ Poissonian uncertainty) to the percentages of optically faint RL HERGs that are X-ray, SED, and MIR AGN (24.1\%, 74.1\%, and 11.2\%, respectively).  Very similar results were found when comparing the RL HERGs with high radio luminosity ($L_{\rm{1.4GHz}} > 10^{24.7}$ W Hz$^{-1}$) and ones with low radio luminosity ($L_{\rm{1.4GHz}} < 10^{24.7}$ W Hz$^{-1}$).  This demonstrates that the classification scheme applied to the XXL-S radio sources is relatively insensitive to signal-to-noise ratio and photometric data quality.

\subsection{Implications of XXL-S radio mode feedback results for galaxy evolution}

\begin{figure}
        \includegraphics[width=\columnwidth]{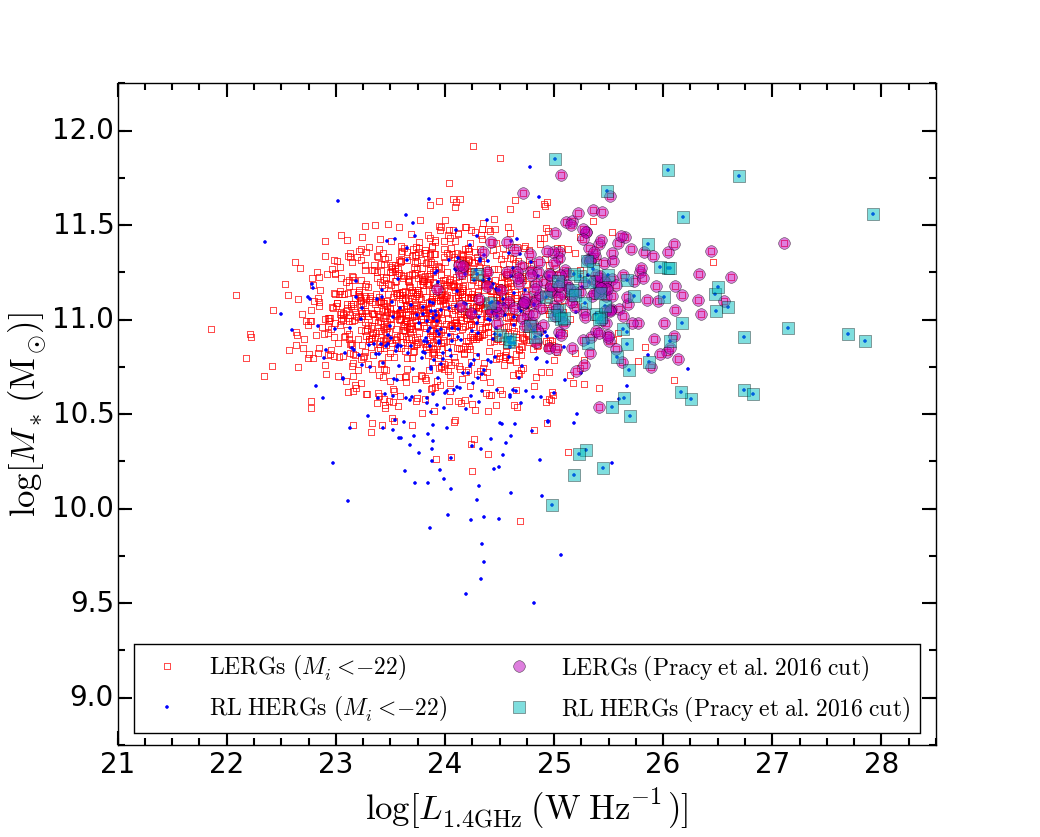}
    \caption[Stellar mass vs 1.4 GHz radio luminosity for XXL-S RL HERGs and LERGs $z<1.3$ with $M_i < -22$, showing the selection bias in the cuts made by Pracy16.]{Stellar mass vs 1.4 GHz radio luminosity for RL HERGs (blue dots) and LERGs (open red squares) at $z<1.3$ with $M_i < -22$.  The semi-transparent cyan squares and magenta circles show the RL HERGs and LERGs, respectively, that would have been selected by the cuts that Pracy16 made ($M_i < -23$ and $S_{\rm{1.4GHz}} > 2.8$ mJy).  The latter galaxies are some of the most massive, most radio luminous in the XXL-S sample.}
    \label{fig:M_star_vs_L_1400MHz_Pracy_cut_comp}
\end{figure}

The Pracy16 sample selected only the most radio luminous, most massive galaxies, which is demonstrated in Figure \ref{fig:M_star_vs_L_1400MHz_Pracy_cut_comp}.  Clearly, making this selection excludes a large number of lower mass ($M_* \lesssim 10^{10.5}$ M$_{\odot}$) and radio faint ($L_{\rm{1.4GHz}} \lesssim 10^{24.5}$ W Hz$^{-1}$) galaxies, and even misses some sources with higher radio luminosity in the XXL-S sample.  As described in Section \ref{sec:comp_IKLDs_to_other_samples}, this results in a different measurement of the evolution of the RL HERG population.  This implies that the evolution measured for RL HERGs depends on the range of optical luminosities included by the sample selection.  This has more of an effect at lower redshift and for higher radio luminosities, since fainter optical sources are more likely to exist at lower redshift and the majority of the feedback power ($L_{\rm{kin}}$) in a given redshift bin comes from sources with luminosities that are close to the knee in the RLF.  For XXL-S RL HERGs with $L_{\rm{1.4GHz}} > 10^{24.5}$ W Hz$^{-1}$ at $z<0.6$, those selected by Pracy16's $M_i < -23$ cut account for $\sim$38\% of the total $L_{\rm{kin}}$ emitted by all RL HERGs with $M_i < -22$ in that volume, whereas those that were added by the deeper XXL-S optical cut ($-23 < M_i < -22$) account for $\sim$49\% of the total $L_{\rm{kin}}$ from all RL HERGs with $M_i < -22$ in that volume (the remaining $\sim$13\% comes from the lower luminosity RL HERGs).  The corresponding percentages for XXL-S RL HERGs with $L_{\rm{1.4GHz}} > 10^{24.5}$ W Hz$^{-1}$ at $0.6 < z< 1.3$ are $\sim$93\% and $\sim$4\%, respectively. This means that the inclusion of the XXL-S sources with $-23 < M_i < -22$ increased the measurement of $\Omega_{\rm{kin}}$ for XXL-S RL HERGs at $z<0.6$ by up to a factor of $\sim$2, while the $\Omega_{\rm{kin}}$ values for $0.6 < z< 1.3$ were virtually unaffected. This is reflected in the different $\Omega_{\rm{kin}}$ slopes between the XXL-S RL HERGs and those from Pracy16 for $z < 0.6$ (see Figure \ref{fig:XXL-S_int_kin_lum_density_RL_HERG_LERG_vs_z_sims_comp}).
Overall, this results in lower PDE and PLE measurements for the evolution of RL HERGs and demonstrates the impact that deeper optical and radio data can have on the calculation of radio mode feedback in RL AGN samples.  Furthermore, Figure \ref{fig:M_i_decam_vs_z} shows that a significant number of even fainter ($M_i > -22$) RL HERGs exist.  If deeper optical samples can be constructed in future surveys, the evolution of the RL HERG population may be found to be weaker still.  Deeper optical data would not affect LERGs as much because their hosts tend to be inherently brighter than RL HERG hosts (only the most shallow flux limited surveys miss a substantial portion of the LERG population).  All these results suggest that RL HERGs contribute more to radio mode feedback at low redshifts ($z \lesssim 0.6$) than previously thought due to the inclusion of optically fainter ($-23 < M_i < -22$) RL HERGs.

\section{Summary and conclusions}
\label{sec:conclusion}

Radio mode feedback is an important element in galaxy evolution because of its influence on a galaxy's ability to form stars.  Measuring the amount of radio mode feedback throughout cosmic history puts important constraints on the amount of power available to prevent star formation, thus allowing an evaluation of its role in limiting the stellar mass of galaxies.  The XXL-S field was observed at 2.1 GHz with ATCA for the purpose of measuring the evolution of radio mode feedback of the RL HERGs and LERGs therein.

The wide area ($\sim$23.3 deg$^2$) and relatively deep radio data ($\sigma$$\sim$41 $\mu$Jy beam$^{-1}$) of XXL-S allowed the construction of the RLFs for both RL HERGs and LERGs across a wide range in radio luminosity out to $z$$\sim$1.3, which was not possible previously due to small sky coverage or shallow radio depths.  Using the RLFs constructed in four redshift bins, the evolution in the RLFs was measured for all RL HERGs and LERGs with $M_i < -22$.  The kinetic luminosity density $(\Omega_{\rm{kin}}$) evolution of all RL AGN in XXL-S is consistent with other samples of RL AGN surveyed at similar optical and radio depths (e.g. \citealp{smolcic2009a,smolcic2017b}) within the uncertainties of the \cite{cavagnolo2010} scaling relation.  The LERGs contributed the majority of the total $\Omega_{\rm{kin}}$ at a given redshift and exhibited positive yet weak evolution ($K_{\rm{D}}~=~0.67~\pm~0.17$ and $K_{\rm{L}}~=~0.84~\pm~0.31$).  This implies that LERGs account for most of the radio mode feedback throughout cosmic history and that accretion onto SMBHs in massive, passively evolving galaxies (which comprise the vast majority of the LERG population) has been steadily decreasing since $z$$\sim$1.3.  On the other hand, the RL HERG RLFs displayed stronger evolution ($K_{\rm{D}} = 1.81 \pm 0.15$ and $K_{\rm{L}} = 3.19 \pm 0.29$).  However, the latter result is weaker than the previously measured strong evolution in the HERG population (e.g. Pracy16).  This implies that radio mode feedback from SMBHs existing in bluer, star-forming hosts is more prominent in recent cosmic history than previously thought.  In turn, this suggests that radio mode feedback in RL HERGs, and not just LERGs, is important for understanding the mechanism behind radio mode feedback and its ability to limit the mass of galaxies in the universe.

The evolution results for RL HERG and LERGs in XXL-S were compared to the predictions of simulations of radio mode feedback.  The latest simulation \citep{croton2016} predicts an approximately constant value (log[$\Omega_{\rm{kin}}$ (W Hz$^{-1}$)] $\approx$ 31.8) out to $z$ $\sim$ 1.3.  If the \cite{willott1999} scaling relation is used, the total $\Omega_{\rm{kin}}$ from all RL AGN in XXL-S is consistent with the prediction from \cite{croton2016} for $0 < z < 1.3$.  Other simulations \citep{merloni2008,kording2008,mocz2013} expressed their predictions for radio mode feedback with separate contributions from LERG and RL HERG equivalent populations.  All three simulations made similar predictions for the LERG equivalent populations: they have similar slopes to but positive normalisation offsets above the XXL-S LERG measurement.  This indicates that models of slowly-accreting SMBHs undergoing advection-dominated accretion flows correspond closely to observations of LERGs, with the difference in normalisation being due to the details of the conversion from radio luminosity to kinetic luminosity.  The $\Omega_{\rm{kin}}$ predictions from \cite{merloni2008} and \cite{kording2008} for the RL HERG equivalent populations correspond more closely with the HERG evolution from Pracy16, which evolved strongly.  On the other hand, the \cite{mocz2013} prediction for the RL HERG equivalent population with a constant AGN duty cycle ($f=0.1$) had a similar slope to the XXL-S RL HERG evolution, but was slightly offset above it in normalisation (by $\sim$0.1 dex) for $0 < z < 1.3$.  This suggests that a constant AGN duty cycle could be responsible for producing a higher local abundance and relatively weak evolution of SMBHs rapidly accreting cold gas, as found for the XXL-S RL HERGs.   However, the mechanism that generates the constant duty cycle is unknown.

\section*{Acknowledgements}

AB acknowledges the University of Western Australia (UWA) for funding support from a University Postgraduate Award PhD scholarship and the International Centre for Radio Astronomy Research (ICRAR) for additional support.  VS acknowledges funding from the European Union's Seventh Framework programme under grant agreement 333654 (CIG, `AGN feedback'), and VS and ID acknowledge funding under grant agreement 337595 (ERC Starting Grant, `CoSMass'). The Saclay group acknowledges long-term support from the Centre National d'Etudes Spatiales (CNES).  All the authors thank the referee for his helpful comments that improved the clarity and quality of the paper. AB also thanks Elizabeth Mahoney, Tom Muxlow, and Tim Heckman for additional comments.
The Australia Telescope Compact Array is part of the Australia Telescope National Facility which is funded by the Australian Government for operation as a National Facility managed by CSIRO.  XXL is an international project based around an XMM Very Large Programme surveying two 25 deg$^2$ extragalactic fields at a depth of $\sim$6 $\times$ 10$^{-15}$ erg cm$^{-2}$ s$^{-1}$ in the [0.5-2] keV band for point-like sources. The XXL website is http://irfu.cea.fr/xxl. Multi-band information and spectroscopic follow-up of the X-ray sources are obtained through a number of survey programmes, summarised at http://xxlmultiwave.pbworks.com/.


\bibliographystyle{aa}
\bibliography{paper_3_aa}

\addcontentsline{toc}{chapter}{Bibliography}
\begin{thebibliography}{101}
\expandafter\ifx\csname natexlab\endcsname\relax\def\natexlab#1{#1}\fi

\bibitem[{{Allen} {et~al.}(2006){Allen}, {Dunn}, {Fabian}, {Taylor}, \&
  {Reynolds}}]{allen2006}
{Allen}, S.~W., {Dunn}, R.~J.~H., {Fabian}, A.~C., {Taylor}, G.~B., \&
  {Reynolds}, C.~S. 2006, \mnras, 372, 21

\bibitem[{{Best} \& {Heckman}(2012)}]{best2012}
{Best}, P.~N. \& {Heckman}, T.~M. 2012, \mnras, 421, 1569

\bibitem[{{Best} {et~al.}(2006){Best}, {Kaiser}, {Heckman}, \&
  {Kauffmann}}]{best2006}
{Best}, P.~N., {Kaiser}, C.~R., {Heckman}, T.~M., \& {Kauffmann}, G. 2006,
  \mnras, 368, L67

\bibitem[{{Best} {et~al.}(2005{\natexlab{a}}){Best}, {Kauffmann}, {Heckman},
  {Brinchmann}, {Charlot}, {Ivezi{\'c}}, \& {White}}]{best2005a}
{Best}, P.~N., {Kauffmann}, G., {Heckman}, T.~M., {et~al.} 2005{\natexlab{a}},
  \mnras, 362, 25

\bibitem[{{Best} {et~al.}(2005{\natexlab{b}}){Best}, {Kauffmann}, {Heckman}, \&
  {Ivezi{\'c}}}]{best2005b}
{Best}, P.~N., {Kauffmann}, G., {Heckman}, T.~M., \& {Ivezi{\'c}}, {\v Z}.
  2005{\natexlab{b}}, \mnras, 362, 9

\bibitem[{{Best} {et~al.}(2014){Best}, {Ker}, {Simpson}, {Rigby}, \&
  {Sabater}}]{best2014}
{Best}, P.~N., {Ker}, L.~M., {Simpson}, C., {Rigby}, E.~E., \& {Sabater}, J.
  2014, \mnras, 445, 955

\bibitem[{{Binney} \& {Tabor}(1995)}]{binney1995}
{Binney}, J. \& {Tabor}, G. 1995, \mnras, 276, 663

\bibitem[{{B{\^i}rzan} {et~al.}(2008){B{\^i}rzan}, {McNamara}, {Nulsen},
  {Carilli}, \& {Wise}}]{birzan2008}
{B{\^i}rzan}, L., {McNamara}, B.~R., {Nulsen}, P.~E.~J., {Carilli}, C.~L., \&
  {Wise}, M.~W. 2008, \apj, 686, 859

\bibitem[{{B{\^i}rzan} {et~al.}(2004){B{\^i}rzan}, {Rafferty}, {McNamara},
  {Wise}, \& {Nulsen}}]{birzan2004}
{B{\^i}rzan}, L., {Rafferty}, D.~A., {McNamara}, B.~R., {Wise}, M.~W., \&
  {Nulsen}, P.~E.~J. 2004, \apj, 607, 800

\bibitem[{{B\"{o}hringer} {et~al.}(1993){B\"{o}hringer}, {Voges}, {Fabian},
  {Edge}, \& {Neumann}}]{bohringer1993}
{B\"{o}hringer}, H., {Voges}, W., {Fabian}, A.~C., {Edge}, A.~C., \& {Neumann},
  D.~M. 1993, \mnras, 264, L25

\bibitem[{{Bower} {et~al.}(2006){Bower}, {Benson}, {Malbon}, {Helly}, {Frenk},
  {Baugh}, {Cole}, \& {Lacey}}]{bower2006}
{Bower}, R.~G., {Benson}, A.~J., {Malbon}, R., {et~al.} 2006, \mnras, 370, 645

\bibitem[{{Butler} {et~al.}(2018{\natexlab{a}}){Butler}, {Huynh}, {Delhaize},
  {Smol{\v c}i{\'c}}, {Kapi{\'n}ska}, {Milakovi{\'c}}, {Novak}, {Baran},
  {O'Brien}, {Chiappetti}, {Desai}, {Fotopoulou}, {Horellou}, {Lidman}, \&
  {Pierre}}]{butler2018_xxl18}
{Butler}, A., {Huynh}, M., {Delhaize}, J., {et~al.} 2018{\natexlab{a}}, \aap,
  620, A3 (XXL Survey, XVIII)

\bibitem[{{Butler} {et~al.}(2018{\natexlab{b}}){Butler}, {Huynh}, {Delvecchio},
  {Kapi{\'n}ska}, {Ciliegi}, {Jurlin}, {Delhaize}, {Smol{\v c}i{\'c}}, {Desai},
  {Fotopoulou}, {Lidman}, {Pierre}, \& {Plionis}}]{butler2018_xxl31}
{Butler}, A., {Huynh}, M., {Delvecchio}, I., {et~al.} 2018{\natexlab{b}}, \aap,
  620, A16 (XXL Survey, XXXI)

\bibitem[{{Cattaneo} {et~al.}(2009){Cattaneo}, {Faber}, {Binney}, {Dekel},
  {Kormendy}, {Mushotzky}, {Babul}, {Best}, {Br{\"u}ggen}, {Fabian}, {Frenk},
  {Khalatyan}, {Netzer}, {Mahdavi}, {Silk}, {Steinmetz}, \&
  {Wisotzki}}]{cattaneo2009a}
{Cattaneo}, A., {Faber}, S.~M., {Binney}, J., {et~al.} 2009, \nat, 460, 213

\bibitem[{{Cavagnolo} {et~al.}(2010){Cavagnolo}, {McNamara}, {Nulsen},
  {Carilli}, {Jones}, \& {B{\^i}rzan}}]{cavagnolo2010}
{Cavagnolo}, K.~W., {McNamara}, B.~R., {Nulsen}, P.~E.~J., {et~al.} 2010, \apj,
  720, 1066

\bibitem[{{Ceraj} {et~al.}(2018){Ceraj}, {Smol{\v c}i{\'c}}, {Delvecchio},
  {Novak}, {Zamorani}, {Delhaize}, {Schinnerer}, {Vardoulaki}, \& {Herrera
  Ruiz}}]{ceraj2018}
{Ceraj}, L., {Smol{\v c}i{\'c}}, V., {Delvecchio}, I., {et~al.} 2018, \aap,
  620, A192

\bibitem[{{Ching} {et~al.}(2017{\natexlab{a}}){Ching}, {Croom}, {Sadler},
  {Robotham}, {Brough}, {Baldry}, {Bland-Hawthorn}, {Colless}, {Driver},
  {Holwerda}, {Hopkins}, {Jarvis}, {Johnston}, {Kelvin}, {Liske}, {Loveday},
  {Norberg}, {Pracy}, {Steele}, {Thomas}, \& {Wang}}]{ching2017b}
{Ching}, J.~H.~Y., {Croom}, S.~M., {Sadler}, E.~M., {et~al.}
  2017{\natexlab{a}}, \mnras, 469, 4584

\bibitem[{{Ching} {et~al.}(2017{\natexlab{b}}){Ching}, {Sadler}, {Croom},
  {Johnston}, {Pracy}, {Couch}, {Hopkins}, {Jurek}, \& {Pimbblet}}]{ching2017a}
{Ching}, J.~H.~Y., {Sadler}, E.~M., {Croom}, S.~M., {et~al.}
  2017{\natexlab{b}}, \mnras, 464, 1306

\bibitem[{{Ciliegi} {et~al.}(2018){Ciliegi}, {Jurlin}, {Butler}, {Delhaize},
  {Fotopoulou}, {Huynh}, {Iovino}, {Smol{\v c}i{\'c}}, {Chiappetti}, \&
  {Pierre}}]{ciliegi2018_xxl26}
{Ciliegi}, P., {Jurlin}, N., {Butler}, A., {et~al.} 2018, \aap, 620, A11 (XXL
  Survey, XXVI)

\bibitem[{{Croton} {et~al.}(2006){Croton}, {Springel}, {White}, {De Lucia},
  {Frenk}, {Gao}, {Jenkins}, {Kauffmann}, {Navarro}, \& {Yoshida}}]{croton2006}
{Croton}, D.~J., {Springel}, V., {White}, S.~D.~M., {et~al.} 2006, \mnras, 365,
  11

\bibitem[{{Croton} {et~al.}(2016){Croton}, {Stevens}, {Tonini}, {Garel},
  {Bernyk}, {Bibiano}, {Hodkinson}, {Mutch}, {Poole}, \&
  {Shattow}}]{croton2016}
{Croton}, D.~J., {Stevens}, A.~R.~H., {Tonini}, C., {et~al.} 2016, \apjs, 222,
  22

\bibitem[{{Daly} {et~al.}(2012){Daly}, {Sprinkle}, {O'Dea}, {Kharb}, \&
  {Baum}}]{daly2012}
{Daly}, R.~A., {Sprinkle}, T.~B., {O'Dea}, C.~P., {Kharb}, P., \& {Baum}, S.~A.
  2012, \mnras, 423, 2498

\bibitem[{{Delvecchio} {et~al.}(2014){Delvecchio}, {Gruppioni}, {Pozzi},
  {Berta}, {Zamorani}, {Cimatti}, {Lutz}, {Scott}, {Vignali}, {Cresci},
  {Feltre}, {Cooray}, {Vaccari}, {Fritz}, {Le Floc'h}, {Magnelli}, {Popesso},
  {Oliver}, {Bock}, {Carollo}, {Contini}, {Le F{\'e}vre}, {Lilly}, {Mainieri},
  {Renzini}, \& {Scodeggio}}]{delvecchio2014}
{Delvecchio}, I., {Gruppioni}, C., {Pozzi}, F., {et~al.} 2014, \mnras, 439,
  2736

\bibitem[{{Desai} {et~al.}(2012){Desai}, {Armstrong}, {Mohr}, {Semler}, {Liu},
  {Bertin}, {Allam}, {Barkhouse}, {Bazin}, {Buckley-Geer}, {Cooper}, {Hansen},
  {High}, {Lin}, {Lin}, {Ngeow}, {Rest}, {Song}, {Tucker}, \&
  {Zenteno}}]{desai2012}
{Desai}, S., {Armstrong}, R., {Mohr}, J.~J., {et~al.} 2012, \apj, 757, 83

\bibitem[{{Desai} {et~al.}(2015){Desai}, {Mohr}, {Henderson}, {K{\"u}mmel},
  {Paech}, \& {Wetzstein}}]{desai2015}
{Desai}, S., {Mohr}, J.~J., {Henderson}, R., {et~al.} 2015, Journal of
  Instrumentation, 10, C06014

\bibitem[{{Dunlop} \& {Peacock}(1990)}]{dunlop1990}
{Dunlop}, J.~S. \& {Peacock}, J.~A. 1990, \mnras, 247, 19

\bibitem[{{English} {et~al.}(2016){English}, {Hardcastle}, \&
  {Krause}}]{english2016}
{English}, W., {Hardcastle}, M.~J., \& {Krause}, M.~G.~H. 2016, \mnras, 461,
  2025

\bibitem[{{Fabian}(2012)}]{fabian2012}
{Fabian}, A.~C. 2012, \araa, 50, 455

\bibitem[{{Flaugher} {et~al.}(2015){Flaugher}, {Diehl}, {Honscheid}, {Abbott},
  {Alvarez}, {Angstadt}, {Annis}, {Antonik}, {Ballester}, {Beaufore},
  {Bernstein}, {Bernstein}, {Bigelow}, {Bonati}, {Boprie}, {Brooks},
  {Buckley-Geer}, {Campa}, {Cardiel-Sas}, {Castander}, {Castilla}, {Cease},
  {Cela-Ruiz}, {Chappa}, {Chi}, {Cooper}, {da Costa}, {Dede}, {Derylo},
  {DePoy}, {de Vicente}, {Doel}, {Drlica-Wagner}, {Eiting}, {Elliott}, {Emes},
  {Estrada}, {Fausti Neto}, {Finley}, {Flores}, {Frieman}, {Gerdes},
  {Gladders}, {Gregory}, {Gutierrez}, {Hao}, {Holland}, {Holm}, {Huffman},
  {Jackson}, {James}, {Jonas}, {Karcher}, {Karliner}, {Kent}, {Kessler},
  {Kozlovsky}, {Kron}, {Kubik}, {Kuehn}, {Kuhlmann}, {Kuk}, {Lahav}, {Lathrop},
  {Lee}, {Levi}, {Lewis}, {Li}, {Mandrichenko}, {Marshall}, {Martinez},
  {Merritt}, {Miquel}, {Mu{\~n}oz}, {Neilsen}, {Nichol}, {Nord}, {Ogando},
  {Olsen}, {Palaio}, {Patton}, {Peoples}, {Plazas}, {Rauch}, {Reil}, {Rheault},
  {Roe}, {Rogers}, {Roodman}, {Sanchez}, {Scarpine}, {Schindler}, {Schmidt},
  {Schmitt}, {Schubnell}, {Schultz}, {Schurter}, {Scott}, {Serrano}, {Shaw},
  {Smith}, {Soares-Santos}, {Stefanik}, {Stuermer}, {Suchyta}, {Sypniewski},
  {Tarle}, {Thaler}, {Tighe}, {Tran}, {Tucker}, {Walker}, {Wang}, {Watson},
  {Weaverdyck}, {Wester}, {Woods}, {Yanny}, \& {DES
  Collaboration}}]{flaugher2015}
{Flaugher}, B., {Diehl}, H.~T., {Honscheid}, K., {et~al.} 2015, \aj, 150, 150

\bibitem[{{Forman} {et~al.}(2005){Forman}, {Nulsen}, {Heinz}, {Owen}, {Eilek},
  {Vikhlinin}, {Markevitch}, {Kraft}, {Churazov}, \& {Jones}}]{forman2005}
{Forman}, W., {Nulsen}, P., {Heinz}, S., {et~al.} 2005, \apj, 635, 894

\bibitem[{{Fotopoulou} {et~al.}(2016){Fotopoulou}, {Pacaud}, {Paltani},
  {Ranalli}, {Ramos-Ceja}, {Faccioli}, {Plionis}, {Adami}, {Bongiorno},
  {Brusa}, {Chiappetti}, {Desai}, {Elyiv}, {Lidman}, {Melnyk}, {Pierre},
  {Piconcelli}, {Vignali}, {Alis}, {Ardila}, {Arnouts}, {Baldry}, {Bremer},
  {Eckert}, {Guennou}, {Horellou}, {Iovino}, {Koulouridis}, {Liske},
  {Maurogordato}, {Menanteau}, {Mohr}, {Owers}, {Poggianti}, {Pompei},
  {Sadibekova}, {Stanford}, {Tuffs}, \& {Willis}}]{fotopoulou2016_xxl6}
{Fotopoulou}, S., {Pacaud}, F., {Paltani}, S., {et~al.} 2016, \aap, 592, A5,
  (XXL Survey, VI)

\bibitem[{{Frank} {et~al.}(1992){Frank}, {King}, \& {Raine}}]{frank1992}
{Frank}, J., {King}, A., \& {Raine}, D. 1992, {Accretion power in
  astrophysics.}

\bibitem[{{Gebhardt} {et~al.}(2000){Gebhardt}, {Bender}, {Bower}, {Dressler},
  {Faber}, {Filippenko}, {Green}, {Grillmair}, {Ho}, {Kormendy}, {Lauer},
  {Magorrian}, {Pinkney}, {Richstone}, \& {Tremaine}}]{gebhardt2000}
{Gebhardt}, K., {Bender}, R., {Bower}, G., {et~al.} 2000, \apjl, 539, L13

\bibitem[{{Godfrey} \& {Shabala}(2013)}]{godfrey2013}
{Godfrey}, L.~E.~H. \& {Shabala}, S.~S. 2013, \apj, 767, 12

\bibitem[{{Godfrey} \& {Shabala}(2016)}]{godfrey2016}
{Godfrey}, L.~E.~H. \& {Shabala}, S.~S. 2016, \mnras, 456, 1172

\bibitem[{{Graham}(2008)}]{graham2008}
{Graham}, A.~W. 2008, \apj, 680, 143

\bibitem[{{G{\"u}rkan} {et~al.}(2014){G{\"u}rkan}, {Hardcastle}, \&
  {Jarvis}}]{gurkan2014}
{G{\"u}rkan}, G., {Hardcastle}, M.~J., \& {Jarvis}, M.~J. 2014, \mnras, 438,
  1149

\bibitem[{{Hardcastle}(2018{\natexlab{a}})}]{hardcastle2018b}
{Hardcastle}, M. 2018{\natexlab{a}}, Nature Astronomy, 2, 273

\bibitem[{{Hardcastle}(2018{\natexlab{b}})}]{hardcastle2018a}
{Hardcastle}, M.~J. 2018{\natexlab{b}}, \mnras, 475, 2768

\bibitem[{{Hardcastle} {et~al.}(2013){Hardcastle}, {Ching}, {Virdee}, {Jarvis},
  {Croom}, {Sadler}, {Mauch}, {Smith}, {Stevens}, {Baes}, {Baldry}, {Brough},
  {Cooray}, {Dariush}, {De Zotti}, {Driver}, {Dunne}, {Dye}, {Eales},
  {Hopwood}, {Liske}, {Maddox}, {Micha{\l}owski}, {Rigby}, {Robotham},
  {Steele}, {Thomas}, \& {Valiante}}]{hardcastle2013a}
{Hardcastle}, M.~J., {Ching}, J.~H.~Y., {Virdee}, J.~S., {et~al.} 2013, \mnras,
  429, 2407

\bibitem[{{Hardcastle} {et~al.}(2006){Hardcastle}, {Evans}, \&
  {Croston}}]{hardcastle2006}
{Hardcastle}, M.~J., {Evans}, D.~A., \& {Croston}, J.~H. 2006, \mnras, 370,
  1893

\bibitem[{{Hardcastle} {et~al.}(2007){Hardcastle}, {Evans}, \&
  {Croston}}]{hardcastle2007}
{Hardcastle}, M.~J., {Evans}, D.~A., \& {Croston}, J.~H. 2007, \mnras, 376,
  1849

\bibitem[{{Hardcastle} {et~al.}(2009){Hardcastle}, {Evans}, \&
  {Croston}}]{hardcastle2009}
{Hardcastle}, M.~J., {Evans}, D.~A., \& {Croston}, J.~H. 2009, \mnras, 396,
  1929

\bibitem[{{Hardcastle} {et~al.}(2018){Hardcastle}, {Williams}, {Best},
  {Croston}, {Duncan}, {Rottgering}, {Sabater}, {Shimwell}, {Tasse},
  {Callingham}, {Cochrane}, {de Gasperin}, {Gurkan}, {Jarvis}, {Mahatma},
  {Miley}, {Mingo}, {Mooney}, {Morabito}, {O'Sullivan}, {Prandoni},
  {Shulevski}, \& {Smith}}]{hardcastle2018c}
{Hardcastle}, M.~J., {Williams}, W.~L., {Best}, P.~N., {et~al.} 2018, arXiv
  e-prints [\eprint[arXiv]{1811.07943}]

\bibitem[{{Heckman} \& {Best}(2014)}]{heckman2014}
{Heckman}, T.~M. \& {Best}, P.~N. 2014, \araa, 52, 589

\bibitem[{{Herbert} {et~al.}(2010){Herbert}, {Jarvis}, {Willott}, {McLure},
  {Mitchell}, {Rawlings}, {Hill}, \& {Dunlop}}]{herbert2010}
{Herbert}, P.~D., {Jarvis}, M.~J., {Willott}, C.~J., {et~al.} 2010, \mnras,
  406, 1841

\bibitem[{{Hickox} {et~al.}(2009){Hickox}, {Jones}, {Forman}, {Murray},
  {Kochanek}, {Eisenstein}, {Jannuzi}, {Dey}, {Brown}, {Stern}, {Eisenhardt},
  {Gorjian}, {Brodwin}, {Narayan}, {Cool}, {Kenter}, {Caldwell}, \&
  {Anderson}}]{hickox2009}
{Hickox}, R.~C., {Jones}, C., {Forman}, W.~R., {et~al.} 2009, \apj, 696, 891

\bibitem[{{Hine} \& {Longair}(1979)}]{hine1979}
{Hine}, R.~G. \& {Longair}, M.~S. 1979, \mnras, 188, 111

\bibitem[{{Hinshaw} {et~al.}(2013){Hinshaw}, {Larson}, {Komatsu}, {Spergel},
  {Bennett}, {Dunkley}, {Nolta}, {Halpern}, {Hill}, {Odegard}, {Page}, {Smith},
  {Weiland}, {Gold}, {Jarosik}, {Kogut}, {Limon}, {Meyer}, {Tucker}, {Wollack},
  \& {Wright}}]{hinshaw2013}
{Hinshaw}, G., {Larson}, D., {Komatsu}, E., {et~al.} 2013, \apjs, 208, 19

\bibitem[{{Hopkins} {et~al.}(2003){Hopkins}, {Miller}, {Nichol}, {Connolly},
  {Bernardi}, {G{\'o}mez}, {Goto}, {Tremonti}, {Brinkmann}, {Ivezi{\'c}}, \&
  {Lamb}}]{hopkins2003}
{Hopkins}, A.~M., {Miller}, C.~J., {Nichol}, R.~C., {et~al.} 2003, \apj, 599,
  971

\bibitem[{{Jackson} \& {Rawlings}(1997)}]{jackson1997}
{Jackson}, N. \& {Rawlings}, S. 1997, \mnras, 286, 241

\bibitem[{{Janssen} {et~al.}(2012){Janssen}, {R{\"o}ttgering}, {Best}, \&
  {Brinchmann}}]{janssen2012}
{Janssen}, R.~M.~J., {R{\"o}ttgering}, H.~J.~A., {Best}, P.~N., \&
  {Brinchmann}, J. 2012, \aap, 541, A62

\bibitem[{{Jarrett} {et~al.}(2011){Jarrett}, {Cohen}, {Masci}, {Wright},
  {Stern}, {Benford}, {Blain}, {Carey}, {Cutri}, {Eisenhardt}, {Lonsdale},
  {Mainzer}, {Marsh}, {Padgett}, {Petty}, {Ressler}, {Skrutskie}, {Stanford},
  {Surace}, {Tsai}, {Wheelock}, \& {Yan}}]{jarrett2011}
{Jarrett}, T.~H., {Cohen}, M., {Masci}, F., {et~al.} 2011, \apj, 735, 112

\bibitem[{{Juneau} {et~al.}(2011){Juneau}, {Dickinson}, {Alexander}, \&
  {Salim}}]{juneau2011}
{Juneau}, S., {Dickinson}, M., {Alexander}, D.~M., \& {Salim}, S. 2011, \apj,
  736, 104

\bibitem[{{Kauffmann} {et~al.}(2003){Kauffmann}, {Heckman}, {Tremonti},
  {Brinchmann}, {Charlot}, {White}, {Ridgway}, {Brinkmann}, {Fukugita}, {Hall},
  {Ivezi{\'c}}, {Richards}, \& {Schneider}}]{kauffmann2003}
{Kauffmann}, G., {Heckman}, T.~M., {Tremonti}, C., {et~al.} 2003, \mnras, 346,
  1055

\bibitem[{{Kewley} {et~al.}(2001){Kewley}, {Dopita}, {Sutherland}, {Heisler},
  \& {Trevena}}]{kewley2001b}
{Kewley}, L.~J., {Dopita}, M.~A., {Sutherland}, R.~S., {Heisler}, C.~A., \&
  {Trevena}, J. 2001, \apj, 556, 121

\bibitem[{{K{\"o}rding} {et~al.}(2008){K{\"o}rding}, {Jester}, \&
  {Fender}}]{kording2008}
{K{\"o}rding}, E.~G., {Jester}, S., \& {Fender}, R. 2008, \mnras, 383, 277

\bibitem[{{Kormendy} \& {Ho}(2013)}]{kormendy2013}
{Kormendy}, J. \& {Ho}, L.~C. 2013, \araa, 51, 511

\bibitem[{{Laigle} {et~al.}(2016){Laigle}, {McCracken}, {Ilbert}, {Hsieh},
  {Davidzon}, {Capak}, {Hasinger}, {Silverman}, {Pichon}, {Coupon}, {Aussel},
  {Le Borgne}, {Caputi}, {Cassata}, {Chang}, {Civano}, {Dunlop}, {Fynbo},
  {Kartaltepe}, {Koekemoer}, {Le F{\`e}vre}, {Le Floc'h}, {Leauthaud}, {Lilly},
  {Lin}, {Marchesi}, {Milvang-Jensen}, {Salvato}, {Sanders}, {Scoville},
  {Smolcic}, {Stockmann}, {Taniguchi}, {Tasca}, {Toft}, {Vaccari}, \&
  {Zabl}}]{laigle2016}
{Laigle}, C., {McCracken}, H.~J., {Ilbert}, O., {et~al.} 2016, \apjs, 224, 24

\bibitem[{{Laing} {et~al.}(1994){Laing}, {Jenkins}, {Wall}, \&
  {Unger}}]{laing1994}
{Laing}, R.~A., {Jenkins}, C.~R., {Wall}, J.~V., \& {Unger}, S.~W. 1994, in
  Astronomical Society of the Pacific Conference Series, Vol.~54, The Physics
  of Active Galaxies, ed. G.~V. {Bicknell}, M.~A. {Dopita}, \& P.~J. {Quinn},
  201

\bibitem[{{Longair}(1966)}]{longair1966}
{Longair}, M.~S. 1966, \mnras, 133, 421

\bibitem[{{Magorrian} {et~al.}(1998){Magorrian}, {Tremaine}, {Richstone},
  {Bender}, {Bower}, {Dressler}, {Faber}, {Gebhardt}, {Green}, {Grillmair},
  {Kormendy}, \& {Lauer}}]{magorrian1998}
{Magorrian}, J., {Tremaine}, S., {Richstone}, D., {et~al.} 1998, \aj, 115, 2285

\bibitem[{{Mateos} {et~al.}(2012){Mateos}, {Alonso-Herrero}, {Carrera},
  {Blain}, {Watson}, {Barcons}, {Braito}, {Severgnini}, {Donley}, \&
  {Stern}}]{mateos2012}
{Mateos}, S., {Alonso-Herrero}, A., {Carrera}, F.~J., {et~al.} 2012, \mnras,
  426, 3271

\bibitem[{{Mauch} \& {Sadler}(2007)}]{mauch2007}
{Mauch}, T. \& {Sadler}, E.~M. 2007, \mnras, 375, 931

\bibitem[{{McNamara} \& {Nulsen}(2007)}]{mcnamara2007}
{McNamara}, B.~R. \& {Nulsen}, P.~E.~J. 2007, \araa, 45, 117

\bibitem[{{McNamara} \& {Nulsen}(2012)}]{mcnamara2012}
{McNamara}, B.~R. \& {Nulsen}, P.~E.~J. 2012, New Journal of Physics, 14,
  055023

\bibitem[{{Merloni} \& {Heinz}(2007)}]{merloni2007}
{Merloni}, A. \& {Heinz}, S. 2007, \mnras, 381, 589

\bibitem[{{Merloni} \& {Heinz}(2008)}]{merloni2008}
{Merloni}, A. \& {Heinz}, S. 2008, \mnras, 388, 1011

\bibitem[{{Mingo} {et~al.}(2014){Mingo}, {Hardcastle}, {Croston}, {Dicken},
  {Evans}, {Morganti}, \& {Tadhunter}}]{mingo2014}
{Mingo}, B., {Hardcastle}, M.~J., {Croston}, J.~H., {et~al.} 2014, \mnras, 440,
  269

\bibitem[{{Miraghaei} \& {Best}(2017)}]{miraghaei2017}
{Miraghaei}, H. \& {Best}, P.~N. 2017, \mnras, 466, 4346

\bibitem[{{Mocz} {et~al.}(2013){Mocz}, {Fabian}, \& {Blundell}}]{mocz2013}
{Mocz}, P., {Fabian}, A.~C., \& {Blundell}, K.~M. 2013, \mnras, 432, 3381

\bibitem[{{Narayan} \& {Yi}(1994)}]{narayan1994}
{Narayan}, R. \& {Yi}, I. 1994, \apjl, 428, L13

\bibitem[{{Narayan} \& {Yi}(1995{\natexlab{a}})}]{narayan1995a}
{Narayan}, R. \& {Yi}, I. 1995{\natexlab{a}}, \apj, 444, 231

\bibitem[{{Narayan} \& {Yi}(1995{\natexlab{b}})}]{narayan1995b}
{Narayan}, R. \& {Yi}, I. 1995{\natexlab{b}}, \apj, 452, 710

\bibitem[{{Newville} {et~al.}(2016){Newville}, {Stensitzki}, {Allen}, {Rawlik},
  {Ingargiola}, \& {Nelson}}]{newville2016}
{Newville}, M., {Stensitzki}, T., {Allen}, D.~B., {et~al.} 2016, {Lmfit:
  Non-Linear Least-Square Minimization and Curve-Fitting for Python},
  Astrophysics Source Code Library

\bibitem[{{Norris} {et~al.}(2011){Norris}, {Hopkins}, {Afonso}, {Brown},
  {Condon}, {Dunne}, {Feain}, {Hollow}, {Jarvis}, {Johnston-Hollitt}, {Lenc},
  {Middelberg}, {Padovani}, {Prandoni}, {Rudnick}, {Seymour}, {Umana},
  {Andernach}, {Alexander}, {Appleton}, {Bacon}, {Banfield}, {Becker}, {Brown},
  {Ciliegi}, {Jackson}, {Eales}, {Edge}, {Gaensler}, {Giovannini}, {Hales},
  {Hancock}, {Huynh}, {Ibar}, {Ivison}, {Kennicutt}, {Kimball}, {Koekemoer},
  {Koribalski}, {L{\'o}pez-S{\'a}nchez}, {Mao}, {Murphy}, {Messias},
  {Pimbblet}, {Raccanelli}, {Randall}, {Reiprich}, {Roseboom},
  {R{\"o}ttgering}, {Saikia}, {Sharp}, {Slee}, {Smail}, {Thompson}, {Urquhart},
  {Wall}, \& {Zhao}}]{norris2011}
{Norris}, R.~P., {Hopkins}, A.~M., {Afonso}, J., {et~al.} 2011, \pasa, 28, 215

\bibitem[{{O'Sullivan} {et~al.}(2011){O'Sullivan}, {Giacintucci}, {David},
  {Gitti}, {Vrtilek}, {Raychaudhury}, \& {Ponman}}]{osullivan2011}
{O'Sullivan}, E., {Giacintucci}, S., {David}, L.~P., {et~al.} 2011, \apj, 735,
  11

\bibitem[{{Pierre} {et~al.}(2016){Pierre}, {Pacaud}, {Adami}, {Alis},
  {Altieri}, {Baran}, {Benoist}, {Birkinshaw}, {Bongiorno}, {Bremer}, {Brusa},
  {Butler}, {Ciliegi}, {Chiappetti}, {Clerc}, {Corasaniti}, {Coupon}, {De
  Breuck}, {Democles}, {Desai}, {Delhaize}, {Devriendt}, {Dubois}, {Eckert},
  {Elyiv}, {Ettori}, {Evrard}, {Faccioli}, {Farahi}, {Ferrari}, {Finet},
  {Fotopoulou}, {Fourmanoit}, {Gandhi}, {Gastaldello}, {Gastaud},
  {Georgantopoulos}, {Giles}, {Guennou}, {Guglielmo}, {Horellou}, {Husband},
  {Huynh}, {Iovino}, {Kilbinger}, {Koulouridis}, {Lavoie}, {Le Brun}, {Le
  Fevre}, {Lidman}, {Lieu}, {Lin}, {Mantz}, {Maughan}, {Maurogordato},
  {McCarthy}, {McGee}, {Melin}, {Melnyk}, {Menanteau}, {Novak}, {Paltani},
  {Plionis}, {Poggianti}, {Pomarede}, {Pompei}, {Ponman}, {Ramos-Ceja},
  {Ranalli}, {Rapetti}, {Raychaudury}, {Reiprich}, {Rottgering}, {Rozo},
  {Rykoff}, {Sadibekova}, {Santos}, {Sauvageot}, {Schimd}, {Sereno}, {Smith},
  {Smol{\v c}i{\'c}}, {Snowden}, {Spergel}, {Stanford}, {Surdej}, {Valageas},
  {Valotti}, {Valtchanov}, {Vignali}, {Willis}, \& {Ziparo}}]{pierre2016_xxl1}
{Pierre}, M., {Pacaud}, F., {Adami}, C., {et~al.} 2016, \aap, 592, A1, (XXL
  Survey, I)

\bibitem[{{Pracy} {et~al.}(2016){Pracy}, {Ching}, {Sadler}, {Croom}, {Baldry},
  {Bland-Hawthorn}, {Brough}, {Brown}, {Couch}, {Davis}, {Drinkwater},
  {Hopkins}, {Jarvis}, {Jelliffe}, {Jurek}, {Loveday}, {Pimbblet}, {Prescott},
  {Wisnioski}, \& {Woods}}]{pracy2016}
{Pracy}, M.~B., {Ching}, J.~H.~Y., {Sadler}, E.~M., {et~al.} 2016, \mnras, 460,
  2

\bibitem[{{Raouf} {et~al.}(2017){Raouf}, {Shabala}, {Croton}, {Khosroshahi}, \&
  {Bernyk}}]{raouf2017}
{Raouf}, M., {Shabala}, S.~S., {Croton}, D.~J., {Khosroshahi}, H.~G., \&
  {Bernyk}, M. 2017, \mnras, 471, 658

\bibitem[{{Romero} {et~al.}(2017){Romero}, {Boettcher}, {Markoff}, \&
  {Tavecchio}}]{romero2017}
{Romero}, G.~E., {Boettcher}, M., {Markoff}, S., \& {Tavecchio}, F. 2017, \ssr,
  207, 5

\bibitem[{{Sadler} {et~al.}(2002){Sadler}, {Jackson}, {Cannon}, {McIntyre},
  {Murphy}, {Bland-Hawthorn}, {Bridges}, {Cole}, {Colless}, {Collins}, {Couch},
  {Dalton}, {De Propris}, {Driver}, {Efstathiou}, {Ellis}, {Frenk},
  {Glazebrook}, {Lahav}, {Lewis}, {Lumsden}, {Maddox}, {Madgwick}, {Norberg},
  {Peacock}, {Peterson}, {Sutherland}, \& {Taylor}}]{sadler2002}
{Sadler}, E.~M., {Jackson}, C.~A., {Cannon}, R.~D., {et~al.} 2002, \mnras, 329,
  227

\bibitem[{{Saunders} {et~al.}(1990){Saunders}, {Rowan-Robinson}, {Lawrence},
  {Efstathiou}, {Kaiser}, {Ellis}, \& {Frenk}}]{saunders1990}
{Saunders}, W., {Rowan-Robinson}, M., {Lawrence}, A., {et~al.} 1990, \mnras,
  242, 318

\bibitem[{{Schinnerer} {et~al.}(2004){Schinnerer}, {Carilli}, {Scoville},
  {Bondi}, {Ciliegi}, {Vettolani}, {Le F{\`e}vre}, {Koekemoer}, {Bertoldi}, \&
  {Impey}}]{schinnerer2004}
{Schinnerer}, E., {Carilli}, C.~L., {Scoville}, N.~Z., {et~al.} 2004, \aj, 128,
  1974

\bibitem[{{Schinnerer} {et~al.}(2010){Schinnerer}, {Sargent}, {Bondi}, {Smol{\v
  c}i{\'c}}, {Datta}, {Carilli}, {Bertoldi}, {Blain}, {Ciliegi}, {Koekemoer},
  \& {Scoville}}]{schinnerer2010}
{Schinnerer}, E., {Sargent}, M.~T., {Bondi}, M., {et~al.} 2010, \apjs, 188, 384

\bibitem[{{Schinnerer} {et~al.}(2007){Schinnerer}, {Smol{\v c}i{\'c}},
  {Carilli}, {Bondi}, {Ciliegi}, {Jahnke}, {Scoville}, {Aussel}, {Bertoldi},
  {Blain}, {Impey}, {Koekemoer}, {Le Fevre}, \& {Urry}}]{schinnerer2007}
{Schinnerer}, E., {Smol{\v c}i{\'c}}, V., {Carilli}, C.~L., {et~al.} 2007,
  \apjs, 172, 46

\bibitem[{{Schmidt}(1968)}]{schmidt1968}
{Schmidt}, M. 1968, \apj, 151, 393

\bibitem[{{Shabala} \& {Godfrey}(2013)}]{shabala2013}
{Shabala}, S.~S. \& {Godfrey}, L.~E.~H. 2013, \apj, 769, 129

\bibitem[{{Shankar} {et~al.}(2009){Shankar}, {Weinberg}, \&
  {Miralda-Escud{\'e}}}]{shankar2009}
{Shankar}, F., {Weinberg}, D.~H., \& {Miralda-Escud{\'e}}, J. 2009, \apj, 690,
  20

\bibitem[{{Smol{\v c}i{\'c}} {et~al.}(2016){Smol{\v c}i{\'c}}, {Delhaize},
  {Huynh}, {Bondi}, {Ciliegi}, {Novak}, {Baran}, {Birkinshaw}, {Bremer},
  {Chiappetti}, {Ferrari}, {Fotopoulou}, {Horellou}, {McGee}, {Pacaud},
  {Pierre}, {Raychaudhury}, {R{\"o}ttgering}, \& {Vignali}}]{smolcic2016_xxl11}
{Smol{\v c}i{\'c}}, V., {Delhaize}, J., {Huynh}, M., {et~al.} 2016, \aap, 592,
  A10, (XXL Survey, XI)

\bibitem[{{Smol{\v c}i{\'c}} {et~al.}(2017{\natexlab{a}}){Smol{\v c}i{\'c}},
  {Novak}, {Bondi}, {Ciliegi}, {Mooley}, {Schinnerer}, {Zamorani}, {Navarrete},
  {Bourke}, {Karim}, {Vardoulaki}, {Leslie}, {Delhaize}, {Carilli}, {Myers},
  {Baran}, {Delvecchio}, {Miettinen}, {Banfield}, {Balokovi{\'c}}, {Bertoldi},
  {Capak}, {Frail}, {Hallinan}, {Hao}, {Herrera Ruiz}, {Horesh}, {Ilbert},
  {Intema}, {Jeli{\'c}}, {Kl{\"o}ckner}, {Krpan}, {Kulkarni}, {McCracken},
  {Laigle}, {Middleberg}, {Murphy}, {Sargent}, {Scoville}, \&
  {Sheth}}]{smolcic2017c}
{Smol{\v c}i{\'c}}, V., {Novak}, M., {Bondi}, M., {et~al.} 2017{\natexlab{a}},
  \aap, 602, A1

\bibitem[{{Smol{\v c}i{\'c}} {et~al.}(2017{\natexlab{b}}){Smol{\v c}i{\'c}},
  {Novak}, {Delvecchio}, {Ceraj}, {Bondi}, {Delhaize}, {Marchesi}, {Murphy},
  {Schinnerer}, {Vardoulaki}, \& {Zamorani}}]{smolcic2017b}
{Smol{\v c}i{\'c}}, V., {Novak}, M., {Delvecchio}, I., {et~al.}
  2017{\natexlab{b}}, \aap, 602, A6

\bibitem[{{Smol{\v c}i{\'c}} {et~al.}(2009){Smol{\v c}i{\'c}}, {Zamorani},
  {Schinnerer}, {Bardelli}, {Bondi}, {B{\^i}rzan}, {Carilli}, {Ciliegi},
  {Elvis}, {Impey}, {Koekemoer}, {Merloni}, {Paglione}, {Salvato}, {Scodeggio},
  {Scoville}, \& {Trump}}]{smolcic2009a}
{Smol{\v c}i{\'c}}, V., {Zamorani}, G., {Schinnerer}, E., {et~al.} 2009, \apj,
  696, 24

\bibitem[{{Tasse} {et~al.}(2008){Tasse}, {Best}, {R{\"o}ttgering}, \& {Le
  Borgne}}]{tasse2008}
{Tasse}, C., {Best}, P.~N., {R{\"o}ttgering}, H., \& {Le Borgne}, D. 2008,
  \aap, 490, 893

\bibitem[{{Trump} {et~al.}(2009){Trump}, {Impey}, {Kelly}, {Elvis}, {Merloni},
  {Bongiorno}, {Gabor}, {Hao}, {McCarthy}, {Huchra}, {Brusa}, {Cappelluti},
  {Koekemoer}, {Nagao}, {Salvato}, \& {Scoville}}]{trump2009b}
{Trump}, J.~R., {Impey}, C.~D., {Kelly}, B.~C., {et~al.} 2009, \apj, 700, 49

\bibitem[{{Turner} {et~al.}(2018){Turner}, {Rogers}, {Shabala}, \&
  {Krause}}]{turner2018a}
{Turner}, R.~J., {Rogers}, J.~G., {Shabala}, S.~S., \& {Krause}, M.~G.~H. 2018,
  \mnras, 473, 4179

\bibitem[{{Urry} \& {Padovani}(1995)}]{urry1995}
{Urry}, C.~M. \& {Padovani}, P. 1995, \pasp, 107, 803

\bibitem[{{Williams} {et~al.}(2018){Williams}, {Calistro Rivera}, {Best},
  {Hardcastle}, {R{\"o}ttgering}, {Duncan}, {de Gasperin}, {Jarvis}, {Miley},
  {Mahony}, {Morabito}, {Nisbet}, {Prandoni}, {Smith}, {Tasse}, \&
  {White}}]{williams2018}
{Williams}, W.~L., {Calistro Rivera}, G., {Best}, P.~N., {et~al.} 2018, \mnras,
  475, 3429

\bibitem[{{Willott} {et~al.}(1999){Willott}, {Rawlings}, {Blundell}, \&
  {Lacy}}]{willott1999}
{Willott}, C.~J., {Rawlings}, S., {Blundell}, K.~M., \& {Lacy}, M. 1999,
  \mnras, 309, 1017

\bibitem[{{Willott} {et~al.}(2001){Willott}, {Rawlings}, {Blundell}, {Lacy}, \&
  {Eales}}]{willott2001}
{Willott}, C.~J., {Rawlings}, S., {Blundell}, K.~M., {Lacy}, M., \& {Eales},
  S.~A. 2001, \mnras, 322, 536

\bibitem[{{Xue} {et~al.}(2011){Xue}, {Luo}, {Brandt}, {Bauer}, {Lehmer},
  {Broos}, {Schneider}, {Alexander}, {Brusa}, {Comastri}, {Fabian}, {Gilli},
  {Hasinger}, {Hornschemeier}, {Koekemoer}, {Liu}, {Mainieri}, {Paolillo},
  {Rafferty}, {Rosati}, {Shemmer}, {Silverman}, {Smail}, {Tozzi}, \&
  {Vignali}}]{xue2011}
{Xue}, Y.~Q., {Luo}, B., {Brandt}, W.~N., {et~al.} 2011, \apjs, 195, 10

\end{thebibliography}


\appendix

\section{Comparison of kinetic luminosity scaling relations}
\label{sec:appendix_scaling_relations}

A number of scaling relations between monochromatic radio luminosity (e.g. $L_{\rm{1.4GHz}}$) and kinetic luminosity ($L_{\rm{kin}}$) have been published in the literature \citep{willott1999,merloni2007,cavagnolo2010,osullivan2011,daly2012,godfrey2016}.  All these log($L_{\rm{kin}}$) versus log($L_{\rm{1.4GHz}}$) relations have similar slopes and intercepts (as demonstrated by \citealp{smolcic2017b} in their Appendix A), but suffer from very large uncertainties due to the small sample sizes (tens of galaxies) used to produce the relations.  Furthermore, \cite{shabala2013} and \cite{godfrey2016} have pointed out that these relations are distance dependent (i.e. are affected by Malmquist bias).  This dependence arises from the necessary range in distance required to probe a large range of radio and X-ray luminosity.

More specifically, \cite{shabala2013} showed that any derived relation between radio-luminosity and jet kinetic power depends sensitively on sample properties -- in particular the size-luminosity correlation inherent in the sample.  Their results indicate that accurate estimates of the integrated kinetic power output of AGNs can only be obtained if a measure of radio source ages, such as size or spectral index, is used in addition to their radio luminosities. The sole use of radio luminosity as a proxy for jet power underpredicts the jet powers for the largest sources and overpredicts the jet powers for the smallest sources.  They conclude that adopting a simple scaling relation between radio luminosity and jet power that does not include a measure of source size or age results in significant errors in jet power estimates.  Consequently, incorrect estimates of AGN jet power will result in mismodelling of AGN feedback processes.   Accurate jet power measurements are required to test whether the AGN heating and gas cooling rates are indeed balanced.  Accordingly, they determined a new expression for jet power that accounts for the source size (see their Equation 8).

In addition, \cite{godfrey2016} found that in a sample of FRI X-ray cavity systems, after accounting for the mutual distance dependence, the jet power and radio luminosity are only weakly correlated, with slope $\beta_L \approx$ 0.3, which is significantly flatter than the slopes of the other scaling relations ($\sim$0.6-0.8).  This flat regression slope implies that a greater amount of mechanical energy is available from lower luminosity radio galaxies than previously thought. This tentative result has strong implications for studies of radio mode feedback because low-luminosity radio galaxies typically deposit energy at smaller radii, and therefore they may be more effective at depositing more energy in the regions where it is most needed to offset cooling.  They also found that in previously used samples of high-power FRII sources, no evidence for an intrinsic correlation is present when the effect of distance is accounted for.  They conclude that the scaling relations remain poorly constrained through observations.

\begin{figure}
        \includegraphics[width=\columnwidth]{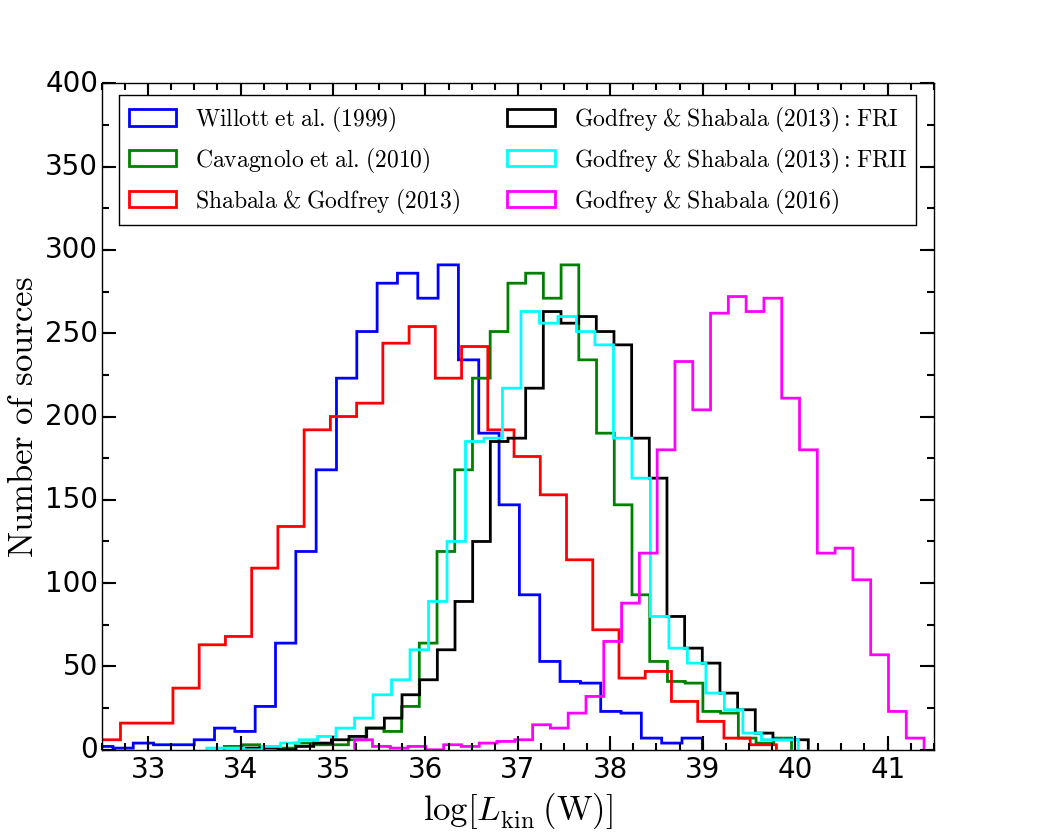}
    \caption[Distribution of kinetic luminosities of all radio loud AGN in XXL-S using different scaling relations.]{Distribution of kinetic luminosities of all RL AGN in XXL-S using different scaling relations.  The relations from \cite{godfrey2013} for FRI and FRII sources are similar to the relation from \cite{cavagnolo2010} (which is similar to \citealp{merloni2007} and \citealp{osullivan2011}), but this apparent agreement is due to the similar distance dependence of jet-power measurement techniques used for FRI and FRII radio galaxies in these samples.}
    \label{fig:L_kin_dists}
\end{figure}

Figure \ref{fig:L_kin_dists} shows the distribution of kinetic luminosities for all RL AGN in XXL-S using various scaling relations between radio and kinetic luminosity, and Table \ref{tab:scaling_relation_equations} shows the equations corresponding to each scaling relation (converted into the form in which a monochromatic radio luminosity is an input variable).  It demonstrates that the scaling relation derived from each sample is highly dependent upon sample properties and how those properties correlate with distance to each radio source.  The reason that the results from \cite{godfrey2013} and \cite{cavagnolo2010} (which is similar to \citealp{merloni2007} and \citealp{osullivan2011}) are so similar is because the jet-power measurement techniques used for FRI and FRII radio galaxies in these samples have similar distance dependences.  For simplicity, the \cite{shabala2013} distribution was constructed by assuming each ATCA XXL-S radio source is unresolved, for which an upper limit to each source's size was used (i.e. no larger than 5.39$''$, the major axis of the ATCA beam, across).  The fact that this distribution not only produces a peak in the kinetic luminosity distribution that is over an order of magnitude less than the peak in the \cite{cavagnolo2010} and \cite{godfrey2013} distributions, but is broader as well (i.e. incorporates a wider range in kinetic luminosity), illustrates the importance of taking into account the source size when computing the kinetic luminosity associated with radio jets.  The \cite{willott1999} relation takes into account distance effects, and this is reflected in the fact that the peak of that distribution is very close to the peak in the \cite{shabala2013} distribution.  Finally, the \cite{godfrey2016} relation clearly produced kinetic luminosities much higher than the other relations.  This is probably due to the fact that it is based on a sample that extends to only $z \leq 0.23$ and the slope of the relation is shallow ($\sim$0.3), which causes galaxies of lower radio luminosity to have higher kinetic luminosities than when using the other relations.  Since most of the XXL-S sample is at $z > 0.23$, the \cite{godfrey2016} relation may not be applicable to the full XXL-S sample.  Regardless of the relative credibility of each scaling relation, Figure \ref{fig:L_kin_dists} demonstrates that there are very large differences in the kinetic luminosity distributions.

\begin{table*}
\centering
\caption[Equations of the scaling relations discussed in Appendix \ref{sec:appendix_scaling_relations}.]{Equations of the scaling relations discussed in Appendix \ref{sec:appendix_scaling_relations}.  All have been converted into the form in which a monochromatic radio luminosity is an input variable.  In all cases, $L_{\rm{kin}}$ has units of W and the monochromatic radio luminosities ($L_{\rm{1.4GHz}}$, $L_{\rm{151MHz}}$) have units of W Hz$^{-1}$.  For the \cite{willott1999} relation (see \citealp{heckman2014}), a value of $f_W$=4 was chosen.  For the \cite{cavagnolo2010} relation, $\nu = 1.4 \times 10^9$ Hz.  For the \cite{shabala2013} and \cite{godfrey2013} relations, $L_{\rm{151MHz}}$ was obtained by converting the $S_{\rm{1.8GHz}}$ values (the XXL-S flux densities measured at the effective frequency) into $S_{\rm{151MHz}}$ assuming the spectral index assigned to each XXL-S radio source (see Appendix A in XXL Paper XXXI), $g$ = 2 is the normalisation factor, and $D$ is the source size in kpc.  For the \cite{godfrey2016} relation, $d_L$ is the luminosity distance in Mpc.}
\begin{adjustbox}{width=\textwidth}
\begin{tabular}{c c}
Reference & Equation\\
\hline
\hline
\cite{willott1999} & log($L_{\rm{kin}}$) = 0.86 $\cdot$ log($L_{\rm{1.4GHz}}$) + 14.08 + 1.5 log($f_W$) \\
\cite{cavagnolo2010} & $L_{\rm{kin}} = (10^{35} \textrm{ W}) \cdot 10^{0.75[\textrm{log}(\nu L_{\textrm{1.4GHz}})] - 22.84}$\\
\cite{shabala2013} & $L_{\rm{kin}}$ = (10$^{36}$ W) $\cdot$ 1.5 $\left(\frac{L_{\rm{151MHz}}}{10^{27} \ \rm{W} \ \rm{Hz}^{-1}}\right)^{0.8}$ $\cdot$ (1 + $z$) $\cdot$ $\left(\frac{D}{\rm{kpc}}\right)^{0.58}$ \\
\cite{godfrey2013} (FRI) & $L_{\rm{kin}}$ = (5 $\cdot$ 10$^{37}$ W) $\left(\frac{L_{\rm{151MHz}}}{10^{25} \ \rm{W} \ \rm{Hz}^{-1}}\right)^{0.64}$ \\
\cite{godfrey2013} (FRII) & $L_{\rm{kin}}$ = $g$ (1.5 $\cdot$ 10$^{37}$ W) $\left(\frac{L_{\rm{151MHz}}}{10^{25} \ \rm{W} \ \rm{Hz}^{-1}}\right)^{0.67}$ \\
\cite{godfrey2016} & log($L_{\rm{kin}}$) = 36.56 + 0.27 log$\left(\frac{L_{\textrm{1.4GHz}}}{10^{24} \ \rm{W} \ \rm{Hz}^{-1}}\right)$ + 1.4 log$\left(\frac{d_L}{100 \ \rm{Mpc}}\right)$ + 0.33 \\
\hline
\end{tabular}
\label{tab:scaling_relation_equations}
\end{adjustbox}
\end{table*}

Expanding on this, \cite{godfrey2016} found that the uncertainty regarding radio lobe dynamics in FRI and FRII sources provides some uncertainty in the predicted scaling relations. Their theoretical modelling showed that $\beta_L$ is expected to be significantly lower in samples of FRI radio galaxies than it is for FRIIs, due to the differing dynamics for these two classes of radio source. For FRI X-ray cavity systems the model predicts $\beta_L \gtrsim$ 0.5, in contrast to the $\beta_L \gtrsim$ 0.8 slope for FRII radio galaxies.  These theoretical results are more consistent with the parameters of the other scaling relations than their own empirical results, and thus considerable uncertainty in the relationship between radio luminosity and kinetic luminosity remains.  Future radio and X-ray observations will be able to construct larger samples of FRI and FRII sources and of LERGs and HERGs, which will determine the driving factors that set up the relationship between radio luminosity and kinetic luminosity.

\clearpage

\section{Effect of rebinning the local RL HERG RLF}
\label{sec:appendix_rebinning_local_RL_HERG_RLF}

\begin{figure}
        \includegraphics[width=\columnwidth]{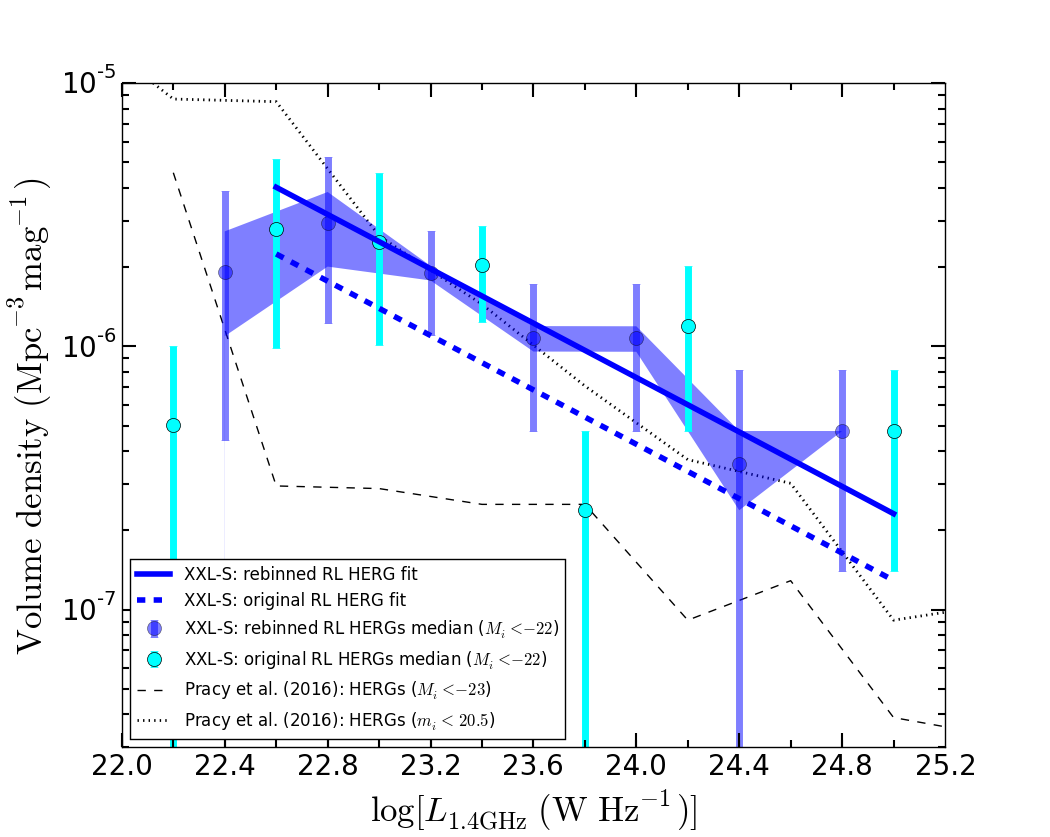}
    \caption[Rebinned local XXL-S RL HERG RLF with $M_i < -22$.]{Rebinned local XXL-S RL HERG RLF with $M_i < -22$ (blue shaded region) and the corresponding fit (solid blue line).  The original local RL HERG RLF with $M_i < -22$ is shown as the cyan circles and its fit is the dashed blue line.  The rebinning caused an increase in the normalisation of the fit (log[$\Phi^*_0$] = $-7.552$).  However, this did not significantly affect the measurement of the evolution of $\Omega_{\rm{kin}}$ for RL HERGs given the large uncertainties in the \cite{cavagnolo2010} scaling relation.  For comparison, the Pracy16 HERG RLFs with $m_i < 20.5$ (including all sources) and $M_i < -23$ are shown.}
    \label{fig:RL_HERG_local_RLF_rebinned}
\end{figure}

Rebinning the luminosity bins of the local RL HERG RLF presented in Section \ref{sec:final_RLFs} was investigated as a means of improving the model fit. This was done by shifting each luminosity bin by 0.2 dex (half of the bin width).  The result was a smoother RLF with no gaps (i.e. no bins with 0 sources) in the range 22.6 < log[$L_{\rm{1.4GHz}}$ (W Hz$^{-1}$)] < 25.0, but it also had a higher normalisation. Figure \ref{fig:RL_HERG_local_RLF_rebinned} shows the rebinned local RL HERG RLF and its model fit (with all the same parameters except log[$\Phi^*_0$] = $-7.552$), and Table \ref{tab:RL_HERG_local_RLF_rebinned} displays the RLF data.

The evolution measured with the new local fit was $K_{\rm{D}} = 0.838 \pm 0.098$ and $K_{\rm{L}} = 1.482 \pm 0.160$. These $K_{\rm{D}}$ and $K_{\rm{L}}$ values are very different to the original $K_{\rm{D}}$ = 1.812 and $K_{\rm{L}}$ = 3.186 values shown in Table \ref{tab:best_fit_PDE_PLE_params}, but the latter values are still used in this paper for the following reasons: 
\begin{enumerate}
\item{The $K_{\rm{D}}$ = 0.838 and $K_{\rm{L}}$ = 1.482 values ultimately result in little difference to the evolution of $\Omega_{\rm{kin}}$ for the RL HERGs. The lower $K_{\rm{D}}$ value caused the uncertainty extrema of the RL HERG $\Omega_{\rm{kin}}$ curve derived using the \cite{cavagnolo2010} scaling relation (Equation \ref{eq:kin_lum_def_unc}) to increase by $\sim$0.2-0.3 dex (a factor of $\sim$1.6-2.0) at $z = 0$ and $\sim$0.1 dex (a factor of $\sim$1.25) at $z = 1.3$ (see Figure \ref{fig:XXL-S_int_kin_lum_density_RL_HERG_LERG_vs_z_sims_comp}). This increase is within the large uncertainty limits of the \cite{cavagnolo2010} scaling relation ($\sim$0.35-0.4 dex).}
\item{Using the original local RL HERG RLF fit (in Table \ref{tab:best_fit_params_local_RLF_HERG_LERG}) allows a more direct comparison to other samples of HERGs selected at GHz frequencies (e.g. Pracy16 and \citealp{ceraj2018}) due to a similar local RLF.}
\item{It is not entirely clear that the RL HERG population is free from contamination, and there is no way to discover if the classification method or the deeper radio and optical data are primarily responsible for the difference between the XXL-S RL HERG RLF and other HERG RLFs until spectra are taken of the remaining XXL-S radio sources without spectra.}
\item{Future radio surveys, such as EMU \citep{norris2011}, will be able to better constrain the local HERG RLF with much wider areas, deeper radio data, and more extensive multi-wavelength data.}
\end{enumerate}
Given all of the above, the parameters given in Table \ref{tab:best_fit_params_local_RLF_HERG_LERG} are used for the final local RL HERG RLF fit, and its corresponding evolution (Table \ref{tab:best_fit_PDE_PLE_params}) can be considered a measurement of the upper limit to the evolution of RL HERGs in XXL-S.

\begin{table}[h]
\centering
\caption[Data for the rebinned local XXL-S RL HERG RLF with $M_i < -22$.]{Data for the rebinned local XXL-S RL HERG RLF with $M_i < -22$, shown in Figure \ref{fig:RL_HERG_local_RLF_rebinned}. See the caption for Table \ref{tab:rlf_z_bin_1_data} for an explanation of the columns.}
\begin{tabular}{c c c}
log($L_{\rm{1.4GHz}}$) & $N$ & log($\Phi$)\\
(W Hz$^{-1}$) & & (mag$^{-1}$ Mpc$^{-3}$)\\
\hline
\hline
22.4 & 4.5 & -5.72$^{+0.31}_{-0.64}$\\
22.8 & 9.0 & -5.53$^{+0.25}_{-0.38}$\\
23.2 & 7.5 & -5.72$^{+0.16}_{-0.24}$\\
23.6 & 4.5 & -5.97$^{+0.21}_{-0.35}$\\
24.0 & 4.5 & -5.97$^{+0.21}_{-0.35}$\\
24.4 & 1.5 & -6.45$^{+0.36}_{-\infty}$\\
24.8 & 2.0 & -6.32$^{+0.23}_{-0.53}$\\
\hline
\end{tabular}
\label{tab:RL_HERG_local_RLF_rebinned}
\end{table}


\label{lastpage}
\end{document}